\documentclass[preprint, aps, prd]{revtex4-1}

\usepackage[dvips]{graphicx}
\usepackage{amssymb}
\usepackage{amssymb,amsmath}
\usepackage{graphicx}
\usepackage{dsfont}
\usepackage{caption}
\usepackage{subcaption}
\usepackage{mathtools}
\usepackage{verbatim}
\usepackage{graphicx}
\usepackage{multirow}
\usepackage[outline]{contour}
\usepackage{xcolor,colortbl}
\makeatother
\begin{document}

\title{Five-branes wrapped on topological disks from 7D $N=2$ gauged supergravity}

\author{Parinya Karndumri} \author{Patharadanai Nuchino} \email[REVTeX Support:
]{parinya.ka@hotmail.com and danai.nuchino@hotmail.com} 
\affiliation{String Theory and
Supergravity Group, Department of Physics, Faculty of Science,
Chulalongkorn University, 254 Phayathai Road, Pathumwan, Bangkok
10330, Thailand}

\date{\today}
\begin{abstract}
We study supersymmetric $AdS_5\times \Sigma$ solutions, for $\Sigma$ being a topological disk with non-trivial $U(1)$ holonomy at the boundary or ``half-spindle", in seven-dimensional $N=2$ gauged supergravity coupled to three vector multiplets. We consider compact and non-compact gauge groups, $SO(4-p,p)$, for $p=0,1,2$, and find a number of $AdS_5\times \Sigma$ solutions with $SO(2)\times SO(2)$ and $SO(2)_{\text{diag}}$ residual symmetry from $SO(4)$ and $SO(2,2)$ gauge groups. We also find an $SO(2)_R\subset SO(3)_R\sim SU(2)_R$ symmetric solution which can be regarded as a solution of pure $N=2$ gauged supergravity with $SO(3)_R$ gauge group and all the fields from vector multiplets vanishing. The solutions preserve $\frac{1}{2}$ of the original $N=2$ supersymmetry and could be interpreted as supergravity duals of $N=1$ superconformal field theories in four dimensions. In particular, some of these solutions can be embedded in ten or eleven dimensions in which a description in terms of five-branes wrapped on a topological disk can be given.   
\end{abstract}
\maketitle

%%%%%%%%%%%%%%%%%%%%%%%%%%%%%%%%%%%%%%%%%%%%%%%%%%%%%%%%%%%%%%%%%%%%%%%%%%%%%%%%%%%%%%%%%%%%%%%%%%%%%%%%%%%%%%%%%%%%%%%%%%%%%%%%%%%%%%%%%
\section{Introduction}
The AdS/CFT correspondence \cite{maldacena,Gubser_AdS_CFT,Witten_AdS_CFT} leads to holographic descriptions of strongly coupled superconformal field theories (SCFTs). Various aspects of these SCFTs including their non-conformal phases can be studied by the corresponding dual gravity solutions in string/M-theory or, at low energy, supergravity theories in ten or eleven dimensions. In this framework, the conformal field theories can be considered as world-volume theories on the branes in the near horizon limit. 
\\
\indent A class of solutions that describes branes wrapping on particular manifolds is of particular interest since these can lead to supersymmetric field theories in lower dimensions arising from world-volume theories of wrapped branes. These configurations describe RG flows across dimensions from higher-dimensional SCFTs to lower-dimensional ones and provide a useful holographic description of the less-known higher-dimensional SCFTs such as $N=(2,0), (1,0)$ SCFTs in six dimensions via four-dimensional SCFTs of which many aspects are better understood. 

At low-energy, these wrapped branes can be described by supersymmetric $AdS_m\times M^n$ solutions of gauged supergravity in $m+n$ dimensions \cite{Maldacena_Twist}. $M^n$ is an $n$-dimensional compact manifold with constant curvature on which the  wrapped $(m+n-2)$-branes leading to an $(m-1)$-dimensional SCFT from a compactification of the dual $(m+n-1)$-dimensional SCFT on $M^n$. The corresponding supergravity solutions preserve some amount of the original supersymmetry by means of a topological twist \cite{Witten_twist}, implemented by turning on some gauge fields to cancel the spin connections on the compact manifold $M^n$. A large number of these solutions have previously been found in various dimensions, see \cite{flow_across_Gauntlett1, flow_across_Gauntlett2, flow_across_Gauntlett3,Kim_AdS_factor,Cucu_AdSD-2,BB,Wraped_D3,3D_CFT_from_LS_point,flow_acrossD_bobev,BBC,
4D_SCFT_from_M5,Wraped_M5,N3_AdS2,flow_across_Betti,AdS2_trisasakian,7D_twist,6D_twist,5D_N4_flow,5Dtwist,BH_microstate_6D2, Minwoo_6D_BH2,Calos_6D_flow1, AdS3_7D_N2, 7D_Max_twist}, for an incomplete list.
\\
\indent Recently, new classes of $AdS\times \Sigma$ solutions, in which unbroken supersymmetry is not realized by a topological twist, have been found for $\Sigma$ being a two-dimensional space with non-constant curvature. These solutions describe supersymmetric branes wrapped on a spindle, which is topologically a two-sphere with orbifold singularities at the poles, \cite{Ferro_D3, Hosseini_D3, Boido_D3,Ferrero_M2, Cassani_M2,Ferrero_M5,D4_spindle}, see also \cite{Gaunlett_super_spindle} for a more recent result, or on a topological disk with non-trivial $U(1)$ holonomy on the boundary or ``half-spindle"  \cite{Bah_AD, Bah_M5,Suh_D3, Suh_D4, Suh_M2,D3_disk}. These lead to new supersymmetric AdS geometries from gauged supergravities which are dual to lower-dimensional SCFTs arising from compactifications of higher-dimensional SCFTs on a spindle or a half-spindle. In particular, supersymmetric $AdS_5\times\Sigma$ solutions from the $U(1)^2$ truncation of the maximal $SO(5)$ gauged supergravity in seven dimensions obtained in \cite{Bah_AD, Bah_M5} have been shown to be dual to four-dimensional $N=2$ SCFTs of Argyres-Douglas (AD) type \cite{AD_origin}. Furthermore, it should be remarked that both the spindle and half-spindle can be obtained from different global extensions of the same local solutions as pointed out recently in \cite{M2-spindle}.
\\
\indent In this work, we are interested in supersymmetric $AdS_5\times \Sigma$ solutions from matter-coupled $N=2$ gauged supergravity in seven dimensions constructed in \cite{Eric_N2_7Dmassive}, see \cite{Pure_N2_7D1,Pure_N2_7D2,Eric_N2_7D,Park_7D,Salam_7DN2} for earlier constructions. We mainly consider $N=2$ gauged supergravity coupled to three vector multiplets with possible gauge groups given by $SO(4)\sim SO(3)\times SO(3)$, $SO(2,2)\sim SO(2,1)\times SO(2,1)$ and $SO(3,1)$. It is well-known that only $SO(4)$ and $SO(3,1)$ gauge groups admit supersymmetric $AdS_7$ vacua dual to $N=(1,0)$ SCFTs in six dimensions with $SO(3)_R$ R-symmetry \cite{7D_flow,7D_noncompact,AdS_7_N2_Jan}. Furthermore, a number of interesting holographic solutions have been found in \cite{7D_twist,AdS3_7D_N2,7D_flow,7D_noncompact,7D_defect}. We will give more solutions to this list by finding supersymmetric $AdS_5\times \Sigma$ solutions within this $N=2$ gauged supergravity for $\Sigma$ being a half-spindle.  
\\
\indent It has been pointed out in \cite{Ferrero_M5} that supersymmetric $AdS_5\times \Sigma$ solutions with $\Sigma$ being a spindle do not exist in minimal or pure $N=2$ gauged supergravity with $SO(3)$ gauge group, see also \cite{Gaunlett_super_spindle}. It turns out that solutions with $\Sigma$ being a half-spindle do exist in pure $N=2$ gauged supergravity. These solutions preserve $\frac{1}{2}$ of the supersymmetry and $SO(2)_R\subset SO(3)_R$ and can be obtained from a truncation of $AdS_5\times \Sigma$ solutions with $SO(2)\times SO(2)$ symmetry in $SO(4)$ gauged supergravity considered in this work. Furthermore, we also find $SO(2)_{\textrm{diag}}$ symmetric solutions that can be mapped to the solution found in \cite{Bah_M5} from $U(1)^2$ truncation of the maximal $SO(5)$ gauged supergravity. 

However, unlike the solutions in \cite{Bah_M5} dual to $N=2$ SCFTs of AD type, these solutions preserve only eight supercharges and should be dual to $N=1$ SCFTs in four dimensions. In addition, most of the solutions found in this paper currently have no known higher dimensional origin. In particular, it has been shown in \cite{Henning_Malek_AdS7_6} that uplifting seven-dimensional $N=2$ gauged supergravity with $AdS_7$ vacua to ten dimensions can be achieved only if there is no vector multiplet or one vector multiplet. On the other hand, the uplift to eleven dimensions can be performed via an $S^4$ truncation if the $N=2$ theories are truncations of the maximal $N=4$ gauged supergravity. The embedding in this case can be obtained from the results of \cite{Pure_Red_Ans} and \cite{7D_from_11D}. Moreover, pure $N=2$ gauged supergravity with $SO(3)$ gauge group can also be uplifted to type IIA supergravity \cite{uni_6D_7D,minimalAdS7_AdS6_Henning}. 
\\
\indent Finally, we will find $AdS_5\times \Sigma$ solutions in $N=2$ gauged supergravity with a non-compact $SO(2,2)\sim SO(2,1)\times SO(2,1)$ gauge group. Since this gauged supergravity does not admit any supersymmetric $AdS_7$ vacua, the maximally supersymmetric vacua are given by half-supersymmetric domain walls dual to $N=(1,0)$ non-conformal field theories in six dimensions according to the DW/QFT correspondence \cite{DW_QFT1,DW_QFT2,DW_QFT3,DW_QFT4}. In this case, the resulting $AdS_5\times \Sigma$ solutions are expected to describe four-dimensional $N=1$ SCFTs arising from six-dimensional $N=(1,0)$ field theories compactified on a topological disk or half-spindle. To the best of our knowledge, these are the first example of $AdS_5\times \Sigma$ solutions involving half-spindles with domain wall asymptotics.
\\
\indent The paper is organized as follows. In Section \ref{gSUGRA_Sec}, we give a brief review of seven-dimensional $N=2$ gauged supergravity coupled to an arbitrary number of vector multiplets. Supersymmetric $AdS_5\times\Sigma$ solutions in $SO(4)$, $SO(2,2)$ and $SO(3,1)$ gauge groups will be considered in Sections \ref{SO(4)_Sec}, \ref{SO(2,2)_Sec}, and \ref{SO(3,1)_Sec}, respectively. Some conclusions and comments will be given in Section \ref{Conclus}. 

%%%%%%%%%%%%%%%%%%%%%%%%%%%%%%%%%%%%%%%%%%%%%%%%%%%%%%%%%%%%%%%%%%%%%%%%%%%%%%%%%%%%%%%%%%%%%%%%%%%%%%%%%%%%%%%%%%%%%%%%%%%%%%%%%%%%%%%%%
\section{Matter-coupled $N=2$ gauged supergravity in seven dimensions}\label{gSUGRA_Sec}
In this section, we give relevant formulae involving bosonic Lagrangian and supersymmetry transformations of fermions to find supersymmetric solutions of matter-coupled $N=2$ gauged supergravity in seven dimensions. We follow most of the conventions and notations in \cite{Eric_N2_7D} in which the detailed construction can be found, see also \cite{Dibietto_7D_embedding_tensor} for gaugings using the embedding tensor formalism in the case of three vector multiplets.

In seven dimensions, the half-maximal $N=2$ supergravity multiplet contains the following component fields
\begin{equation}
(e^{\hat{\mu}}_\mu, \psi^a_\mu,A^i_\mu,\chi^a,B_{\mu\nu},\sigma) \nonumber \\
\end{equation}
given by the graviton $e^{\hat{\mu}}_\mu$, two gravitini $\psi^a_\mu$, three vectors $A^i_\mu$, two spin-$\frac{1}{2}$ fields $\chi^a$, a two-form field $B_{\mu\nu}$ and the scalar field or dilaton $\sigma$. We respectively denote curved and flat space-time indices by $\mu,\nu,\ldots$ and $\hat{\mu},\hat{\nu},\ldots$. Indices $i, j=1, 2, 3$ and $a,b=1,2$ respectively label triplet and doublet of $SO(3)_R\sim SU(2)_R$ $R$-symmetry.

The supergravity multiplet can couple to an arbitrary number $n$ of vector multiplets with the field content
\begin{equation}
(A_\mu,\lambda^a,\phi^i)^r\, .
\end{equation}
Each vector multiplet, labelled by an index $r=1,\ldots,n$, consists of a vector field $A_\mu$, two gaugini $\lambda^a$, and three scalars $\phi^{i}$. Together with the supergravity multiplet, there are $3+n$ vector fields transforming in a fundamental representation of the global symmetry $SO(3,n)$, collectively denoted by $A^I_\mu=(A^i_\mu,A^r_\mu)$. The $SO(3,n)$ fundamental indices $I,J=1,\ldots,3+n$ are lowered and raised respectively by the $SO(3,n)$ invariant tensor and its inverse
\begin{equation}\label{eta}
\eta_{IJ}= \eta^{IJ}=\text{diag}(-1,-1,-1,\underbrace{1,\ldots ,1}_n).
\end{equation} 
\indent Similar to the dilaton $\sigma$ described by a coset manifold $SO(1,1)\sim \mathbb{R}^+$, the $3n$ scalars $\phi^{ir}$ from the $n$ vector multiplets are parametrized by $SO(3,n)/SO(3)\times SO(n)$ coset manifold. With $A=(i,r)$ being an $SO(3)\times SO(n)$ index, the associated coset representative can be written as
\begin{equation}\label{DefL}
{L_I}^A=({L_I}^i,{L_I}^r).
\end{equation}
${L_I}^A$ transforms under the global $SO(3,n)$ and the local $SO(3)\times SO(n)$ by left and right multiplications, respectively. The inverse of ${L_I}^A$ will be denoted by ${L_A}^{I}=({L_i}^I,{L_r}^I)$ and satisfies the following relations
\begin{equation}
{L_j}^I{L_I}^i=\delta^i_j,\qquad {L_s}^I{L_I}^r=\delta^r_s\,,\qquad\eta_{IJ}= -{L_I}^i{L_J}^i+{L_I}^r{L_J}^r\, .
\end{equation}
Note also that indices $i,j$ and $r,s$ are raised and lowered by $\delta_{ij}$ and $\delta_{rs}$, respectively.

Gaugings of the matter-coupled $N=2$ supergravity can be obtained by promoting a subgroup $G\subset SO(3,n)$ to be a local symmetry. The embedding of $G$ in $SO(3,n)$ is described by the $SO(3,n)$ tensor ${f_{IJ}}^K$ identified with structure constants of the gauge group $G$ via the gauge algebra
\begin{equation}
[T_I,T_J]={f_{IJ}}^KT_K
\end{equation}
where $T_I$ are the gauge generators. In the embedding tensor formalism, ${f_{IJ}}^K$ is one of the components of the embedding tensor, see \cite{Dibietto_7D_embedding_tensor} for more detail.  
\\
\indent In order for the gauging to be consistent with supersymmetry, ${f_{IJ}}^K$ must satisfy the conditions
\begin{equation}
f_{IJK}=\eta_{KL}{f_{IJ}}^L=f_{[IJK]}\qquad \textrm{and}\qquad {f_{[IJ}}^L{f_{K]L}}^M=0\, .
\end{equation}
Since $\eta_{IJ}$ has only three negative eigenvalues, any gauge group can have at most three compact or three non-compact generators. Therefore, the allowed semi-simple gauge groups are of the form $G\sim G_0\times H\subset SO(3,n)$ with $H$ being a compact group of dimension $\text{dim}H\leq (n+3-\text{dim}G_0)$ \cite{Eric_N2_7D}. On the other hand, $G_0$ can only be one of the six possibilities: $SO(3)$, $SO(3,1)$, $SL(3,\mathbb{R})$, $SO(2,1)$, $SO(2,2)$, and $SO(2,2)\times SO(2,1)$. 

Apart from the usual gaugings, there is also a massive deformation given by adding a topological mass term to the three-form field $C_{\mu\nu\rho}$ dual to the two-form field $B_{\mu\nu}$. This additional deformation is crucial for the gauged supergravity to admit $AdS_7$ vacua. With both of these deformations, the bosonic Lagrangian of the matter-coupled $N=2$ gauged supergravity are given in differential form language by
\begin{eqnarray}\label{Lag}
\mathcal{L}&=&\ \frac{1}{2}R\ast\mathbf{1}-\frac{1}{2}e^{\sigma}a_{IJ}\ast F^I_{(2)}\wedge F^J_{(2)}-\frac{1}{2}e^{-2\sigma}\ast H_{(4)}\wedge H_{(4)}-\frac{5}{8}\ast d\sigma\wedge d\sigma\nonumber \\
& &-\frac{1}{2}\ast P^{ir}_{(1)}\wedge P^{ir}_{(1)}+\frac{1}{\sqrt{2}}H_{(4)}\wedge\omega_{(3)}-4hH_{(4)}\wedge C_{(3)}-\mathbf{V}\ast\mathbf{1}\, .
\end{eqnarray}
The constant $h$ describes the topological mass term for the three-form $C_{(3)}$ with the field strength $H_{(4)}=dC_{(3)}$. The gauge field strength is defined by
\begin{equation}
F^I_{(2)}=dA^I_{(1)}+\frac{1}{2}{f_{JK}}^IA^J_{(1)}\wedge A^K_{(1)}.
\end{equation}
The scalar matrix $a_{IJ}$ appearing in the kinetic term of vector fields is given in terms of the coset representative as follows
\begin{equation}\label{aIJ}
a_{IJ}={L_I}^i{L_J}^i+{L_I}^r{L_J}^r\, .
\end{equation}
The Chern-Simons three-form satisfying $d\omega_{(3)}=F^I_{(2)}\wedge F^I_{(2)}$ is defined by
\begin{equation}
\omega_{(3)}=F^I_{(2)}\wedge A^I_{(1)}-\frac{1}{6}{f_{IJ}}^KA^I_{(1)}\wedge A^J_{(1)}\wedge A_{(1)K}\, .
\end{equation}
\indent The scalar potential is given by
\begin{equation}\label{Pot}
\mathbf{V}=\frac{1}{4}e^{-\sigma}\left(C^{ir}C_{ir}-\frac{1}{9}C^2\right)+16h^2e^{4\sigma}-\frac{4\sqrt{2}}{3}he^{\frac{3\sigma}{2}}C
\end{equation}
where $C$-functions, or fermion-shift matrices, are defined as
\begin{eqnarray}\label{CFn}
C&=&-\frac{1}{\sqrt{2}}{f_{IJ}}^K{L_i}^I{L_j}^JL_{Kk}\varepsilon^{ijk},\\
C^{ir}&=& \frac{1}{\sqrt{2}}{f_{IJ}}^K{L_j}^I{L_k}^J{L_K}^r\varepsilon^{ijk},\\[5pt]
C_{rsi}&=& {f_{IJ}}^K{L_r}^I{L_s}^JL_{Ki}\, .
\end{eqnarray}
The scalar kinetic term is defined in terms of the vielbein on the $SO(3,n)/SO(3)\times SO(n)$ coset as
\begin{equation}\label{P^ir}
P^{ir}_{(1)}=L^{rI}\left(\delta^K_Id+{f_{IJ}}^KA^J_{(1)}\right){L_K}^i.
\end{equation}
\indent Supersymmetry transformations of fermionic fields read
\begin{eqnarray}\label{SUSY}
\delta\psi_\mu^a&\hspace{-0.3cm}=&\hspace{-0.3cm} 2D_\mu\epsilon^a-\frac{\sqrt{2}}{30}e^{-\frac{\sigma}{2}}C\Gamma_\mu\epsilon^a-\frac{4}{5}he^{2\sigma}\Gamma_\mu\epsilon^a-\frac{i}{20}e^{\frac{\sigma}{2}}F^i_{\rho\sigma}{(\sigma^i)^a}_b(3\Gamma_\mu\Gamma^{\rho\sigma}-5\Gamma^{\rho\sigma}\Gamma_\mu)\epsilon^b\nonumber\\&&-\frac{1}{240\sqrt{2}}e^{-\sigma}H_{\rho\sigma\lambda\tau}(\Gamma_\mu\Gamma^{\rho\sigma\lambda\tau}+5\Gamma^{\rho\sigma\lambda\tau}\Gamma_\mu)\epsilon^a,\\
\delta\chi^a&\hspace{-0.2cm}=&\hspace{-0.2cm}-\frac{1}{2}\Gamma^\mu\partial_\mu\sigma\epsilon^a+\frac{\sqrt{2}}{30}e^{-\frac{\sigma}{2}}C\epsilon^a-\frac{16}{5}e^{2\sigma}h\epsilon^a-\frac{i}{10}e^{\frac{\sigma}{2}}F^i_{\mu\nu}{(\sigma^i)^a}_b\Gamma^{\mu\nu}\epsilon^b\nonumber\\&&-\frac{1}{60\sqrt{2}}e^{-\sigma}H_{\mu\nu\rho\sigma}\Gamma^{\mu\nu\rho\sigma}\epsilon^a,\\
\delta\lambda^{ar}&\hspace{-0.2cm}=&\hspace{-0.2cm}i\Gamma^\mu P^{ir}_\mu {(\sigma^i)^a}_b\epsilon^b-\frac{1}{2}e^{\frac{\sigma}{2}}F^r_{\mu\nu}\Gamma^{\mu\nu}\epsilon^a-\frac{i}{\sqrt{2}}e^{-\frac{\sigma}{2}}C^{ir}{(\sigma^i)^a}_b\epsilon^b.\label{dlambda}
\end{eqnarray}
In these equations, ${(\sigma^i)^a}_b$ are the usual Pauli matrices and $\Gamma_\mu=e^{\hat{\mu}}_\mu\Gamma_{\hat{\mu}}$ in which $\Gamma_{\hat{\mu}}$ are seven-dimensional space-time gamma matrices satisfying the $SO(1,6)$ Clifford algebra 
\begin{equation}
\{\Gamma_{\hat{\mu}},\Gamma_{\hat{\nu}}\}=2\eta_{\hat{\mu}\hat{\nu}}, \qquad\eta_{\hat{\mu}\hat{\nu}}=\text{diag}(-++++++).
\end{equation} 
The dressed field strengths $F^i_{(2)}$ and $F^r_{(2)}$ are defined by
\begin{equation}
F^i_{(2)}={L_I}^iF^I_{(2)}\qquad  \textrm{and}\qquad F^r_{(2)}={L_I}^rF^I_{(2)}\, .
\end{equation}
The covariant derivative of the supersymmetry parameter $\epsilon^a$ is given by
\begin{equation}\label{Depsilon}
D_\mu\epsilon^a=\partial_\mu\epsilon^a+\frac{1}{4}{\omega_\mu}^{\hat{\nu}\hat{\rho}}\Gamma_{\hat{\nu}\hat{\rho}}\epsilon^a+\frac{1}{2\sqrt{2}}Q^i_\mu{(\sigma^i)^a}_b\epsilon^b
\end{equation}
where $Q^i_\mu$ is defined in terms of the $SO(3)_R$ composite connection $Q^{ij}_\mu$ as
\begin{equation}
Q^i_\mu=\frac{i}{\sqrt{2}}\varepsilon^{ijk}Q^{jk}_\mu
\end{equation}
with
\begin{equation}\label{DefQ}
Q^{ij}_\mu=L^{jI}\left(\delta^K_I\partial_\mu+{f_{IJ}}^KA^J_\mu\right){L_K}^i\, .
\end{equation}
\indent For convenience, we also give all the bosonic field equations derived from the Lagrangian \eqref{Lag}
\begin{eqnarray}
0&=& d(e^{-2\sigma}\ast H_{(4)})+8hH_{(4)}-\frac{1}{\sqrt{2}}F^I_{(2)}\wedge F^I_{(2)},\label{C3_eq}\\
0&=&D(e^\sigma a_{IJ}\ast F^I_{(2)})-\sqrt{2}H_{(4)}\wedge F^J_{(2)}+\ast P^{ir}_{(1)}{f_{IJ}}^K{L_r}^IL_{Ki},\phantom{\frac{1}{\sqrt{2}}}\label{Vec_eq}\\
0&=&D(\ast P^{ir}_{(1)})-2e^\sigma{L_I}^i{L_J}^r\ast F^I_{(2)}\wedge F^J_{(2)}\nonumber \\&&-\left(\frac{1}{\sqrt{2}}e^{-\sigma}C^{js}C_{rsk}\varepsilon^{ijk}+4\sqrt{2}he^{\frac{3\sigma}{2}}C^{ir}\right)\varepsilon_{(7)},\label{phiEQ}\\
0&=&\frac{5}{4}d(\ast d\sigma)-\frac{1}{2}e^\sigma a_{IJ}\ast F^I_{(2)}\wedge F^J_{(2)}+e^{-2\sigma}\ast H_{(4)}\wedge H_{(4)}\nonumber\\ &&+\left[\frac{1}{4}e^{-\sigma}\left(C^{ir}C_{ir}-\frac{1}{9}C^2\right)+2\sqrt{2}he^{\frac{3\sigma}{2}}C-64h^2e^{4\sigma}\right]\varepsilon_{(7)},\label{DilEq}\\
0&=&R_{\mu\nu}-\frac{5}{4}\partial_\mu \sigma\partial_\nu \sigma-a_{IJ}e^\sigma\left(F^I_{\mu\rho}{F^J_\nu}^\rho-\frac{1}{10}g_{\mu\nu}F^I_{\rho\sigma}F^{J\ \rho\sigma}\right)\nonumber \\&& -P^{ir}_\mu P^{ir}_\nu-\frac{2}{5}g_{\mu\nu}\mathbf{V}-\frac{1}{6}e^{-2\sigma}\left(H_{\mu\rho\sigma\lambda}{H_\nu}^{\rho\sigma\lambda}-\frac{3}{20}g_{\mu\nu}H_{\rho\sigma\lambda\tau}H^{\rho\sigma\lambda\tau}\right).\quad\label{EinsteinEQ}
\end{eqnarray}
\indent We finally give a general parametrization of the $SO(3, n)/SO(3)\times SO(n)$ coset which is useful for finding explicit solutions. We first introduce $(n+3)^2$ basis elements of a general $(n + 3)\times (n + 3)$ matrix as follow
\begin{equation}
(e_{IJ})_{KL}=\delta_{IK}\delta_{JL}\, .
\end{equation}
The composite $SO(3)\times SO(n)$ generators are given by
\begin{eqnarray}
SO(3):\qquad J^{(1)}_{ij}&=&e_{ji}-e_{ij},\qquad\qquad\qquad\ i, j = 1, 2, 3,\nonumber\\
SO(n):\qquad J^{(2)}_{rs}&=&e_{s+3,r+3}-e_{r+3,s+3},\qquad r, s = 1,\ldots, n\, .
\end{eqnarray}
The non-compact generators corresponding to the $3n$ scalars are given by
\begin{equation}
Y_{ir}=e_{i,r+3}+e_{r+3,i}\, .
\end{equation}

%%%%%%%%%%%%%%%%%%%%%%%%%%%%%%%%%%%%%%%%%%%%%%%%%%%%%%%%%%%%%%%%%%%%%%%%%%%%%%%%%%%%%%%%%%%%%%%%%%%%%%%%%%%%%%%%%%%%%%%%%%%%%%%%%%%%%%%%%
\section{$SO(4)$ gauge group}\label{SO(4)_Sec}
We first consider $N=2$ gauged supergravity with $SO(4)\sim SO(3)\times SO(3)$ gauge group obtained by coupling the gravity multiplet to $n=3$ vector multiplets. The first $SO(3)$ factor is identified with the $SO(3)_R\sim SU(2)_R$ R-symmetry. The corresponding structure constants are given by
\begin{equation}
f_{IJK}=(\widetilde{g}_1\varepsilon_{ijk},-\widetilde{g}_2\varepsilon_{rst}),\qquad r,s,\ldots=1,2,3
\end{equation}
in which $\widetilde{g}_1$ and $\widetilde{g}_2$ are coupling constants of $SO(3)_R$ and $SO(3)$ generated by $J^{(1)}_{ij}$ and $J^{(2)}_{rs}$, respectively.  

We are interested in supersymmetric solutions in the form of a product space between an $AdS_5$ and a topological disk $\Sigma$ with a non-trivial $U(1)$ holonomy at the boundary. Following \cite{Bah_M5}, we take the ansatz for the seven-dimensional metric to be
\begin{equation}\label{7Dmetrix}
ds^2_7=f(r)ds^2_{AdS_5}+g_1(r)dr^2+g_2(r)dz^2
\end{equation}
where the metric on $AdS_5$ with unit radius is given by
\begin{equation}
ds^2_{AdS_5}=\frac{1}{\rho^2}(dx^2_{1,3}+d\rho^2)
\end{equation}
with $dx^2_{1,3}=\eta_{mn}dx^m dx^n$, $m,n= 0,\ldots, 3$ being the flat metric on the four-dimensional Minkowski space $Mkw_4$. $r$ and $z$ are respectively radial and angular coordinates on the topological disk $\Sigma$ whose ranges will be determined later on. The seven-dimensional curved and flat space-time indices will be split into $\mu=(m,\rho,r,z)$ and $\hat{\mu}=(\hat{m},\hat{\rho},\hat{r},\hat{z})$, respectively. 

With the following vielbein one-forms
\begin{eqnarray}
& &e^{\hat{m}}_{(1)}=\frac{\sqrt{f(r)}}{\rho}dx^{m},\qquad e^{\hat{\rho}}_{(1)}=\frac{\sqrt{f(r)}}{\rho}d\rho,\nonumber \\
& & e^{\hat{r}}_{(1)}=\sqrt{g_1(r)}dr,\qquad e^{\hat{z}}_{(1)}=\sqrt{g_2(r)}dz, \label{7Dvielbein}
\end{eqnarray}
we can straightforwardly compute all non-vanishing components of the spin connections
\begin{eqnarray}
& &\omega_{(1)}^{\hat{m}\hat{\rho}}=-\frac{e^{\hat{m}}}{\sqrt{f}},\qquad\omega_{(1)}^{\hat{m}\hat{r}}=\frac{f'e^{\hat{m}}}{2f\sqrt{g_1}},\nonumber \\
& &\omega_{(1)}^{\hat{\rho}\hat{r}}=\frac{f'e^{\hat{\rho}}}{2f\sqrt{g_1}},\qquad\omega_{(1)}^{\hat{z}\hat{r}}=\frac{g'_2e^{\hat{z}}}{2g_2\sqrt{g_1}}\, .\label{spinCon}
\end{eqnarray}
From now on, we will use primes to denote $r$-derivatives and mostly suppress arguments of the $r$-dependent functions for convenience.

\subsection{$SO(2)_R$ and $SO(2)\times SO(2)$ symmetric solutions}
We now move to the ansatz for gauge fields in the case of $SO(2)_R\subset SO(3)_R$ and $SO(2)\times SO(2)$ symmetry. Since the former can be obtained as a truncation of the latter, we will firstly consider the case of $SO(2)\times SO(2)$ symmetry and perform a suitable truncation to obtain $SO(2)_R$ symmetric solutions. For $SO(2)\times SO(2)$ symmetric solutions, we will choose the $SO(2)\times SO(2)$ subgroup generated by $J^{(1)}_{12}$ and $J^{(2)}_{12}$ and turn on the following gauge fields
\begin{equation}\label{SO(2)xSO(2)_Gaugefield}
A^{I}_{(1)}=\left[A_1(r)\delta^I_3+A_2(r)\delta^I_6\right]dz\, .
\end{equation}

The corresponding two-form field strengths are given by 
\begin{equation}
F^{I}_{(2)}=(A'_1\delta^I_3+A'_2\delta^I_6)\,dr\wedge dz\, .
\end{equation}
This ansatz leads to $F^I_{(2)}\wedge F^I_{(2)}=0$. According to the field equation of the three-form field given in \eqref{C3_eq}, we can consistently set $C_{(3)}=0$.

Among the nine scalars from the $SO(3,3)/SO(3)\times SO(3)$ coset, there is only one $SO(2)\times SO(2)$ singlet scalar. Following \cite{7D_flow}, this scalar field corresponds to the non-compact generator $Y_{33}$, and the coset representative can be written as
\begin{equation}
L=e^{\phi Y_{33}}\, .\label{L_SO3d}
\end{equation}
This singlet scalar $\phi$ and the dilaton $\sigma$ depend only on the radial coordinate. It is now straightforward to compute the $C$-functions appearing in the supersymmetry transformations
\begin{equation}\label{SO(2)xSO(2)_Cfns}
C=3\sqrt{2}\,\widetilde{g}_1\cosh\phi,\qquad C^{ir}=-\sqrt{2}\,\widetilde{g}_1\sinh{\phi}\,\delta^i_3\delta^r_3\, .
\end{equation}
The scalar vielbein and $SO(2)_R$ composite connection have the following non-vanishing components
\begin{equation}\label{SO(4)_SO(2)xSO(2)_P_Q}
P^{ir}_{(1)}=\phi'\,\delta^i_3\delta^r_3dr\qquad\text{ and }\qquad Q^{ij}_{(1)}=\widetilde{g}_1A_1\varepsilon^{ij3}dz\, .
\end{equation}
It should be noted here that only $A_1$ appears in the composite connection because $A^3_{(1)}$ is the vector field that gauges $SO(2)_R\subset SO(3)_R$ under which the gravitini and supersymmetry parameters are charged. With all these and a similar analysis as in \cite{Bah_M5}, we can determine all the BPS equations from the supersymmetry transformations of fermionic fields. The detailed analysis and relevant results can be found in the appendix. In the following, we will separately consider solutions with $SO(2)_R$ and $SO(2)\times SO(2)$ symmetries.

%%%%%%%%%%%%%%%%%%%%%%%%%%%%%%%%%%%%%%%%%%%%%%
\subsubsection{$SO(2)_R$ symmetric solution in pure $N=2$ gauged supergravity}\label{pSUGRA_Soln_Sec}
We first consider a simple case of $SO(2)_R$ symmetric solutions which can be obtained by setting $A^6_{(1)}=0$. Equation \eqref{Gen_SO(2)xSO(2)_Ansatz} then gives
\begin{equation}\label{SO(4)_SO(2)R_dressF}
\mathbf{F}_1=be^{-\sigma}\cosh{\phi}f^{-\frac{5}{2}}\qquad\text{ and }\qquad \mathbf{F}_2=-be^{-\sigma}\sinh{\phi}f^{-\frac{5}{2}}
\end{equation}
in which we have written $a_1=a_2=b$. 

With this explicit form of $\mathbf{F}_1$ and $\mathbf{F}_2$, the BPS condition in \eqref{AABPS6} implies that
\begin{equation}
bh\sinh\phi=0\, .
\end{equation}
For $h=0$, the gauged supergravity does not admit any supersymmetric $AdS_7$ vacua, so we will keep $h\neq 0$. With $b=0$, all the gauge fields vanish. This clearly does not lead to any solutions of the form $AdS_5\times \Sigma$. Therefore, to find possible $AdS_5\times \Sigma$ solutions with $AdS_7$ asymptotics, we need to set $\phi=0$. Effectively, all the fields from vector multiplets are truncated out. The resulting solutions can be then considered as solutions of the minimal or pure $N=2$ gauged supergravity with $SO(3)_R$ gauge group.

With $\phi=0$, equation \eqref{Gen_SO(2)xSO(2)_Ansatz} implies that $A'_2=0$. All the BPS conditions from \eqref{keyBPS4} and \eqref{A44_BPS_con} to \eqref{SBCBPS6} as well as the field equation \eqref{phiEQ} for scalars from the vector multiplets are automatically satisfied. With all these, we are left with the algebraic conditions \eqref{DABPS1} to \eqref{SBCBPS3}. Firstly, we solve for $f$ from equation \eqref{ABCBPS2} with the solution given by
\begin{equation}\label{fsoln}
f=\frac{2\sqrt{b\,h}\,e^\sigma}{\sqrt{s(16he^{\frac{5\sigma}{2}}-\widetilde{g}_1)}}\, .
\end{equation}
With this result, equation \eqref{DABPS3} can be solved for $g_1$ giving rise to
\begin{equation}
g_1=\frac{200\sqrt{b\,h^5}\,e^{\sigma}\,(\sigma')^2}{(16he^{\frac{5\sigma}{2}}-\widetilde{g}_1)^2\left[32\sqrt{b\,h^5}-e^{-5\sigma}\sqrt{s(16he^{\frac{5\sigma}{2}}-\widetilde{g}_1)}\right]}.
\end{equation}
\indent The condition \eqref{ABCBPS3} together with the solution for $A'_1=\widehat{A}'_1$ in \eqref{Gen_SO(2)xSO(2)_Ansatz} give an ordinary differential equation of the form
\begin{equation}
\widehat{A}'_1=\frac{5\widetilde{g}_1\,\sigma'}{2(12he^{\frac{5\sigma}{2}}-\widetilde{g}_1)}\widehat{A}_1
\end{equation}
which can be readily solved by
\begin{equation}
\widehat{A}_1=c\left(12h-\widetilde{g}_1e^{-\frac{5\sigma}{2}}\right)
\end{equation}
for an integration constant $c$. Substituting all these results in \eqref{AABPS1} or \eqref{ABCBPS3} leads to the following solution for $g_2$ of the form 
\begin{equation}
g_2=\frac{c^2\widetilde{g}^2_1e^\sigma\left[32\sqrt{b\,h^5}-e^{-5\sigma}\sqrt{s(16he^{\frac{5\sigma}{2}}-\widetilde{g}_1)}\right]}{\sqrt{s(16he^{\frac{5\sigma}{2}}-\widetilde{g}_1)}}.
\end{equation}
Finally, it can be verified that all the BPS conditions \eqref{DABPS1} to \eqref{SBCBPS3} as well as the dilaton's and Einstein's field equations, \eqref{DilEq} and \eqref{EinsteinEQ}, are satisfied by these solutions provided that
\begin{equation}\label{SO(2)xSO(2)_sign_Con}
\text{sign}(c\widetilde{g}_1\sigma')=+1\, .
\end{equation}
\indent Furthermore, the BPS equations \eqref{keyBPS1} to \eqref{keyBPS3} are satisfied by the following form of the two-component spinor 
\begin{equation}
\eta=e^{iqz}Ye^{\frac{5\sigma}{4}}\begin{pmatrix} \sqrt{8(h^5b)^{\frac{1}{4}}+\sqrt{2}se^{-\frac{5\sigma}{2}}\left[s(16he^{\frac{5\sigma}{2}}-\widetilde{g}_1)\right]^{\frac{1}{4}}} \\ -\sqrt{8(h^5b)^{\frac{1}{4}}-\sqrt{2}se^{-\frac{5\sigma}{2}}\left[s(16he^{\frac{5\sigma}{2}}-\widetilde{g}_1)\right]^{\frac{1}{4}}}\end{pmatrix}
\end{equation}
with the function $Y$ being the solution of an ordinary differential equation given in \eqref{nBPS3}. The explicit form of the solution for $Y$ can be written as
\begin{equation}
Y=\frac{Y_0\,e^{-\sigma}}{\left[s(16he^{\frac{5\sigma}{2}}-\widetilde{g}_1)\right]^{\frac{1}{8}}}
\end{equation}
in which $Y_0$ is an integration constant. It should be noted that the solution is characterized by a set of functions that are determined in terms of the dilaton $\sigma$ together with its derivative. However, the $r$-dependent function $\sigma(r)$ is not determined by the BPS equations. This is very similar to the solutions obtained in \cite{Bah_M5, Suh_D3, Suh_D4, Suh_M2}. 

To further analyse the solution, we first define the following parameters
\begin{equation}
B=8h^2\sqrt{b},\qquad m=\frac{\widetilde{g}_1}{16h},\qquad \mathcal{C}=2\,\widetilde{g}_1hc
\end{equation}
together with the function
\begin{equation}\label{Def_W}
W=B-e^{-5\sigma}\sqrt{s(e^{\frac{5\sigma}{2}}-m)}\, .
\end{equation}
In terms of these quantities, the seven-dimensional metric reads
\begin{equation}
ds^2_7=\frac{B\,e^\sigma}{16h^2\sqrt{s(e^{\frac{5\sigma}{2}}-m)}}ds^2_{AdS_5}+\frac{25B\,e^{\sigma}\,(\sigma')^2}{4^5h^2W(e^{\frac{5\sigma}{2}}-m)^2}dr^2+\frac{\mathcal{C}^2We^\sigma}{4h^2\sqrt{s(e^{\frac{5\sigma}{2}}-m)}}dz^2
\end{equation}
which is singular when $W\rightarrow0$. It turns out that the analysis near $W=0$ is simpler if we fix the solution of $\sigma$ to 
\begin{equation}\label{2/5_Dil_Soln}
\sigma=-\frac{2}{5}\ln r\, .
\end{equation}
This choice implies $r>0$, and the sign condition \eqref{SO(2)xSO(2)_sign_Con} requires $\text{sign}(c\widetilde{g}_1)=-1$. Accordingly, the constant $\mathcal{C}$ must be negative for $h>0$. 

In terms of the radial coordinate, the seven-dimensional metric is given by
\begin{eqnarray}
ds^2_7&=&\frac{B\,r^{1/10}}{16h^2\sqrt{s(1-mr)}}\left[ds^2_{AdS_5}+ds^2_\Sigma\right]\nonumber\\
\textrm{with}\qquad ds^2_\Sigma&=&\frac{r^{-1/2}}{16W\left[s(1-mr)\right]^{3/2}}dr^2+\frac{4\mathcal{C}^2W}{B}dz^2\label{Sigma_Met}
\end{eqnarray}
and
\begin{equation}
W=B-r^{3/2}\sqrt{s(1-mr)}\, .
\end{equation}
\indent The equation $W=0$ admits four roots given by
\begin{equation}\label{r_Roots}
r_{(\pm_1,\pm_2)}=\frac{1}{4m}\left[1\pm_12m\sqrt{X}\pm_22m\sqrt{\frac{3}{4m^2}-X\pm_1\frac{1}{4m^3\sqrt{X}}}\,\right]
\end{equation}
with
\begin{equation}
X=\frac{1}{4m^2}+\frac{4(\frac{2}{3})^{1/3}sB^{4/3}}{\left[9s+\sqrt{81-768sB^2m^3}\right]^{\frac{1}{3}}}+\frac{B^{2/3}\left[9s+\sqrt{81-768sB^2m^3}\right]^{\frac{1}{3}}}{18^{1/3}m}\, .
\end{equation}
These roots are all distinct due to different sign choices $\pm_1$ and $\pm_2$ appearing in \eqref{r_Roots}.

\begin{figure}[h!]
  \centering
  \begin{subfigure}[b]{0.326\linewidth}
    \includegraphics[width=\linewidth]{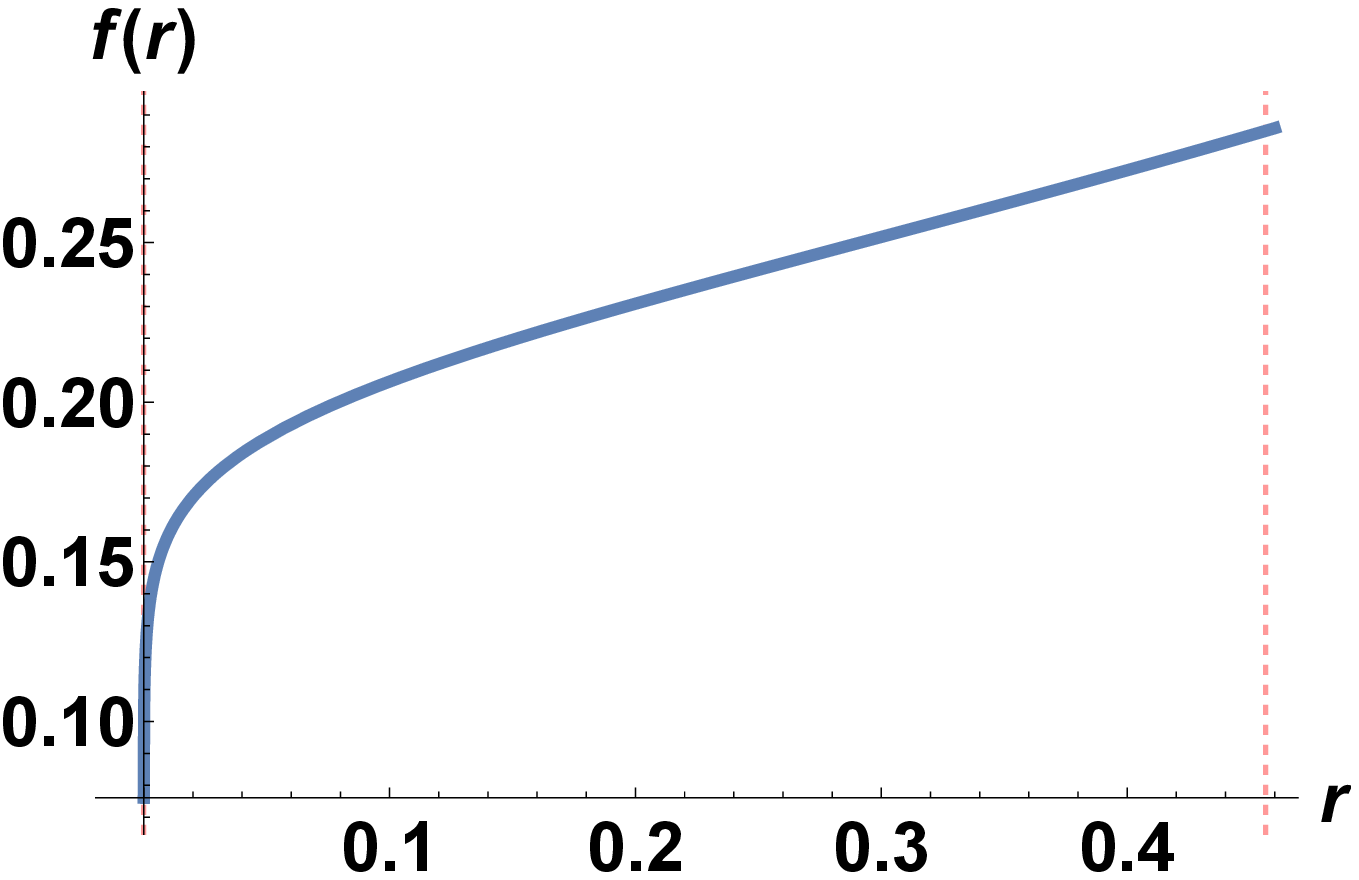}
\caption{$f$ solution}
  \end{subfigure}
  \begin{subfigure}[b]{0.326\linewidth}
    \includegraphics[width=\linewidth]{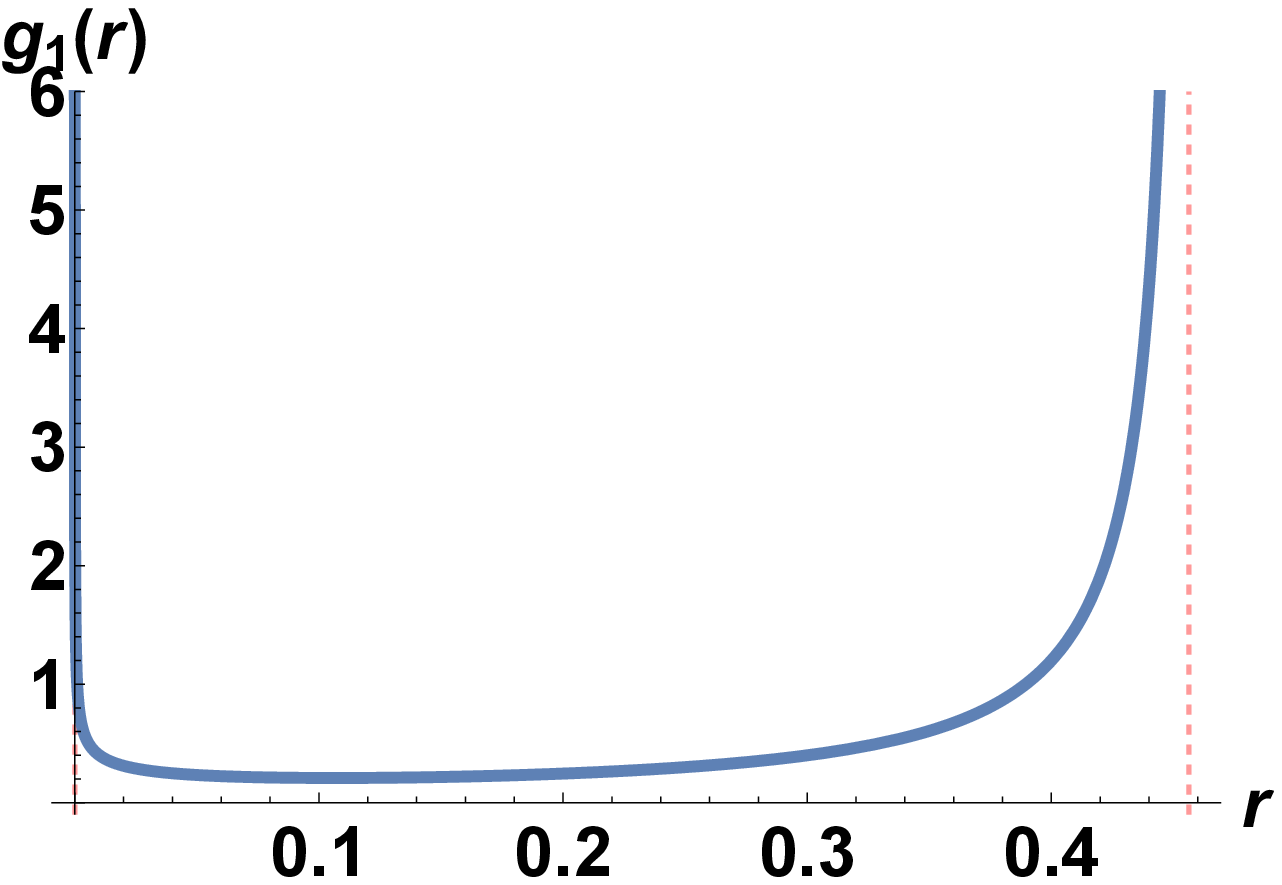}
\caption{$g_1$ solution}
  \end{subfigure}
  \begin{subfigure}[b]{0.326\linewidth}
    \includegraphics[width=\linewidth]{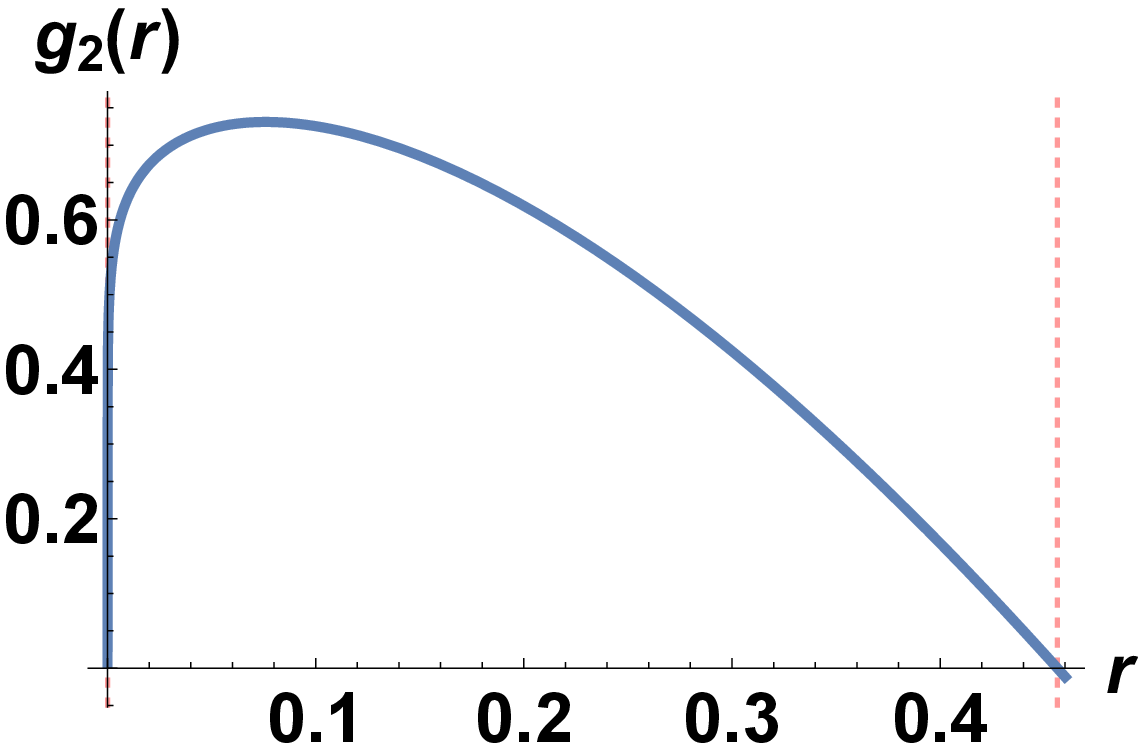}
\caption{$g_2$ solution}
  \end{subfigure}
  \caption{Numerical plots of the warp factors for the $SO(2)$ symmetric solution in pure $N=2$ gauged supergravity with $s=1$, $m=\frac{3}{4}$, $B=h=\frac{1}{4}$, and $\mathcal{C}=-1$. All the warp factors are positive in the range $0<r<r_{(+,-)}=0.456$ with the two vertical red dashed lines representing the two boundaries.}
  \label{Case1_Soln}
\end{figure}

Using \eqref{AandA}, we find the explicit form of the $SO(2)_R$ vector field
\begin{equation}\label{SO(2)R_soln}
A_1=\frac{1}{8mh}\left[\,|\mathcal{C}|(4mr-3)-q\,\right].
\end{equation}
In terms of the radial coordinate, the two-component spinor $\eta$ is given by
\begin{equation}\label{eta_Soln}
\eta=Y_0e^{iqz}\frac{2^{1/4}r^{1/40}}{\left[s(1-mr)\right]^{\frac{1}{8}}}\begin{pmatrix} \sqrt{\sqrt{B}+sr^{3/4}\left[s(1-mr)\right]^{\frac{1}{4}}} \\ -\sqrt{\sqrt{B}-sr^{3/4}\left[s(1-mr)\right]^{\frac{1}{4}}} \end{pmatrix}.
\end{equation}
\indent The allowed ranges of the radial coordinate $r$ for a regular solution are constrained by requiring that the dilaton scalar is real ($r>0$) and all the seven-dimensional metric functions are positive. There are seven possibilities depending on the values of the parameters $s$, $m$, and $B$. It should also be noted that these ranges together with the corresponding behaviours of the solution are very similar to those considered in \cite{Bah_M5}. We now discuss these possibilities in detail. \\
%%%%%
\textbf{Case I:} $s=1$, $m>0$, $0<B<\frac{3\sqrt{3}}{16m\sqrt{m}}$, $0<r<r_{(+,-)}$\\
\indent For clarity, we plot a representative solution of the warp factors with $s=1$, $m=\frac{3}{4}$, $B=h=\frac{1}{4}$, and $\mathcal{C}=-1$ in Figure \ref{Case1_Soln}. As seen from the figure, as $r\rightarrow0$, both $f$ and $g_2$ approach zero while $g_1$ diverges to $+\infty$. This is a curvature singularity of the seven-dimensional metric as pointed out in \cite{Bah_M5}. Setting $r=R^{4/3}$, we find that the seven-dimensional metric becomes  conformal to a product of $AdS_5$ and a cylinder near $R=0$
\begin{equation}\label{CaseIw0_metric}
ds_7^2\approx\frac{B\,R^{2/15}}{16h^2}\left[ds^2_{AdS_5}+\frac{1}{9B}dR^2+4\mathcal{C}^2dz^2\right].
\end{equation}

\begin{figure}[h!]
  \centering
    \includegraphics[width=0.56\linewidth]{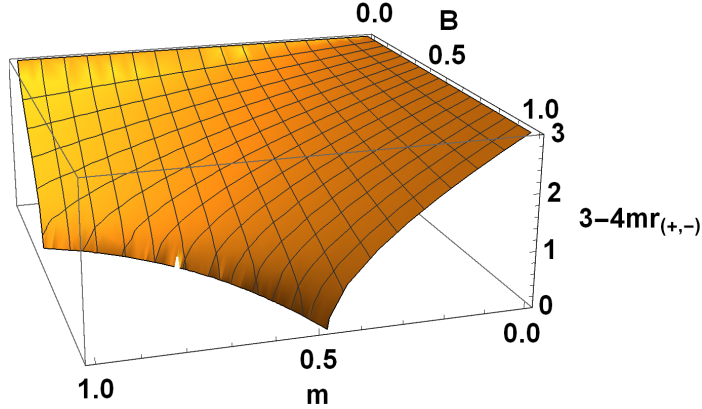}
  \caption{A numerical plot of the function $3-4mr_{(+,-)}$ appearing in \eqref{CaseI_smooth_con}. Note that $3-4mr_{(+,-)}\rightarrow 3$ and $3-4mr_{(+,-)}\rightarrow 0$ as $B\rightarrow 0$ or $m\rightarrow 0$ and $B\rightarrow\frac{3\sqrt{3}}{16m\sqrt{m}}$, respectively.}
  \label{3-4m+-_sur}
\end{figure}

As $r\rightarrow r_{(+,-)}$, the $AdS_5$ warp factor is smooth and the $z$ circle shrinks due to $W(r_{(+,-)})=B-r^{3/2}_{(+,-)}\sqrt{1-mr_{(+,-)}}=0$. By introducing a new radial coordinate $R=\sqrt{r_{(+,-)}-r}$, we find that $g_2$ is finite as $r\rightarrow r_{(+,-)}$. The seven-dimensional metric near this endpoint takes the form of
\begin{equation}\label{spindle_metric}
ds_7^2\approx\frac{B\,r_{(+,-)}^{1/10}}{16h^2\sqrt{1-mr_{(+,-)}}}\left[ds^2_{AdS_5}+\frac{dR^2+4\mathcal{C}^2\left[3-4mr_{(+,-)}\right]^2R^2dz^2}{-4W'(r_{(+,-)})\sqrt{r_{(+,-)}}(1-mr_{(+,-)})^{3/2}}\right].
\end{equation}
The $R-z$ surface becomes locally an $\mathbb{R}^2/\mathbb{Z}_l$ orbifold near the point $r= r_{(+,-)}$ if we set
\begin{equation}\label{CaseI_smooth_con}
|\mathcal{C}|=\frac{1}{2l\left[3-4mr_{(+,-)}\right]},\qquad l=1,2,3,\ldots \, .
\end{equation}
In this case, the function $3-4mr_{(+,-)}$ depends on two constants, $m$ and $B$. However, its explicit form obtained from equation \eqref{r_Roots} is highly complicated, so we will only show that $3-4mr_{(+,-)}$ is strictly positive in the range of the $r$ coordinate under consideration by giving a numerical plot of $3-4mr_{(+,-)}$ in Figure \ref{3-4m+-_sur}.

By using the Gauss-Bonnet theorem and the same computation as in \cite{Bah_M5}, we can calculate the Euler characteristic of $\Sigma$ 
\begin{eqnarray}
\chi(\Sigma)&=&\frac{1}{4\pi}\int_{\Sigma}R_{\Sigma}\text{vol}_{\Sigma}\ =\ \frac{2|\mathcal{C}|\left[3-4mr_{(+.-)}\right]r_{(+.-)}^{3/4}\left[1-mr_{(+,-)}\right]^{1/4}}{\sqrt{B}}\nonumber\\&=&2|\mathcal{C}|\left[3-4mr_{(+.-)}\right]\ =\ \frac{1}{l}
\end{eqnarray}
which is a natural result for a disk with an $\mathbb{R}^2/\mathbb{Z}_l$ orbifold singularity at $r= r_{(+,-)}$. Noted that the integration has been performed on the interval $0<r<r_{(+.-)}$ and $0<z<2\pi$, and we have used $B=r^{3/2}_{(+,-)}\sqrt{1-mr_{(+,-)}}$ obtained from $W(r_{(+,-)})=0$ with $s=+1$.

\begin{figure}[h!]
  \centering
  \begin{subfigure}[b]{0.326\linewidth}
    \includegraphics[width=\linewidth]{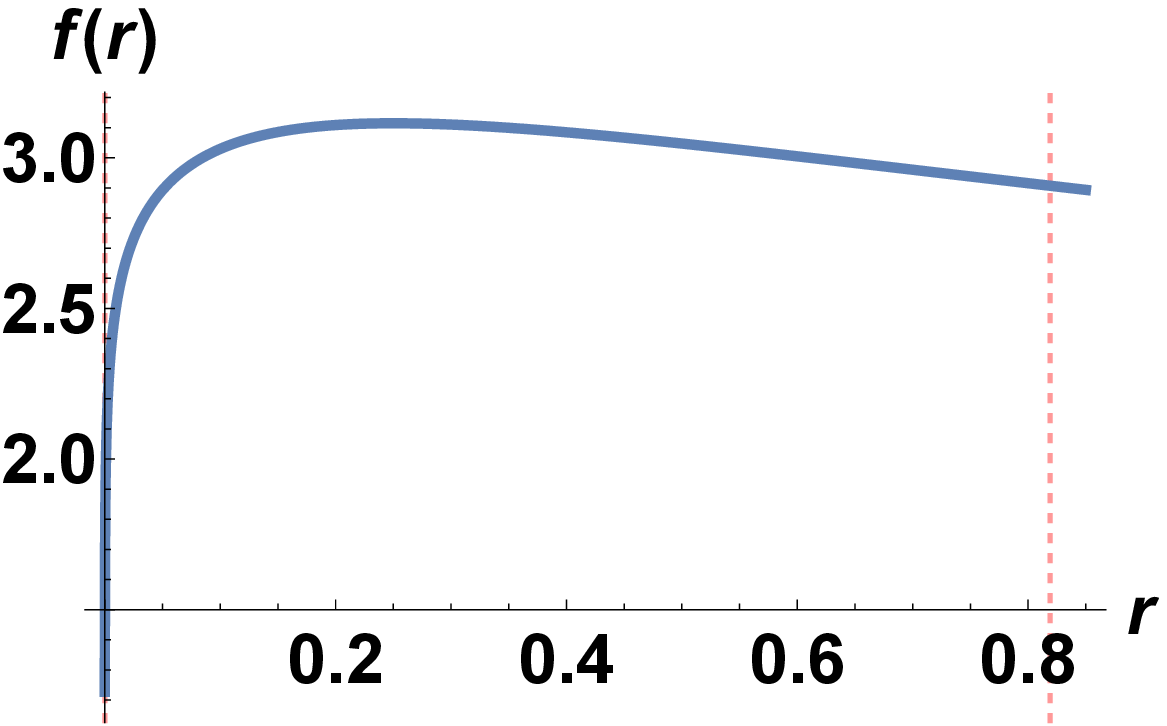}
\caption{$f$ solution}
  \end{subfigure}
  \begin{subfigure}[b]{0.326\linewidth}
    \includegraphics[width=\linewidth]{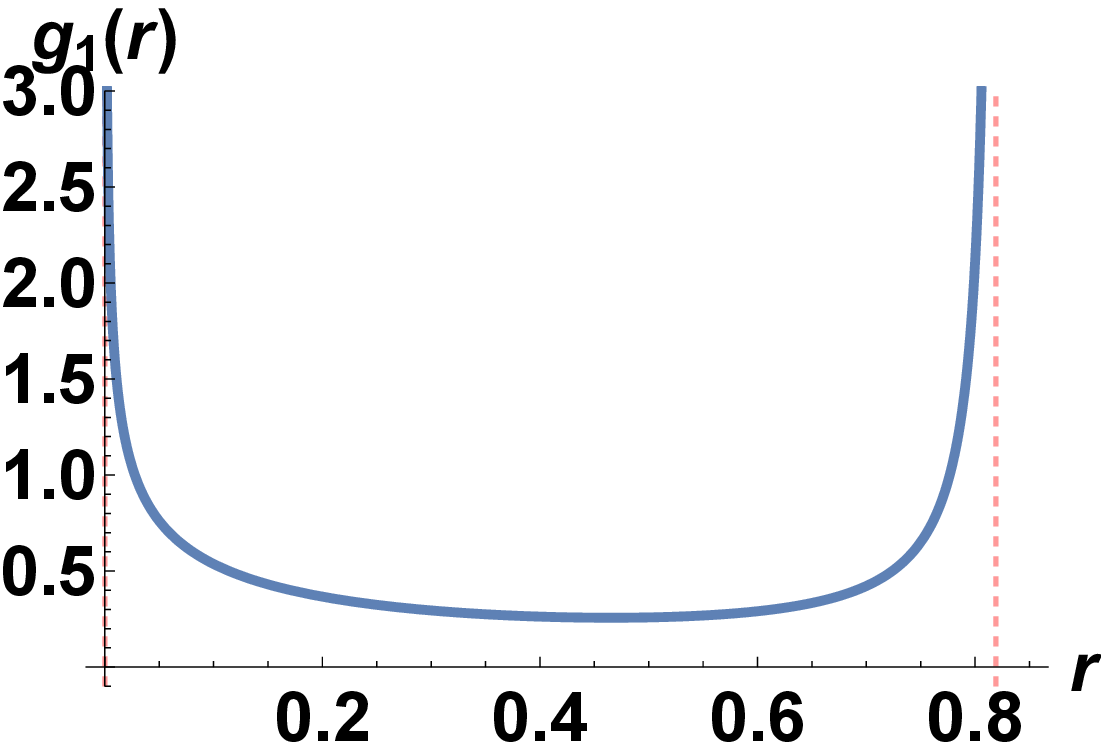}
\caption{$g_1$ solution}
  \end{subfigure}
  \begin{subfigure}[b]{0.326\linewidth}
    \includegraphics[width=\linewidth]{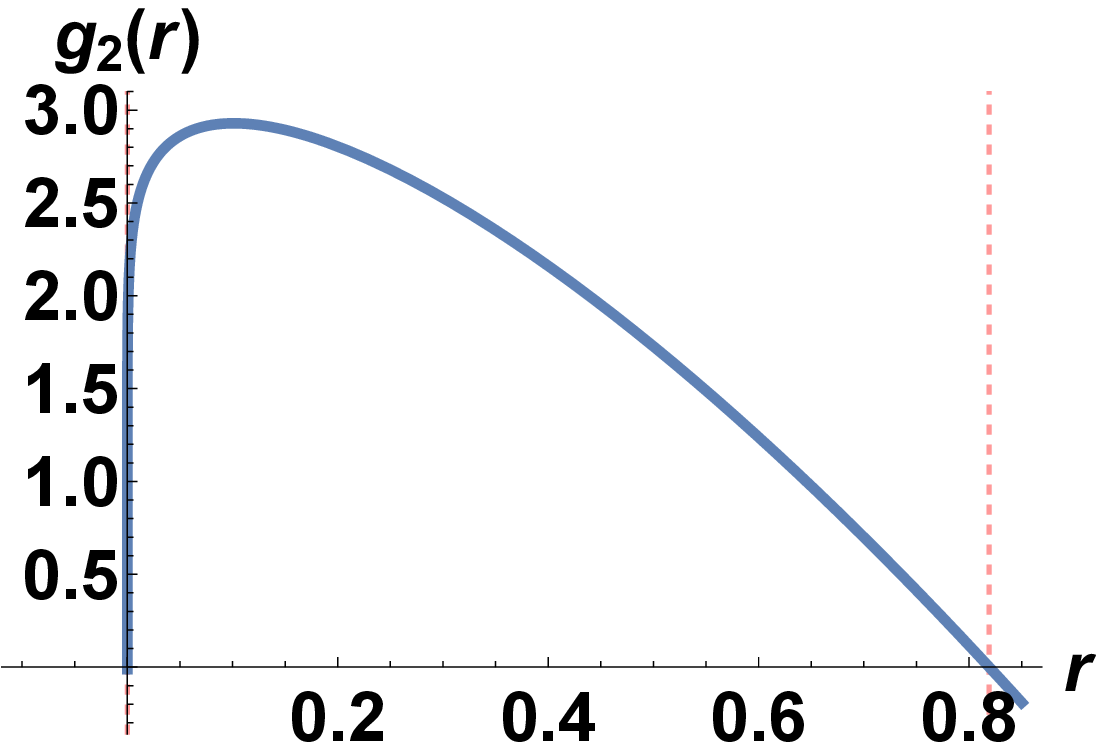}
\caption{$g_2$ solution}
  \end{subfigure}
  \caption{Numerical plots of the warp factors for the $SO(2)$ symmetric solution in pure $N=2$ gauged supergravity with $s=1$, $m=-1$, $B=1$, $h=\frac{1}{8}$, and $\mathcal{C}=-\frac{1}{2}$. The warp factors are positive in the range $0<r<r_{(-,+)}=0.819$ with the two vertical red dashed lines representing the two boundaries.}
  \label{Case2_Soln}
\end{figure}

\begin{figure}[h!]
  \centering
    \includegraphics[width=0.56\linewidth]{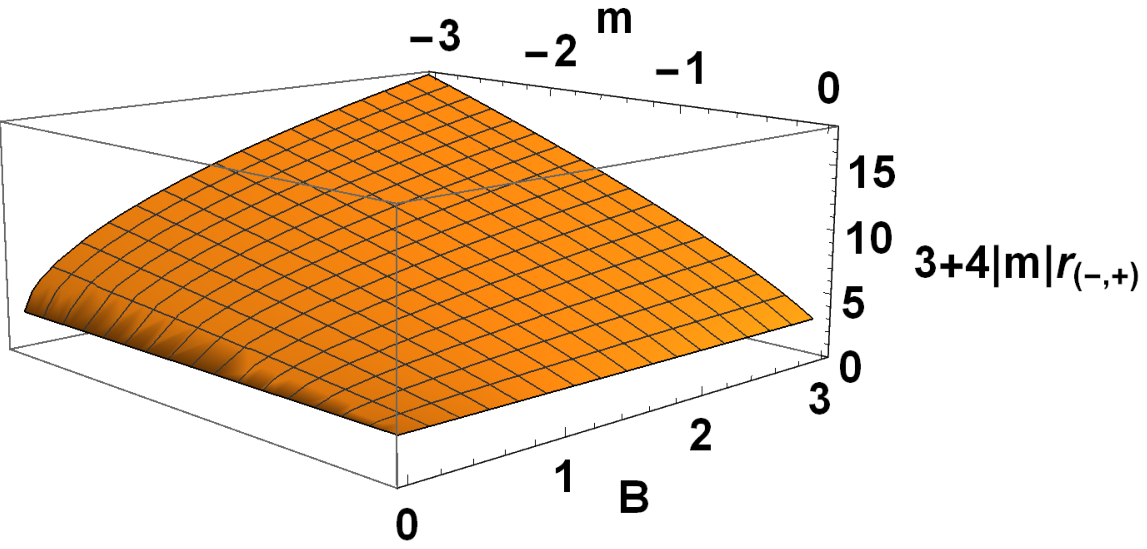}
  \caption{A numerical plot of the function $3+4|m|r_{(-,+)}$ appearing in the condition \eqref{CaseII_smooth_con}. We also note that $3+4|m|r_{(-,+)}\rightarrow 3$ as $B\rightarrow 0$ or $m\rightarrow 0$.}
  \label{3-4m-+_sur}
\end{figure}

In order for the $SO(2)_R$ gauge field to vanish at $r= r_{(+,-)}$, at which the $z$ circle shrinks, we fix the constant $q$ to be
\begin{equation}
q=-|\mathcal{C}|\left[3-4mr_{(+,-)}\right]=-\frac{1}{2l}
\end{equation}
giving rise to the $SO(2)_R$ gauge field of the form
\begin{equation}\label{Fin_SO(2)R_soln}
A_1=\frac{|\mathcal{C}|}{2h}[r-r_{(+,-)}].
\end{equation}
\indent The explicit form of the Killing spinor at $r= r_{(+,-)}$ is given by
\begin{equation}\label{spindle_eta}
\eta=2^{3/4}Y_0e^{-\frac{iz}{2l}}r^{2/5}_{(+,-)}\begin{footnotesize}\begin{pmatrix} 1 \\ 0 \end{pmatrix}\end{footnotesize}.
\end{equation}
Since only the upper component is non-vanishing, only $\frac{1}{4}$ of the original supersymmetry or $4$ supercharges are preserved at $r=r_{(+,-)}$. Moreover, $\eta$ is also well-behaved near $r=r_{(+,-)}$ and hence globally defined on the disk $\Sigma$ as in \cite{Bah_M5}.\\
%%%%%
\textbf{Case II:} $s=1$, $m<0$, $B>0$, $0<r<r_{(-,+)}$\\
\indent An example of numerical solutions with $s=1$, $m=-1$, $B=1$, $h=\frac{1}{8}$, and $\mathcal{C}=-\frac{1}{2}$ is given in Figure \ref{Case2_Soln}. This case is very similar to Case I. The previous analysis at both $r=0$ and $r=r_{(-,+)}$ endpoints can be repeated by formally replacing $m\rightarrow-|m|$ and $r_{(+,-)}\rightarrow r_{(-,+)}$. However, in order for the $z$ circle to shrink smoothly at $r=r_{(-,+)}$, instead of \eqref{CaseI_smooth_con}, we have to impose
\begin{equation}\label{CaseII_smooth_con}
|\mathcal{C}|=\frac{1}{2l\left[3+4|m|r_{(-,+)}\right]},\qquad l=1,2,3,\ldots\,,\, .
\end{equation}
A numerical plot of $3+4|m|r_{(-,+)}$ is shown in Figure \ref{3-4m-+_sur}. The function is very different from $3-4mr_{(+,-)}$ in Case I since the condition on the constant $B$ in this case is less stringent.

\begin{figure}[h!]
  \centering
  \begin{subfigure}[b]{0.326\linewidth}
    \includegraphics[width=\linewidth]{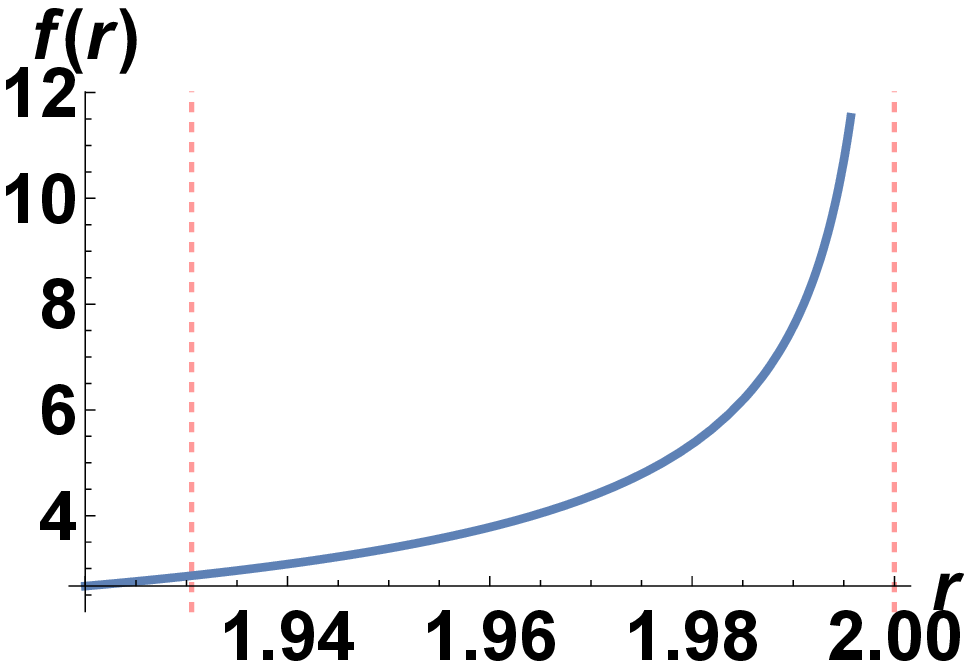}
\caption{$f$ solution}
  \end{subfigure}
  \begin{subfigure}[b]{0.326\linewidth}
    \includegraphics[width=\linewidth]{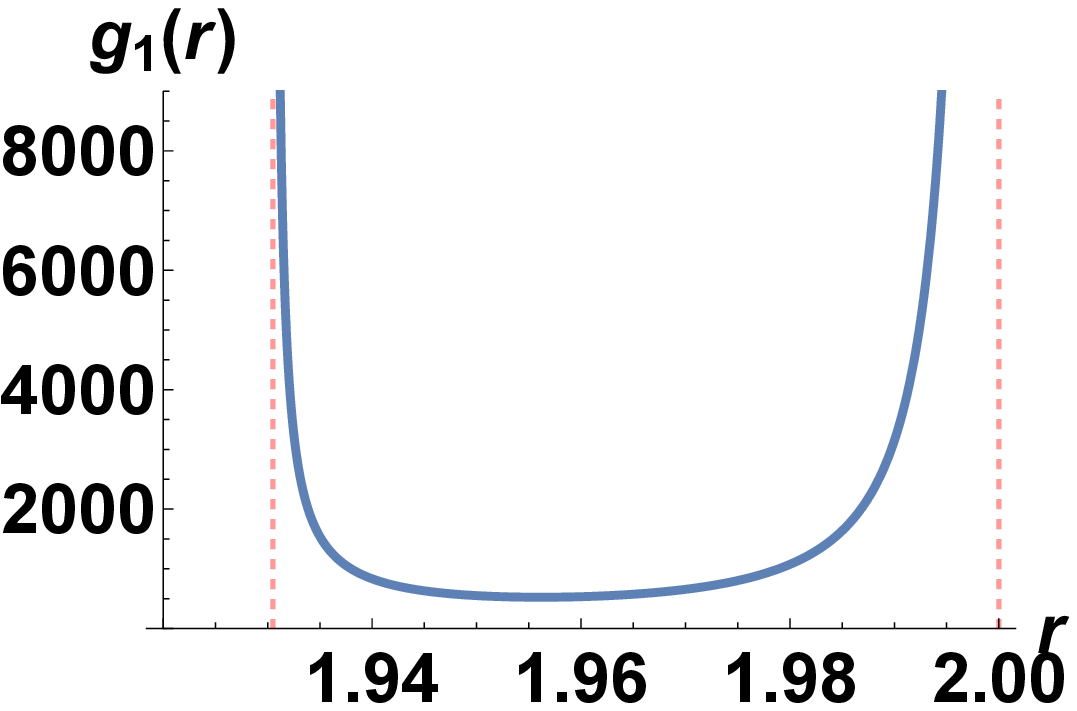}
\caption{$g_1$ solution}
  \end{subfigure}
  \begin{subfigure}[b]{0.326\linewidth}
    \includegraphics[width=\linewidth]{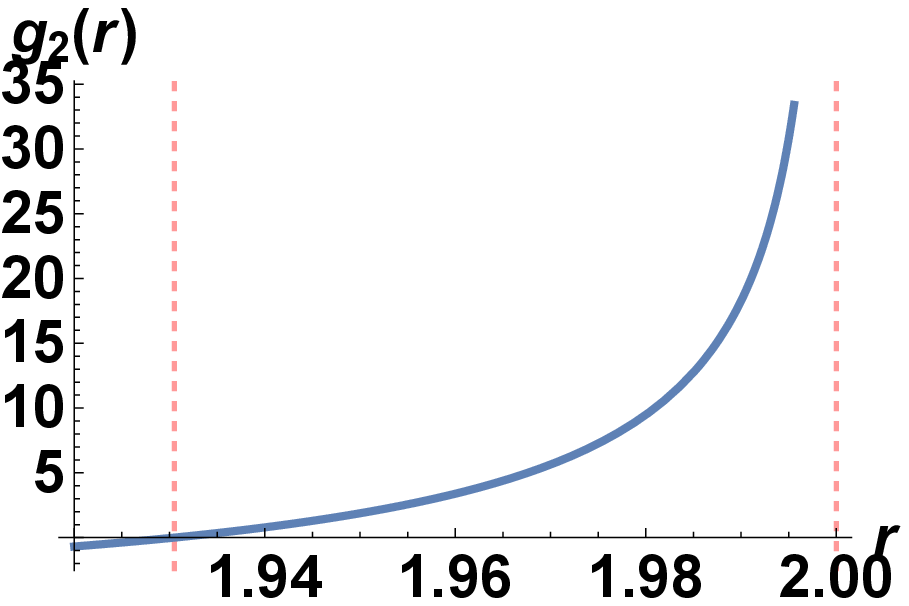}
\caption{$g_2$ solution}
  \end{subfigure}
  \caption{Numerical plots of the warp factors for the $SO(2)$ symmetric solution in pure $N=2$ gauged supergravity with $s=1$, $m=\frac{1}{2}$, $B=\frac{1}{2}$, $h=\frac{1}{4}$, and $\mathcal{C}=-1$. The warp factors are all positive in the range $r_{(+,+)}=1.93<r<\frac{1}{m}=2$ with the two vertical red dashed lines representing the two boundaries.}
  \label{Case3_Soln}
\end{figure}

\noindent\textbf{Case III:} $s=1$, $m>0$, $0<B<\frac{3\sqrt{3}}{16m\sqrt{m}}$, $r_{(+,+)}<r<\frac{1}{m}$\\
\indent As $r\rightarrow\frac{1}{m}$, the $AdS_5$ warp factor goes to $+\infty$ as can be seen from Figure \ref{Case3_Soln}. Setting $r=\frac{1}{m}-16\mathcal{C}^4m^2B^2R^4$ , we find the seven dimensional-metric of the form
\begin{equation}\label{case3_metric}
ds^2_7\approx\frac{1}{16h^2m^{8/5}R^2}\left[\frac{1}{4\mathcal{C}^2}ds^2_{AdS_5}+dR^2+dz^2\right],\qquad R\rightarrow0\, .
\end{equation}
This metric is again conformally related to a product of $AdS_5$ and a cylinder.

\begin{figure}[h!]
  \centering
    \includegraphics[width=0.56\linewidth]{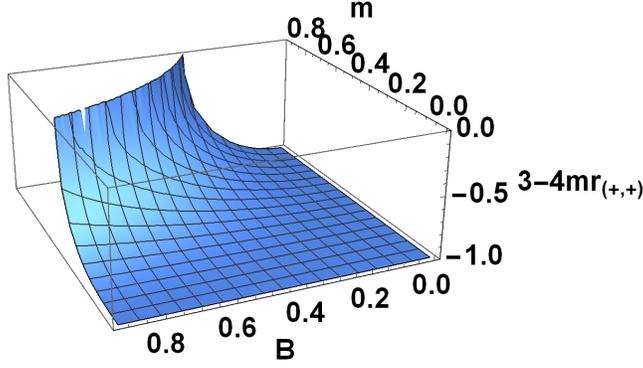}
  \caption{A numerical plot of the function $3-4mr_{(+,+)}$ appearing in the condition \eqref{CaseIII_smooth_con}. The function approaches $-1$ as $B\rightarrow 0$ or $m\rightarrow 0$. As $B\rightarrow\frac{3\sqrt{3}}{16m\sqrt{m}}$, we find that $3-4mr_{(+,-)}\rightarrow 0$.}
  \label{(3-4mr++)_sur}
\end{figure}

\begin{figure}[h!]
  \centering
    \includegraphics[width=0.56\linewidth]{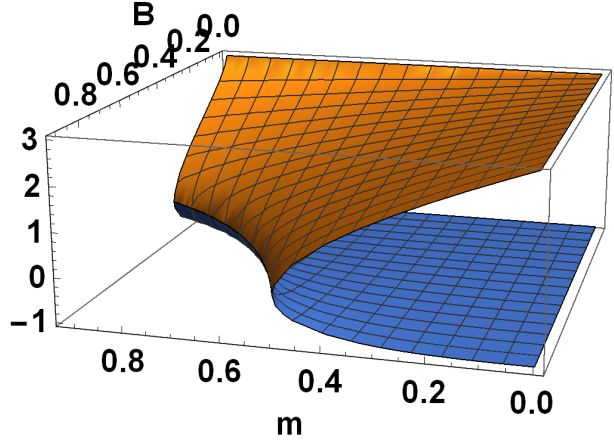}
  \caption{A numerical plot of $3-4mr_{(+,-)}$ (orange surface) and $3-4mr_{(+,+)}$ (blue surface) which are connected to each other at $B=\frac{3\sqrt{3}}{16m\sqrt{m}}$.}
  \label{(3-4mr+-)_(3-4mr++)_sur}
\end{figure}

By changing the radial coordinate to $\rho=\sqrt{r_{(+,+)}-r}$, we find the seven-dimensional metric, as $r\rightarrow r_{(+,+)}$, of the form
\begin{equation}\label{CaseIII_spindle_metric}
ds_7^2\approx\frac{B\,r_{(+,+)}^{1/10}}{16h^2\sqrt{1-mr_{(+,+)}}}\left[ds^2_{AdS_5}+\frac{dR^2+4\mathcal{C}^2\left[3-4mr_{(+,+)}\right]^2R^2dz^2}{-4W'(r_{(+,+)})\sqrt{r_{(+,+)}}(1-mr_{(+,+)})^{3/2}}\right]
\end{equation}
which is similar to \eqref{spindle_metric} with $r_{(+,-)}$ replaced by $r_{(+,+)}$. The $z$ circle also shrinks smoothly near the endpoint at $r=r_{(+,+)}$, and the $R-z$ surface is locally an $\mathbb{R}^2/\mathbb{Z}_l$ if we choose
\begin{equation}\label{CaseIII_smooth_con}
\mathcal{C}=\frac{1}{2l\left[3-4mr_{(+,+)}\right]},\qquad l=1,2,3,\ldots\, .
\end{equation}
Unlike all previous cases, $3-4mr_{(+,+)}$ is negative in the regularity ranges of $B$ and $m$ as can be seen from the numerical plot given in Figure \ref{(3-4mr++)_sur}. In Figure \ref{(3-4mr+-)_(3-4mr++)_sur}, we also plot $3-4mr_{(+,+)}$ (blue surface) and $3-4mr_{(+,-)}$ (orange surface) in Case I. The two joint smoothly at $B=\frac{3\sqrt{3}}{16m\sqrt{m}}$ where $3-4mr_{(+,-)}=0=3-4mr_{(+,+)}$.


To obtain the $SO(2)_R$  gauge field that vanishes at $r=r_{(+,+)}$, we choose
\begin{equation}
q=\mathcal{C}\left[3-4mr_{(+,+)}\right]=\frac{1}{2l}
\end{equation}
resulting in $A_1$ of the form
 \begin{equation}\label{CaseIII_SO(2)R_soln}
A_1=\frac{|\mathcal{C}|}{2h}[r-r_{(+,+)}].
\end{equation}
Since $q=\frac{1}{2l}$ in this case, the Killing spinor $\eta$ near $r=r_{(+,+)}$ is the same \eqref{spindle_eta} with $r_{(+,-)}$ and $z$ replaced by $r_{(+,+)}$ and $-z$. \\
%%%%%
\textbf{Case IV:} $s=1$, $m>0$, $B=\frac{3\sqrt{3}}{16m\sqrt{m}}$, $0<r<r_{(+,-)}$\\
\indent With $B=\frac{3\sqrt{3}}{16m\sqrt{m}}$, the quantity $X$ in the four roots of $W=0$ reduces to 
\begin{equation}
X=\frac{1}{m^2}
\end{equation}
giving rise to a much simple form of $r_{(\pm_1,\pm_2)}$ solution
\begin{equation}
r_{(\pm_1,\pm_2)}=\frac{1}{4m}\left[1\pm_12\pm_2\sqrt{-1\pm_11}\right].
\end{equation}
In this case, there is only one real root of $W=0$ given by $r_\ast=r_{(+,\pm_2)}=\frac{3}{4m}$.

\begin{figure}[h!]
  \centering
  \begin{subfigure}[b]{0.326\linewidth}
    \includegraphics[width=\linewidth]{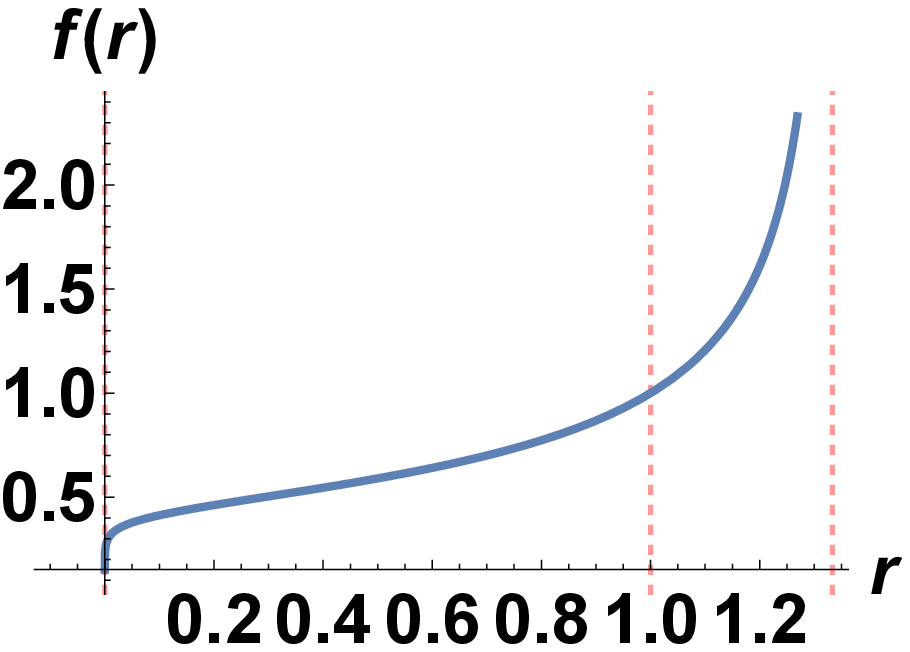}
\caption{$f$ solution}
  \end{subfigure}
  \begin{subfigure}[b]{0.326\linewidth}
    \includegraphics[width=\linewidth]{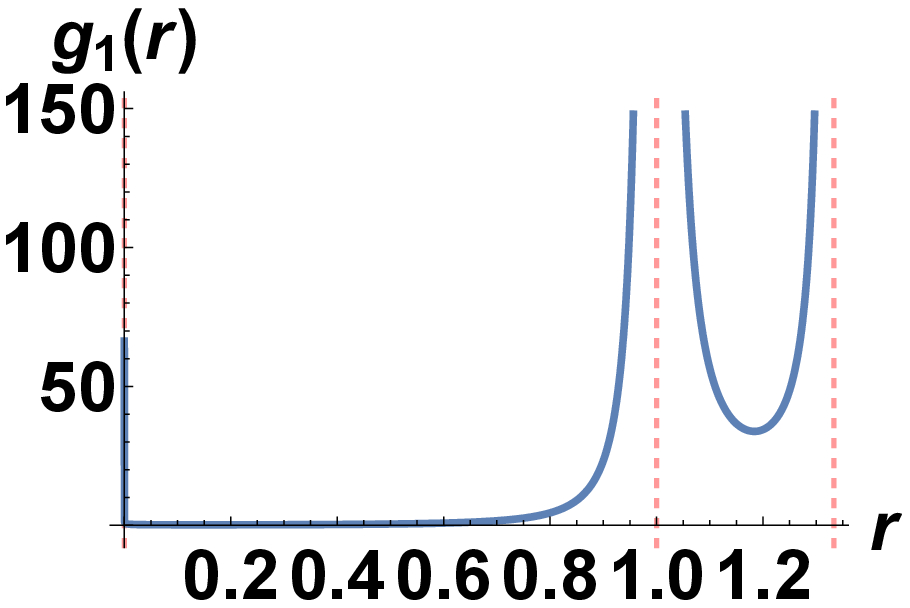}
\caption{$g_1$ solution}
  \end{subfigure}
  \begin{subfigure}[b]{0.326\linewidth}
    \includegraphics[width=\linewidth]{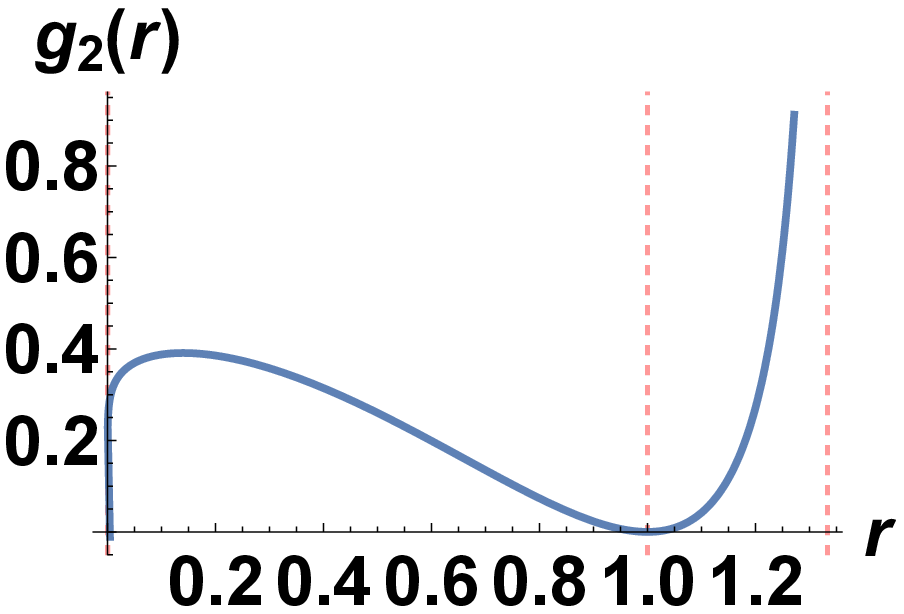}
\caption{$g_2$ solution}
  \end{subfigure}
  \caption{Numerical plots of the warp factors for the $SO(2)$ symmetric solution in pure $N=2$ gauged supergravity with $s=1$, $m=\frac{3}{4}$, $B=\frac{1}{2}$, $h=\frac{1}{4}$, and $\mathcal{C}=-\frac{1}{2}$. The warp factors are positive in the ranges $0<r<r_\ast=1$ (Case IV) and $r_\ast=1<r<\frac{1}{m}=\frac{4}{3}$ (Case V). The three vertical red dashed lines represent the three boundaries.}
  \label{Case4+5_Soln}
\end{figure}

However, as pointed out in \cite{Bah_M5}, the $z$ circle does not shrink smoothly at $r=r_\ast$ for any value of $\mathcal{C}$, due to the function $W(r=r_\ast)$ having a double zero. This can also be seen explicitly in the present work by considering Case I with $B\rightarrow\frac{3\sqrt{3}}{16m\sqrt{m}}$. In this limit, we cannot impose the condition \eqref{CaseI_smooth_con} since $|\mathcal{C}|$ diverges as $3-4mr_{(+,-)}\rightarrow0$. If we set $r=\frac{3}{4m}-\frac{1}{R}$, the seven-dimensional metric is approximately given by
\begin{equation}\label{case4_metric}
ds^2_7\approx\frac{3^{3/5}}{128(2^{1/5})h^2m^{8/5}}\left[3ds^2_{AdS_5}+\frac{dR^2+64\mathcal{C}^2m^2dz^2}{R^2}\right]
\end{equation}
as $R\rightarrow+\infty$. On the other hand, for $r\rightarrow0$, we find the seven-dimensional metric given in \eqref{CaseIw0_metric} as in Case I.\\
%%%%%
\textbf{Case V:} $s=1$, $m>0$, $B=\frac{3\sqrt{3}}{16m\sqrt{m}}$, $r_{(+,+)}<r<\frac{1}{m}$\\
\indent Since $r_{(+,+)}=r_{(+,-)}=r_\ast=\frac{3}{4m}$ for $B=\frac{3\sqrt{3}}{16m\sqrt{m}}$, Case IV and Case V are connected at $r=r_\ast$. Therefore, we will give a representative numerical solution for these two cases collectively in Figure \ref{Case4+5_Soln}. For $r\rightarrow r_\ast$ both from the left and from the right, the behaviour of the metric is given by \eqref{case4_metric}. On the other side, as $r\rightarrow \frac{1}{m}$, the metric becomes \eqref{case3_metric} as in Case III.

\begin{figure}[h!]
  \centering
  \begin{subfigure}[b]{0.326\linewidth}
    \includegraphics[width=\linewidth]{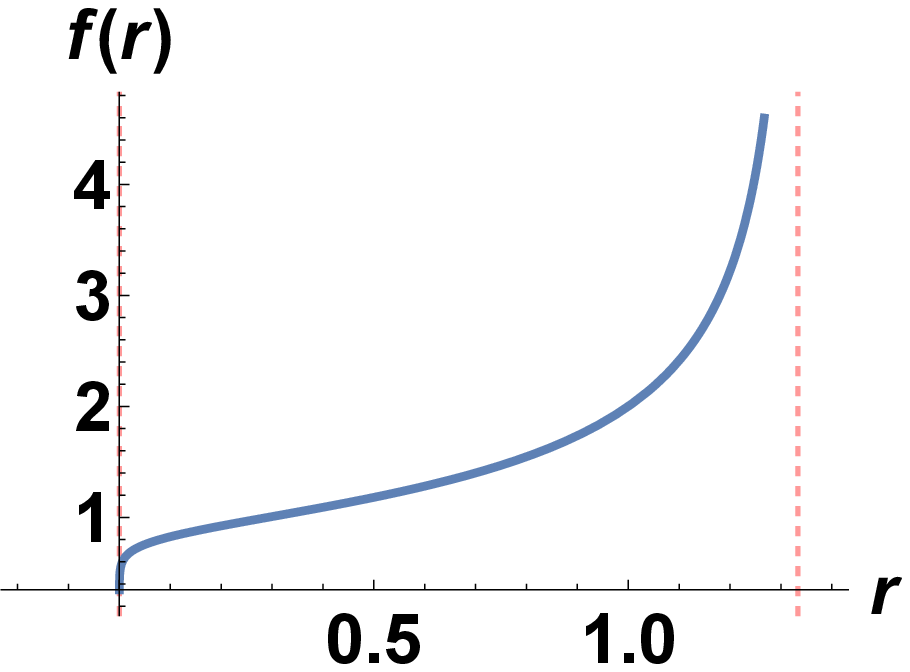}
\caption{$f$ solution}
  \end{subfigure}
  \begin{subfigure}[b]{0.326\linewidth}
    \includegraphics[width=\linewidth]{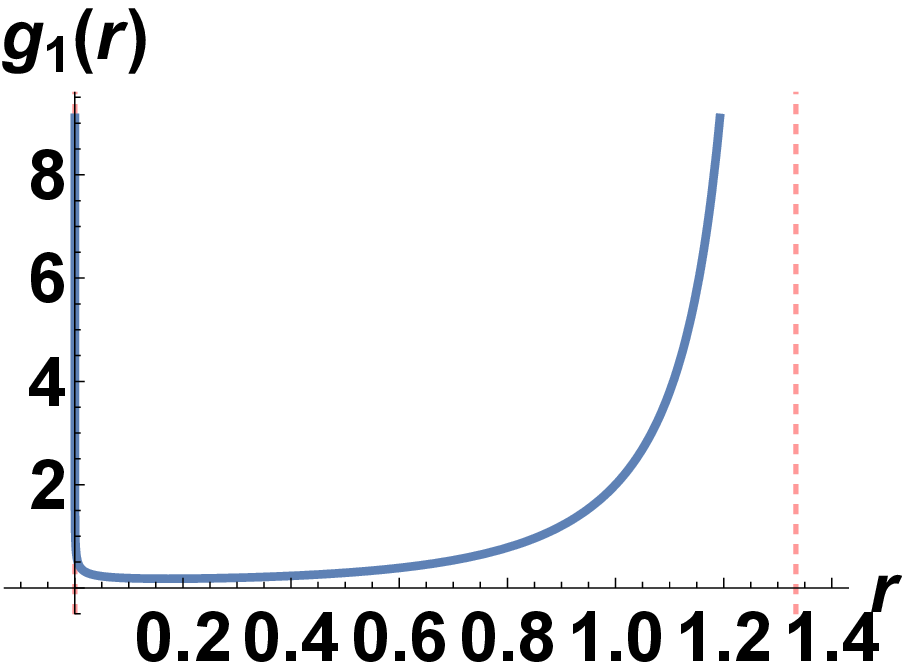}
\caption{$g_1$ solution}
  \end{subfigure}
  \begin{subfigure}[b]{0.326\linewidth}
    \includegraphics[width=\linewidth]{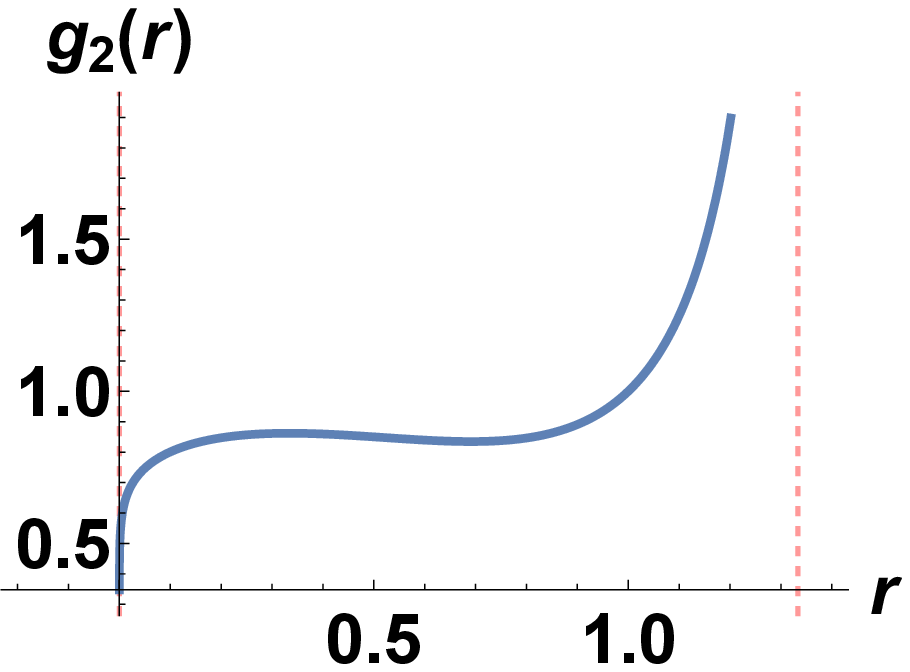}
\caption{$g_2$ solution}
  \end{subfigure}
  \caption{Numerical plots of the warp factors for the $SO(2)$ symmetric solution in pure $N=2$ gauged supergravity with $s=1$, $m=\frac{3}{4}$, $B=1$, $h=\frac{1}{4}$, and $\mathcal{C}=-\frac{1}{2}$. The warp factors are positive in the range $0<r<\frac{1}{m}=\frac{4}{3}$ with the two vertical red dashed lines representing the two boundaries.}
  \label{Case6_Soln}
\end{figure}

For the $SO(2)_R$ vector, we find that taking $r=r_\ast=\frac{3}{4m}$ in \eqref{SO(2)R_soln} results in the $SO(2)_R$ gauge field
\begin{equation}
A_1(r_\ast)=-\frac{q}{8mh}\, .
\end{equation}
Therefore, we will set $q=0$ in order to make the $SO(2)_R$ gauge field vanish at $r=r_\ast$.\\
%%%%%
\textbf{Case VI:} $s=1$, $m>0$, $B>\frac{3\sqrt{3}}{16m\sqrt{m}}$, $0<r<\frac{1}{m}$\\
\indent This case combines the behaviour near $r=0$ in Case I and the feature near $r=\frac{1}{m}$ in Case III. An example of numerical solution with $m=\frac{3}{4}$, $B=1$, $h=\frac{1}{4}$, and $\mathcal{C}=-\frac{1}{2}$ is given in Figure \ref{Case6_Soln}.

\begin{figure}[h!]
  \centering
  \begin{subfigure}[b]{0.326\linewidth}
    \includegraphics[width=\linewidth]{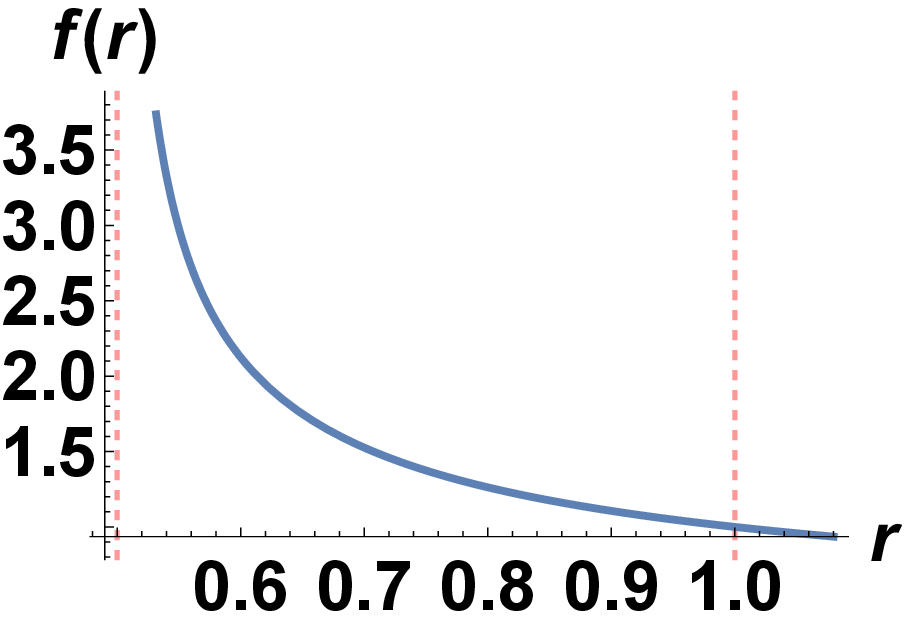}
\caption{$f$ solution}
  \end{subfigure}
  \begin{subfigure}[b]{0.326\linewidth}
    \includegraphics[width=\linewidth]{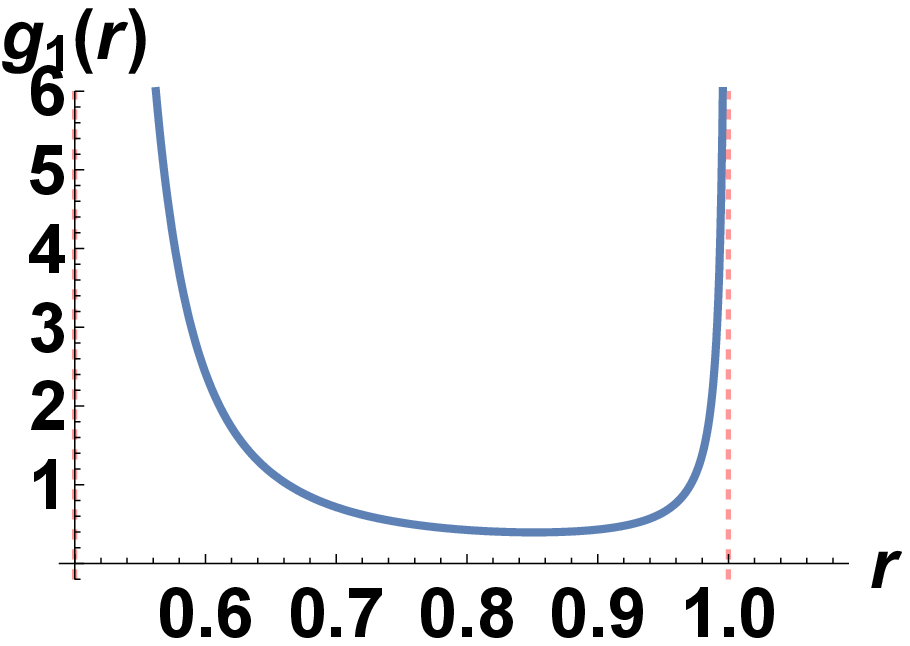}
\caption{$g_1$ solution}
  \end{subfigure}
  \begin{subfigure}[b]{0.326\linewidth}
    \includegraphics[width=\linewidth]{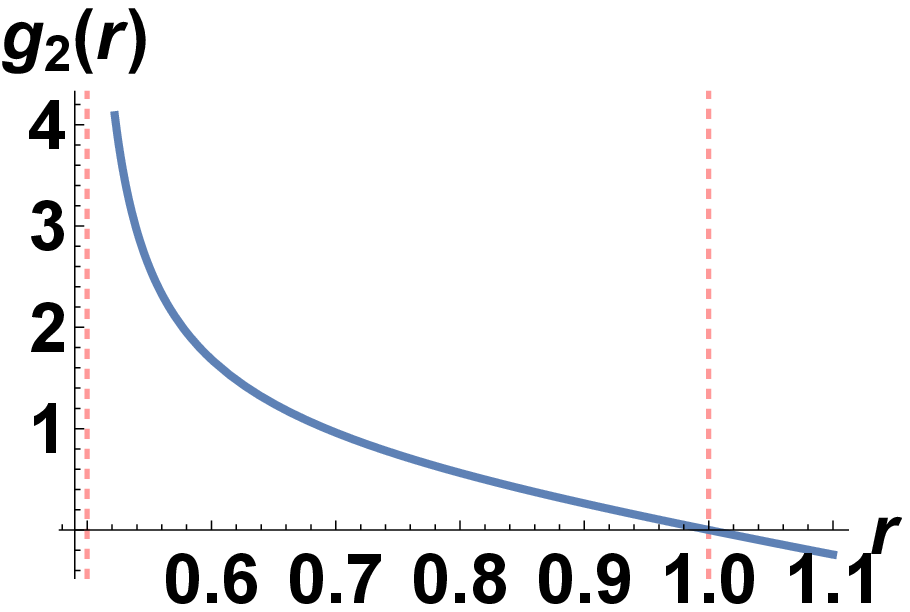}
\caption{$g_2$ solution}
  \end{subfigure}
  \caption{Numerical plots of the warp factors for the $SO(2)$ symmetric solution in pure $N=2$ gauged supergravity with $s=-1$, $m=2$, $B=1$, $h=\frac{1}{4}$, and $\mathcal{C}=-\frac{1}{2}$. The warp factors are positive in the range $\frac{1}{m}=\frac{1}{2}<r<r_{(+,+)}=1$ with the two vertical red dashed lines representing the two boundaries.}
  \label{Case7_Soln}
\end{figure}

\noindent \textbf{Case VII:} $s=-1$, $m>0$, $B>0$, $\frac{1}{m}<r<r_{(+,+)}$\\
\indent This is the only case with $s=-1$ with a numerical solution given in Figure \ref{Case7_Soln}. In comparison with other cases, we find that this case is very similar to Case III with the two boundaries interchanged. As $r\rightarrow\frac{1}{m}$, setting $r=\frac{1}{m}+16\mathcal{C}^4m^2B^2R^4$ , we find the seven-dimensional metric given in \eqref{case3_metric}. Thus, the seven-dimensional space-time is conformal to a product of $AdS_5$ and a cylinder as in Case III.

On the other side, as $r\rightarrow r_{(+,+)}$, we find the following form of the seven-dimensional metric
\begin{equation}\label{CaseVII_spindle_metric}
ds_7^2\approx\frac{B\,r_{(+,+)}^{1/10}}{16h^2\sqrt{mr_{(+,+)}-1}}\left[ds^2_{AdS_5}+\frac{dR^2+4\mathcal{C}^2\left[4mr_{(+,+)}-3\right]^2R^2dz^2}{-4W'(r_{(+,+)})\sqrt{r_{(+,+)}}(mr_{(+,+)}-1)^{3/2}}\right]
\end{equation}
after changing the radial coordinate to $\rho=\sqrt{r_{(+,+)}-r}$ as in Case III. The $z$ circle shrinks smoothly if we impose
\begin{equation}\label{CaseVII_smooth_con}
|\mathcal{C}|=\frac{1}{2l\left[4mr_{(+,+)}-3\right]},\qquad l=1,2,3,\ldots\, .
\end{equation}
The $SO(2)_R$ gauge field is given by \eqref{CaseIII_SO(2)R_soln} with the constant $q$ chosen to be
\begin{equation}
q=|\mathcal{C}|\left[4mr_{(+,+)}-3\right]=\frac{1}{2l}\, .
\end{equation}
Unlike all the previous cases with $s=1$, only the lower component of the Killing spinor $\eta$ is non-vanishing as $r\rightarrow r_{(+,+)}$
\begin{equation}\label{CaseVII_spindle_eta}
\eta=-2^{3/4}Y_0e^{\frac{iz}{2l}}r^{2/5}_{(+,+)}\begin{footnotesize}\begin{pmatrix} 0 \\ 1 \end{pmatrix}\end{footnotesize}.
\end{equation}
\indent We end this section by pointing out that in the matter-coupled $N=2$ gauged supergravity with $SO(4)$ gauge group, there are three $SO(2)_R$ singlet scalars corresponding to $SO(3,3)$ non-compact generators $Y_{31}$, $Y_{32}$, and $Y_{33}$. We have also looked for $SO(2)_R$ symmetric solutions in this case. However, in order to satisfy the vector field equation \eqref{Vec_eq}, it turns out that either the $SO(2)_R$ gauge field needs to be constant or all three vector-multiplet scalars must vanish. This implies that the $SO(2)_R$ symmetric $AdS_5\times \Sigma$ solution can only exist in pure $N=2$ gauged supergravity.

%%%%%%%%%%%%%%%%%%%%%%%%%%%%%%%%%%%%%%%%%%%%%%%%%%%%%%%%%
\subsubsection{$SO(2)\times SO(2)$ symmetric solution}\label{SO(4)_SO(2)xSO(2)_Soln_Sec}
We now repeat the same procedure for more complicated solutions with $SO(2)\times SO(2)$ symmetry. With the explicit form of $A'_1$ and $A'_2$ given in \eqref{Gen_SO(2)xSO(2)_Ansatz}, non-vanishing components of the dressed field strength tensors read
\begin{equation}
\mathbf{F}_1=\frac{e^{-\sigma-\phi}}{2}(a_1+a_2e^{2\phi})f^{-\frac{5}{2}},\qquad \mathbf{F}_2=\frac{e^{-\sigma-\phi}}{2}(a_1-a_2e^{2\phi})f^{-\frac{5}{2}}\, .
\end{equation}
We can now solve equation \eqref{AABPS6} and find the solution for the $SO(2)\times SO(2)$ singlet scalar $\phi$ of the form
\begin{equation}
\phi=\ln\left[\frac{e^{-\frac{5\sigma}{2}}}{32a_2h}\left[(a_2-\lambda a_1)\widetilde{g}_1+K\right]\right]\label{phi_gen_sol}
\end{equation}
in which
\begin{equation}\label{Gen_K_fn}
K=\kappa\sqrt{1024a_1a_2h^2e^{5\sigma}+(a_2-\lambda a_1)^2\widetilde{g}_1^2}
\end{equation}
and $\kappa=\pm1$. We have included a sign factor $\lambda=+1,-1$ for later convenience. In the present case of $SO(4)$ gauge group, we can set the constant $\lambda=1$. In the next section, we will consider non-compact $SO(2,2)$ gauge group and find that the solution takes the same form with $\lambda=-1$. 

With the solution for $\phi$ given above, we can solve equations \eqref{ABCBPS2} and \eqref{DABPS3} resulting in the following solutions for $f$ and $g_1$ of the form
\begin{eqnarray}
f&\hspace{-0.2cm}=&\hspace{-0.2cm}\sqrt{2}\sqrt{\frac{he^{2\sigma}\left[(a_2+\lambda a_1)\widetilde{g}_1+K\right]}{s(256h^2e^{5\sigma}-\lambda\widetilde{g}_1^2)}},\label{f_gen_sol}\\
g_1&\hspace{-0.2cm}=&\hspace{-0.2cm}\frac{25,600\sqrt{2}h^{9/2}e^{6\sigma}K^{-2}(256h^2e^{5\sigma}-\lambda\widetilde{g}_1^2)^{-2}\left[(a_2+\lambda a_1)\widetilde{g}_1+K\right]^{5/2}(\sigma')^2}{\left[16\sqrt{2}h^{9/2}\sqrt{(a_2+\lambda a_1)\widetilde{g}_1+K}+\lambda e^{-5\sigma}\sqrt{s(256h^2e^{5\sigma}-\lambda\widetilde{g}_1^2)}\right]}.\nonumber\\ \label{g1_gen_sol}
\end{eqnarray}
\indent We now determine the function $\widehat{A}_1$ defined in \eqref{AandA}. With all the previous results, the first equation in \eqref{Gen_SO(2)xSO(2)_Ansatz} becomes
\begin{equation}
\widehat{A}'_1=\frac{10 \widetilde{g}_1 \widehat{A} \sigma' \left[K^2 (a_2+\lambda a_1)-2 K \widetilde{g}_1 (a_2-\lambda a_1)^2+\widetilde{g}_1^2 (a_2+\lambda a_1) (a_2-\lambda a_1)^2\right]}{K \left[3 K^2-4 K \widetilde{g}_1 (a_2+\lambda a_1)+\widetilde{g}_1^2 (a_2-\lambda a_1)^2\right]}\, .
\end{equation}
Solving this equation leads to the following solution 
\begin{eqnarray}
\widehat{A}_1\hspace{-0.2cm}&=&\hspace{-0.2cm}\frac{c}{8}e^{-5\sigma}\sqrt{\frac{9 K^4-2 K^2 \widetilde{g}_1^2 \left(5 a_1^2+22\lambda a_1 a_2+5 a_2^2\right)+\widetilde{g}_1^4 (a_2-\lambda a_1)^4}{a_1 a_2}}\nonumber\\&&\times\exp \left[-\frac{\left(a_1^2+2a_1( 7\lambda a_2-y)+a_2(a_2-2\lambda  y)\right) \tan ^{-1}\left(\frac{3 K}{u \widetilde{g}_1}\right)}{\lambda u y}\right.\nonumber\\&&\qquad\quad\left.-\frac{\left(a_1^2+2 a_1 (7\lambda a_2+y)+a_2 (a_2+2\lambda y)\right) \tanh ^{-1}\left(\frac{3 K}{v \widetilde{g}_1}\right)}{\lambda v y}\right]\quad\label{Gen_A1_soln}
\end{eqnarray}
with an integration constant $c$ and
\begin{eqnarray}
u&=&\sqrt{4 a_1 y+4\lambda a_2 y-5 a_1^2-22\lambda a_1a_2-5 a_2^2},\\
v&=&\sqrt{4 a_1 y+4\lambda a_2 y+5 a_1^2+22\lambda a_1 a_2+5 a_2^2},\\
y&=&\sqrt{a_1^2+14\lambda a_1 a_2+a_2^2}\, .
\end{eqnarray}
\indent Similarly, we can find the solution for $g_2$ as
\begin{eqnarray}
g_2\hspace{-0.2cm}&=&\hspace{-0.2cm}\frac{400 \widetilde{g}_1^2 f^5 \widehat{A}_1^2 e^{2 \sigma} \left[K^2-(a_2-\lambda a_1)^2\widetilde{g}_1^2\right](\sigma')^2 }{K^2 g_1 \left[3 K^2-4 K \widetilde{g}_1 (a_2+\lambda a_1)+\widetilde{g}_1^2 (a_2-\lambda a_1)^2\right]^2}\, .\label{g2_gen_sol}
\end{eqnarray}
Finding the solution for $A_2$ in the other $SO(2)$ gauge field is much more difficult since, with all the previous results, the second equation in \eqref{Gen_SO(2)xSO(2)_Ansatz} leads to a highly complicated differential equation. It should also be noted that $A_2$ does not appear in all of the BPS equations which depend only on $A_2'$ and $A_1'$. In particular, we have verified that all the BPS conditions are satisfied by the above solution without the explicit solution for $A_2$. 
\\
\indent To make the subsequent analysis of the solutions more traceable, we will further simplify the ansatz for vector fields by setting 
\begin{equation}
a_1=-a_2=b\, .
\end{equation}
In this case, we find the non-vanishing components of the dressed field strength tensors given by
\begin{equation}
\mathbf{F}_1=-be^{-\sigma}\sinh{\phi}f^{-\frac{5}{2}}\qquad\text{ and }\qquad \mathbf{F}_2=be^{-\sigma}\cosh{\phi}f^{-\frac{5}{2}}\, .
\end{equation}
The solution found above now takes a much simpler form, with $\lambda=1$,
\begin{eqnarray}
\phi&=&\kappa\cosh^{-1}\left[\frac{\widetilde{g}_1e^{-\frac{5\sigma}{2}}}{16h}\right],\\
f&=&\frac{2\sqrt{\kappa\,b\,h}e^\sigma}{\sqrt{s}(-256h^2e^{5\sigma}+\widetilde{g}_1^2)^{1/4}},\label{SO(4)_SO(2)xSO(2)_f_Soln}\\
g_1&=&\frac{51,200\sqrt{\kappa\,b\,h^9}\,e^{6\sigma}\,(\sigma')^2}{(-256h^2e^{5\sigma}+\widetilde{g}_1^2)^2\left[32\sqrt{\kappa\,b\,h^5}-\sqrt{s}e^{-5\sigma}(-256h^2e^{5\sigma}+\widetilde{g}_1^2)^{1/4}\right]},\qquad\\
g_2&=&\frac{256c_1^2h^2\widetilde{g}_1^2e^\sigma\left[32\sqrt{\kappa\,b\,h^5}-\sqrt{s}e^{-5\sigma}(-256h^2e^{5\sigma}+\widetilde{g}_1^2)^{1/4}\right]}{\sqrt{s}(-256h^2e^{5\sigma}+\widetilde{g}_1^2)^{1/4}},\\
\widehat{A}_1&=&c_1\left(192h^2-\widetilde{g}_1^2e^{-5\sigma}\right).
\end{eqnarray}
It can also be verified that all of the BPS conditions and the field equations are satisfied if 
\begin{equation}\label{alter_SO(2)xSO(2)_sign_Con}
\text{sign}(c_1\widetilde{g}_1\sigma')=-1\, .
\end{equation}
\indent In addition, the second equation in \eqref{Gen_SO(2)xSO(2)_Ansatz} gives the following differential equation
\begin{equation}
A'_2=\frac{5c_1\widetilde{g}_1(128h^2-\widetilde{g}_1^2e^{-5\sigma})\sigma'}{\sqrt{-256e^{5\sigma}h^2+\widetilde{g}_1^2}}
\end{equation}
which can be readily solved by
\begin{equation}
A_2=c_1\widetilde{g}_1e^{-5\sigma}\sqrt{-256h^2e^{5\sigma}+\widetilde{g}_1^2}+c_2
\end{equation}
with an integration constant $c_2$. We also have an explicit form of the $A_1$ solution 
\begin{equation}
A_1=c_1\left(192h^2-\widetilde{g}_1^2e^{-5\sigma}\right)-\frac{2q}{\widetilde{g}_1}
\end{equation}
together with the Killing spinor
\begin{equation}
\eta=\frac{Y_0\,e^{iqz-\frac{\sigma}{4}}}{s^{1/8}(-256h^2e^{5\sigma}+\widetilde{g}_1^2)^{\frac{1}{16}}}\begin{pmatrix} \sqrt{4\sqrt{2}(\kappa\,b\,h^5)^{\frac{1}{4}}+s^{\frac{1}{4}}e^{-\frac{5\sigma}{2}}(-256h^2e^{5\sigma}+\widetilde{g}_1^2)^{\frac{1}{8}}} \\ -\sqrt{4\sqrt{2}(\kappa\,b\,h^5)^{\frac{1}{4}}-s^{\frac{1}{4}}e^{-\frac{5\sigma}{2}}(-256h^2e^{5\sigma}+\widetilde{g}_1^2)^{\frac{1}{8}}}\end{pmatrix}
\end{equation}
with $Y_0$ being a constant. 

We now turn to the regularity of the solution. From the $f$ solution in \eqref{SO(4)_SO(2)xSO(2)_f_Soln}, it is immediately seen that we have to impose the conditions $-256e^{5\sigma}h^2+\widetilde{g}_1^2>0$ and $\frac{\kappa\,b}{s}>0$. The latter is clearly consistent with the condition sign$(\kappa bs)=+1$. As in the previous section, we define the following parameters
\begin{equation}
B=\frac{8h^{2}\sqrt{\kappa\, b}}{\sqrt{s}},\qquad m=\frac{\widetilde{g}_1}{16h},\qquad \mathcal{C}=32\,\widetilde{g}_1h^2c_1\, .
\end{equation}
The complete solution then reads
\begin{eqnarray}
f&=&\frac{Be^\sigma}{16h^2(-e^{5\sigma}+m^2)^{1/4}},\\
g_1&=&\frac{25Be^{6\sigma}\,(\sigma')^2}{1024h^2(-e^{5\sigma}+m^2)^2\left[B-e^{-5\sigma}(-e^{5\sigma}+m^2)^{1/4}\right]},\qquad\\
g_2&=&\frac{\mathcal{C}^2e^\sigma\left[B-e^{-5\sigma}(-e^{5\sigma}+m^2)^{1/4}\right]}{4h^2(-e^{5\sigma}+m^2)^{1/4}},\\
A_1&=&\frac{1}{8mh}\left[\mathcal{C}(3-4m^2e^{-5\sigma})-q\right],\\
A_2&=&\frac{\mathcal{C}e^{-5\sigma}}{2h}\sqrt{-e^{5\sigma}+m^2}+c_2,\\
\phi&=&\kappa\cosh^{-1}\left[me^{-\frac{5\sigma}{2}}\right].
\end{eqnarray}
\indent We will also fix the form of $\sigma$ solution to
\begin{equation}\label{1/5sigma_soln}
\sigma=-\frac{1}{5}\ln r
\end{equation}
for $r>0$. This form of $\sigma$ implies that the constant $\mathcal{C}$ can only be a positive integer due to \eqref{alter_SO(2)xSO(2)_sign_Con}. 

\begin{figure}[h!]
  \centering
  \begin{subfigure}[b]{0.326\linewidth}
    \includegraphics[width=\linewidth]{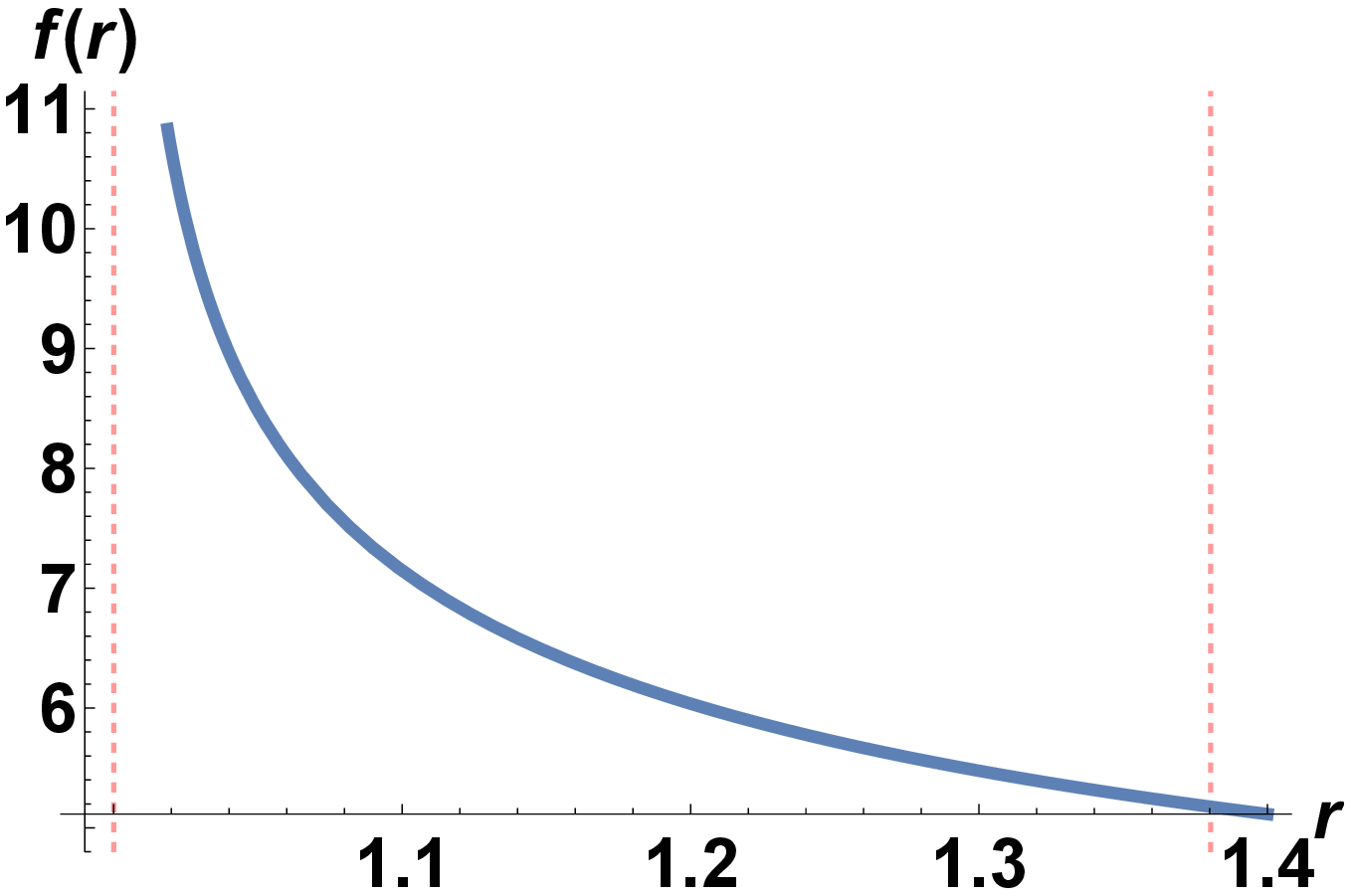}
\caption{$f$ solution}
  \end{subfigure}
  \begin{subfigure}[b]{0.326\linewidth}
    \includegraphics[width=\linewidth]{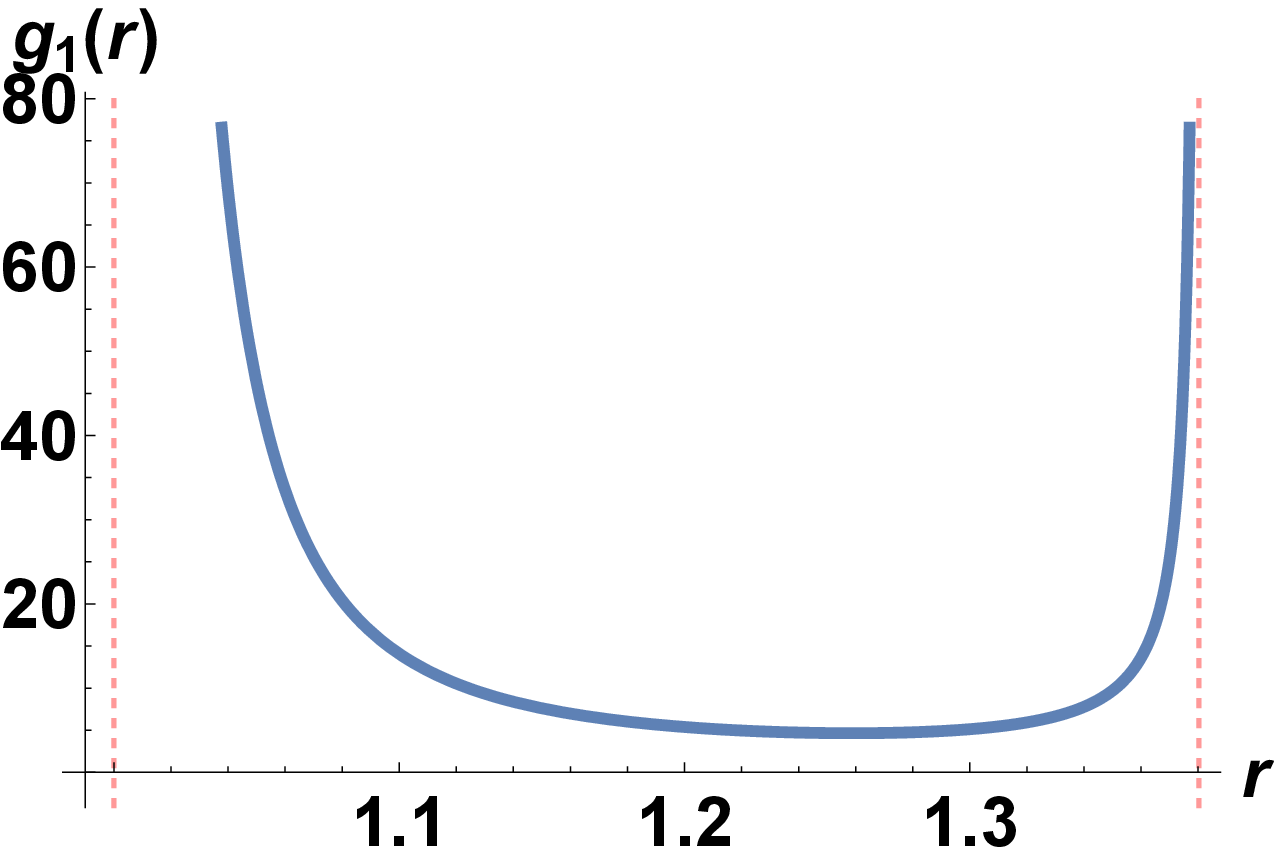}
\caption{$g_1$ solution}
  \end{subfigure}
  \begin{subfigure}[b]{0.326\linewidth}
    \includegraphics[width=\linewidth]{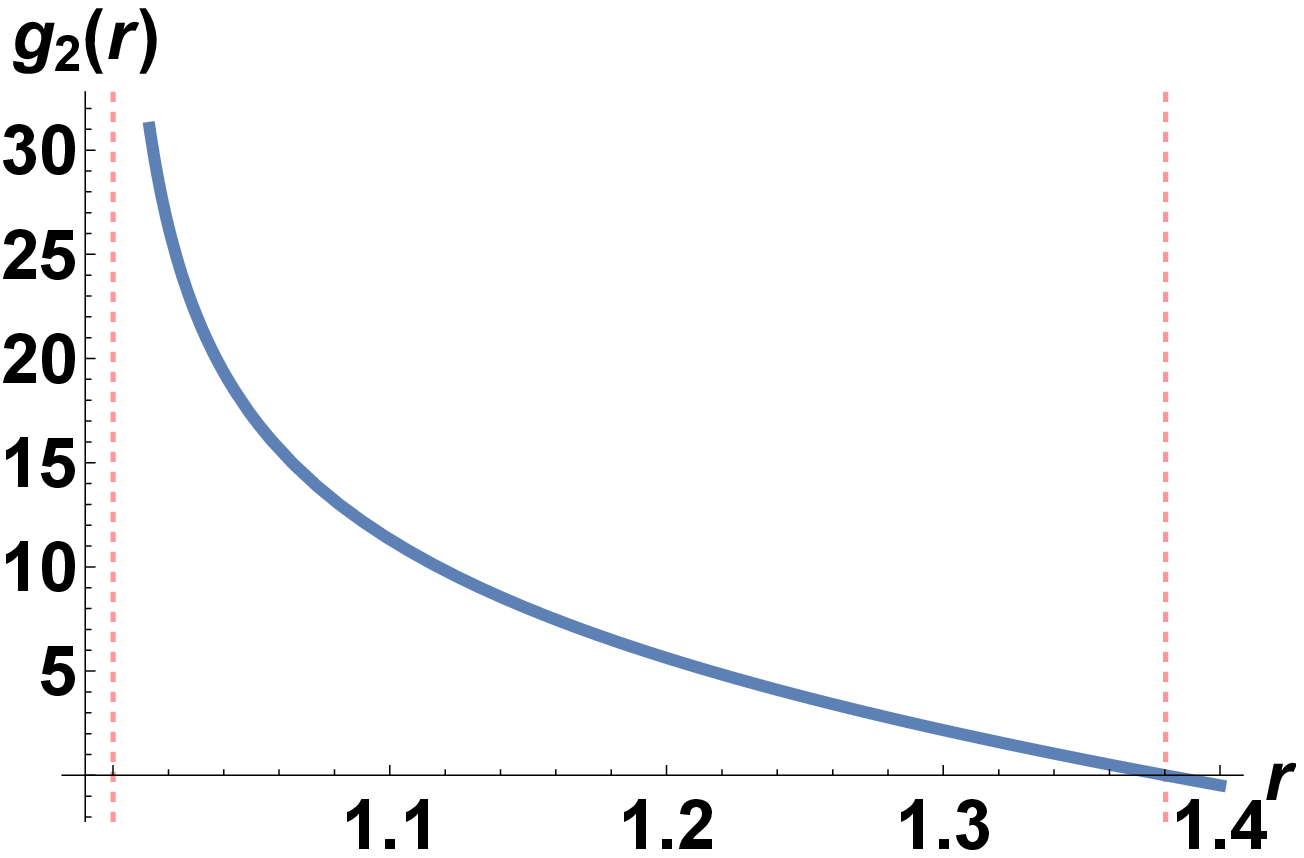}
\caption{$g_2$ solution}
  \end{subfigure}
  \caption{Numerical plots of the warp factors for $SO(2)\times SO(2)$ symmetric solution in $SO(4)$ gauged supergravity with $m=1$, $B=1$, $h=\frac{1}{8}$, and $\mathcal{C}=1$. The solution is regular in the range $\frac{1}{m^2}=1<r<r_1=1.38$ with the two vertical red dashed lines representing the two boundaries.}
  \label{SO(4)_SO(2)xSO(2)_plots}
\end{figure}

In terms of the radial coordinate, the seven-dimensional metric can be written as
\begin{equation}
ds^2_7=\frac{Br^{1/20}}{16h^2(m^2r-1)^{1/4}}\left[ds^2_{AdS_5}+\frac{r^{-5/4}}{64W(m^2r-1)^{7/4}}dr^2+\frac{4\mathcal{C}^2W}{B}dz^2\right]\label{SO(4)_SO(2)xSO(2)_Met}
\end{equation}
with
\begin{equation}\label{SO(4)_SO(2)xSO(2)_Def_W}
W=B-r^{3/4}(m^2r-1)^{1/4}
\end{equation}
while the solutions for $\phi$, $A_1$, and $A_2$ are respectively given by
\begin{eqnarray}
\phi&=&\kappa \cosh^{-1}\left[m\sqrt{r}\right],\\
A_1&=&\frac{1}{8mh}\left[\mathcal{C}(3-4m^2r)-q\right],\\
A_2&=&\frac{\mathcal{C}}{2h}\sqrt{r(m^2r-1)}+c_2\, .
\end{eqnarray}
Finally, the Killing spinor $\eta$ takes the form
\begin{equation}\label{SO(4)_SO(2)xSO(2)_eta_Soln}
\eta=Y_0e^{iqz}\frac{r^{1/80}}{(m^2r-1)^{\frac{1}{16}}}\begin{pmatrix} \sqrt{\sqrt{B}+r^{3/8}(m^2r-1)^{\frac{1}{8}}} \\ -\sqrt{\sqrt{B}-r^{3/8}(m^2r-1)^{\frac{1}{8}}} \end{pmatrix}.
\end{equation}
\indent In the present case, we find only one possible range of the radial coordinate $r$ that leads to a regular solution
\begin{equation}\label{SO(4)_SO(2)xSO(2)_Pos_Range}
m\neq0,\qquad B>0,\qquad \frac{1}{m^2}<r<r_1
\end{equation}
with $r_1$ determined from $W(r_1)=0$. The explicit form of $r_1$ is given by
\begin{equation}\label{SO(4)_SO(2)xSO(2)_r1}
r_1=\frac{1}{4m^2}+\frac{1}{2}\sqrt{X_1}+\frac{1}{2}\sqrt{\frac{3}{4m^4}-X_1+\frac{1}{4m^6\sqrt{X_1}}}
\end{equation}
in which 
\begin{equation}
X_1=\frac{1}{4m^4}-\frac{4(\frac{2}{3})^{1/3}B^{8/3}}{(-9+\sqrt{3(27+256B^4m^6)})^{1/3}}+\frac{B^{4/3}(-9+\sqrt{3(27+256B^4m^6)})^{1/3}}{18^{1/3}m^2}\, .
\end{equation}
Note also that the two possibilities with $m>0$ and $m<0$ are equivalent. For definiteness, we will only consider $m>0$ case. An example of numerical plots of the three warp factors for $m=1$, $B=1$, $h=\frac{1}{8}$, and $\mathcal{C}=1$ is given in Figure \ref{SO(4)_SO(2)xSO(2)_plots}.

\begin{figure}[h!]
  \centering
    \includegraphics[width=0.56\linewidth]{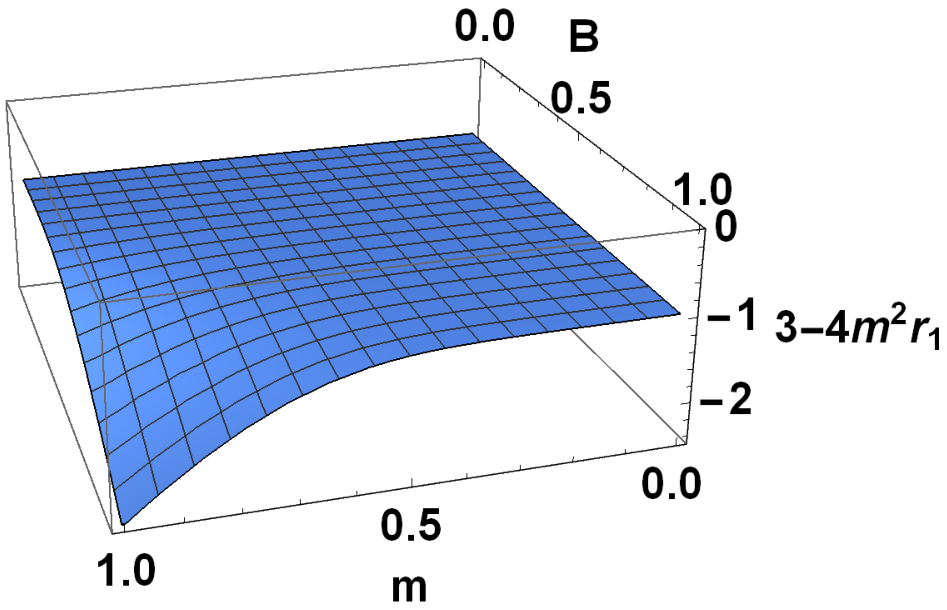}
  \caption{A numerical plot of the function $3-4m^2r_1$ appearing in the condition \eqref{SO(4)_SO(2)xSO(2)_smooth_con}. Note also that $3-4m^2r_1\leq -1$ in the regularity range \eqref{SO(4)_SO(2)xSO(2)_Pos_Range}.}
  \label{SO(4)_SO(2)xSO(2)_sur}
\end{figure}

The result is very similar to Case VII in the previous section. As $r\rightarrow \frac{1}{m^2}$, the seven-dimensional metric becomes conformal to a product of $AdS_5$ and a cylinder. With the new radial coordinate $R$ given by $r=\frac{1}{m^2}+256C^8m^4B^4R^8$, the metric near $R=0$ is given by \eqref{case3_metric}. On the other side, as $r\rightarrow r_1$, the seven-dimensional metric is approximately given by
\begin{equation}\label{SO(4)_SO(2)xSO(2)_spindle_metric}
ds_7^2\approx\frac{B\,r_1^{1/20}}{16h^2(m^2r_1-1)^{1/4}}\left[ds^2_{AdS_5}+\frac{dR^2+4\mathcal{C}^2\left[3-4m^2r_1\right]^2R^2dz^2}{-16W'(r_1)r_1^{5/4}(m^2r_1-1)^{7/4}}\right]
\end{equation}
in which $R$ is the new radial coordinate defined by $R=\sqrt{r_1-r}$. The $z$ circle shrinks smoothly giving rise to an $\mathbb{R}^2/\mathbb{Z}_l$ orbifold at $r=r_1$ if we impose
\begin{equation}\label{SO(4)_SO(2)xSO(2)_smooth_con}
\mathcal{C}=-\frac{1}{2l\left[3-4m^2r_1\right]},\qquad l=1,2,3,\ldots\, .
\end{equation}
As in the previous section, the explicit form of $3-4m^2r_1$ is very complicated, so we will not give it here but only show that this function is negative and less than -1 in the regularity range by a numerical plot in Figure \ref{SO(4)_SO(2)xSO(2)_sur}.

In order for the $SO(2)\times SO(2)$ gauge fields to vanish at $r=r_1$, we fix the constants $q$ and $c_2$ to be
\begin{eqnarray}
& &q=-\mathcal{C}\left[3-4m^2r_1\right]=\frac{1}{2l}\nonumber \\
\textrm{and}\qquad& & c_2=-\frac{\mathcal{C}}{2h}\sqrt{r_1(1+m^2r_1)}=-\frac{\mathcal{C}B}{2h}
\end{eqnarray}
leading to
\begin{eqnarray}
& &A_1=\frac{m\,\mathcal{C}}{2h}(r_1-r)\nonumber \\
\textrm{and}\qquad & & A_2=\frac{\mathcal{C}}{2h}\left[\sqrt{r(m^2r-1)}-\sqrt{r_1(m^2r_1-1)}\right].
\end{eqnarray}
We also note that $A_1$ is well-defined for all values of $r$ while $A_2$ is complex for $r<\frac{1}{m^2}$. 

The Killing spinor $\eta$ at the endpoint $r=r_1$ is given by
\begin{equation}
\eta=\sqrt{2}Y_0e^{\frac{iz}{2l}}r_1^{1/5}\begin{footnotesize}\begin{pmatrix} 1 \\ 0 \end{pmatrix}\end{footnotesize}.
\end{equation}

%%%%%%%%%%%%%%%%%%%%%%%%%%%%%%%%%%%%%%%%%%%%%%%%%%%%%%%%%%%%%%%%%%%%%%%%%%%%%%%%
\subsection{$SO(2)_{\text{diag}}$ symmetric solution}\label{SO(4)_SO(2)diag_Sec}
We now consider $AdS_5\times \Sigma$ solutions with $SO(2)_{\text{diag}}\subset SO(2)\times SO(2)$ symmetry generated by $J^{(1)}_{12}+ J^{(2)}_{12}$. In this case, the two $SO(2)$ gauge fields are related by
\begin{equation}\label{SO(2)diag_Vec_Con}
\widetilde{g}_2A_2=\widetilde{g}_1A_1
\end{equation}
resulting in the following ansatz for the gauge fields
\begin{equation}\label{SO(2)diag_Gaugefield}
A^{I}_{(1)}=A_1\left(\delta^I_3+\frac{\widetilde{g}_1}{\widetilde{g}_2}\delta^I_6\right)dz\, .
\end{equation}
The corresponding two-form field strength is given by
\begin{equation}
F^{I}_{(2)}=A'\left(\delta^I_3+\frac{\widetilde{g}_1}{\widetilde{g}_2}\delta^I_6\right)\,dr\wedge dz\, .
\end{equation}
We can also consistently set $C_{(3)}=0$ as in the previous cases.

Apart from the $SO(2)\times SO(2)$ singlet scalar $\phi$ corresponding to the non-compact-generator $Y_{33}$, there are two additional scalars from $SO(3,3)/SO(3)\times SO(3)$ coset that are invariant under $SO(2)_{\text{diag}}$. These scalars correspond to the non-compact generators
\begin{equation}\label{hatY12}
\hat{Y}_1=Y_{11}+Y_{22}\qquad\text{ and }\qquad \hat{Y}_2=Y_{12}-Y_{21}\, .
\end{equation}
The coset representative then takes the form of
\begin{equation}
L=e^{\varphi_1\hat{Y}_1}e^{\phi Y_{33}}e^{\varphi_2\hat{Y}_2}\, .\label{SO2d_scalar}
\end{equation}
This leads to the following non-vanishing $C$-functions 
\begin{eqnarray}
C&\hspace{-0.2cm}=&\hspace{-0.2cm}\frac{3}{\sqrt{2}}\left[\widetilde{g}_1\cosh{\phi}-\widetilde{g}_2\sinh{\phi}+\cosh{2\varphi_1}\cosh{2\varphi_2}\left(\widetilde{g}_1\cosh{\phi}+\widetilde{g}_2\sinh{\phi}\right)\right],\nonumber\\
C^{11}&\hspace{-0.2cm}=&\hspace{-0.2cm}C^{22}=-\frac{1}{\sqrt{2}}\left(\widetilde{g}_1\cosh{\phi}+\widetilde{g}_2\sinh{\phi}\right)\sinh{2\varphi_1},\nonumber\\
C^{12}&\hspace{-0.2cm}=&\hspace{-0.2cm}-C^{21}=-\frac{1}{\sqrt{2}}\left(\widetilde{g}_1\cosh{\phi}+\widetilde{g}_2\sinh{\phi}\right)\cosh{2\varphi_1}\sinh{2\varphi_2},\nonumber\\
C^{33}&\hspace{-0.2cm}=&\hspace{-0.2cm}\frac{1}{\sqrt{2}}\left[\widetilde{g}_2\cosh{\phi}-\widetilde{g}_1\sinh{\phi}-\cosh{2\varphi_1}\cosh{2\varphi_2}\left(\widetilde{g}_2\cosh{\phi}+\widetilde{g}_1\sinh{\phi}\right)\right].\nonumber\\\label{SO(4)_SO(2)diag_Cfn}
\end{eqnarray}
The scalar vielbein is given by
\begin{equation}
P^{ir}_{(1)}=\begin{pmatrix}	\varphi'_1\cosh{2\varphi_2} & \varphi'_2 & 0 \\
						-\varphi'_2 & \varphi'_1\cosh{2\varphi_2} & 0 \\
						0 & 0 & \phi'
			\end{pmatrix} dr
\end{equation}
while the composite connection takes the form of
\begin{equation}
Q^{ij}_{(1)}=-\varepsilon^{ij3}(\varphi'_1\sinh{2\varphi_2}dr-\widetilde{g}_1A_1dz).
\end{equation}

With all these, the vector field equation \eqref{Vec_eq} gives rise to the following equations
\begin{eqnarray}
\varphi'_2&=&\frac{\varphi'_1}{2}\text{coth}{2\varphi_1}\sinh{4\varphi_2},\\
A''_1&\hspace{-0.2cm}=&\hspace{-0.2cm}-\left[\frac{5f'}{2f}-\frac{g'_1}{2g_1}-\frac{g'_2}{2g_2}+\sigma'+\frac{2(\widetilde{g}_1\cosh{2\phi}+\widetilde{g}_2\sinh{2\phi})\phi'}{\widetilde{g}_2\cosh{2\phi}+\widetilde{g}_1\sinh{2\phi}}\right]A'_1,\\
A''_1&\hspace{-0.2cm}=&\hspace{-0.2cm}-\left[\frac{5f'}{2f}-\frac{g'_1}{2g_1}-\frac{g'_2}{2g_2}+\sigma'+\frac{2(\widetilde{g}_2\cosh{2\phi}+\widetilde{g}_1\sinh{2\phi})\phi'}{\widetilde{g}_1\cosh{2\phi}+\widetilde{g}_2\sinh{2\phi}}\right]A'_1.\qquad
\end{eqnarray}
Consistency between the last two equations for $\phi'\neq0$ requires
\begin{equation}\label{SO(2)diag_Con}
\widetilde{g}_2=\pm\widetilde{g}_1
\end{equation}
leading to a differential equation for the $SO(2)_{\text{diag}}$ gauge field
\begin{equation}
A''_1=-\left(\frac{5f'}{2f}-\frac{g'_1}{2g_1}-\frac{g'_2}{2g_2}+\sigma'\pm2\phi'\right)A'_1\, .
\end{equation}
The plus/minus sign arises from that in the relation $\widetilde{g}_2=\pm\widetilde{g}_1$. From now on, we will choose $\widetilde{g}_2=\widetilde{g}_1$ for definiteness and find
\begin{equation}
A'_1=A'_2=be^{-\sigma-2\phi}\sqrt{g_1g_2}f^{-\frac{5}{2}}\, .
\end{equation}
Non-vanishing components of the dressed field strength tensors are given by
\begin{equation}\label{SO(4)_SO(2)diag_F1F2}
\mathbf{F}_1=\mathbf{F}_2=be^{-\sigma-\phi}f^{-\frac{5}{2}}\, .
\end{equation}
\indent Unlike the previous cases, there are additional BPS conditions arising from the transformations $\delta\lambda^{a1}=0$ and $\delta\lambda^{a2}=0$ resulting in
\begin{equation}
0=\left[\frac{\varphi'_1\cosh{2\varphi_2}}{\sqrt{g_1}}\Gamma^{\hat{r}}-\frac{1}{\sqrt{2}}e^{-\frac{\sigma}{2}}C^{11}\right]{(\sigma^1)^a}_b\epsilon^b+\left[\frac{\varphi'_2}{\sqrt{g_1}}\Gamma^{\hat{r}}-\frac{1}{\sqrt{2}}e^{-\frac{\sigma}{2}}C^{12}\right]{(\sigma^2)^a}_b\epsilon^b\, .
\end{equation}
We note here that the projector in \eqref{ProjCon} implies
\begin{equation}
 {(\sigma^2)^a}_b\epsilon^b=i{(\sigma^1)^a}_b\epsilon^b\, . 
\end{equation} 
Therefore, with the supersymmetry parameter \eqref{7DKilling}, we can rewrite the above condition as
\begin{equation}\label{Addi_SO(4)_SO(2)diag_BPS_eq}
0=\frac{1}{\sqrt{g_1}}\left(\varphi'_1\cosh{2\varphi_2}+i \varphi'_2\right)\sigma^3\eta -\frac{1}{\sqrt{2}}e^{-\frac{\sigma}{2}}\left(C^{11}+i C^{12}\right)(i\sigma^2\eta).
\end{equation}
There are seventeen BPS conditions obtained from the vanishing of the matrices $\mathcal{A}$, $\mathcal{B}$, and $\mathcal{C}$ derived from this BPS equation. Two of these are given by
\begin{eqnarray}
0&=&\frac{s}{\sqrt{f\,g_1}}(\varphi'_1\cosh{2\varphi_2}+i \varphi'_2),\\
0&=&-\frac{\widetilde{g}_1}{\sqrt{2}}e^{\phi}A'_1(\sinh{2\varphi_1}+i\cosh{2\varphi_1}\sinh{2\varphi_2}).\quad
\end{eqnarray}
For non-vanishing $\widetilde{g}_1$ and $A'_1$, the second condition is satisfied only by setting $\varphi_1=\varphi_2=0$. All the remaining conditions are also satisfied by this result. 

In addition, repeating the same procedure as in the previous sections with $\varphi_1=\varphi_2=0$, we find the same set of the BPS conditions \eqref{DABPS1} to \eqref{SBCBPS6}. The resulting $SO(2)_{\text{diag}}$ symmetric solution is then given by
\begin{eqnarray}
f&=&\frac{2\sqrt{2}\sqrt{\widetilde{g}_1b\,h}e^\sigma}{\sqrt{s(256h^2e^{5\sigma}-\widetilde{g}_1^2)}},\label{SO(4)_SO(2)diag_Soln1}\\
g_1&=&\frac{204,800\sqrt{2}\sqrt{\widetilde{g}_1b\,h^9}\,e^{6\sigma}\,(\sigma')^2}{(256h^2e^{5\sigma}-\widetilde{g}_1^2)^2\left[32\sqrt{2}\sqrt{b\,h^5}-e^{-5\sigma}\sqrt{s(256h^2e^{5\sigma}-\widetilde{g}_1^2)}\right]},\\
g_2&=&\frac{1,024c^2\widetilde{g}_1^2h^2e^\sigma\left[32\sqrt{2}\sqrt{\widetilde{g}_1b\,h^5}-e^{-5\sigma}\sqrt{s(256h^2e^{5\sigma}-\widetilde{g}_1^2)}\right]}{\sqrt{s(256h^2e^{5\sigma}-\widetilde{g}_1^2)}},\\
\phi&=&\frac{5\sigma}{2}+\ln\left[\frac{16h}{\widetilde{g}_1}\right],\label{SO(4)_SO(2)diag_phi2_vector_Soln}\\
A_1&=&c\left(128h^2-\widetilde{g}_1^2e^{-5\sigma}\right)-\frac{2q}{\widetilde{g}_1}\label{SO(4)_SO(2)diag_Soln4}
\end{eqnarray}
with the usual sign condition $\text{sign}(c\widetilde{g}_1\sigma')=+1$. The Killing spinor is given by
\begin{equation}
\eta=\frac{Y_0\,e^{iqz+\frac{\sigma}{4}}}{\left[s(256h^2e^{5\sigma}-\widetilde{g}_1^2)\right]^{\frac{1}{8}}}\begin{pmatrix} \sqrt{8(\widetilde{g}_1b\,h^5)^{\frac{1}{4}}+se^{-\frac{5\sigma}{2}}\left[2s(256h^2e^{5\sigma}-\widetilde{g}_1^2)\right]^{\frac{1}{4}}} \\ -\sqrt{8(\widetilde{g}_1b\,h^5)^{\frac{1}{4}}-se^{-\frac{5\sigma}{2}}\left[2s(256h^2e^{5\sigma}-\widetilde{g}_1^2)\right]^{\frac{1}{4}}}\end{pmatrix}.\label{SO(4)_SO(2)diag_Soln5}
\end{equation}
\indent
It should be remarked that the half-spindle solution found in \cite{Bah_M5} from the $U(1)^2$ truncation of the maximal seven-dimensional gauged supergravity can also be recovered from the above solution by the following identification 
\begin{eqnarray}
\sigma&=&-\frac{2}{3}\lambda,\qquad \widetilde{g}_1=2m,\qquad b=\frac{8B^2}{m^4},\nonumber \\ 
c&=&\frac{\mathcal{C}^2}{2m^2}, \qquad h=\frac{m}{8} 
\end{eqnarray}
where $\lambda$, $B$, $m$, and $\mathcal{C}$ are the scalar field and constants used in \cite{Bah_M5}. In more detail, the non-vanishing gauge field in the solution of \cite{Bah_M5} is given by
\begin{equation}
A^{(1)}=A^{12}=\frac{1}{2}(A^3+A^6)=A^3
\end{equation} 
with $A^6=A^3$. The scalar fields are related by the following form of the unimodular matrix
\begin{equation}
T_{ab}=\text{diag}(-\frac{\sigma}{2}-\phi,-\frac{\sigma}{2}-\phi,-\frac{\sigma}{2}+\phi,-\frac{\sigma}{2}+\phi,\,2\sigma).\end{equation} 
with $\sigma=-2(\lambda_1+\lambda_2)$ and $\phi=\lambda_2-\lambda_1$. In this equation, $\lambda_1$ and $\lambda_2$ are the two scalars of the $U(1)^2$ truncation of $N=4$ gauged supergravity used in \cite{Bah_M5}. Therefore, in a sense, the $SO(2)_{\text{diag}}$ symmetric solution found in this work contains the solution of \cite{Bah_M5} for a special value of the gauge coupling constant $\widetilde{g}_1=16h$. However, the present solution preserves only eight supercharges corresponding to $N=1$ superconformal symmetry in four dimensions.   

To further analyze the solution, we introduce the following constants, as in the previous cases,
\begin{equation}\label{SO(2)diag_SO(4)_Cons}
B=2\sqrt{2}h^{3/2}\sqrt{\widetilde{g}_1b},\qquad m=\frac{\widetilde{g}_1}{16h},\qquad \mathcal{C}=32\widetilde{g}_1 h^2c\, .
\end{equation}
We will also take the solution for $\sigma$ as given in \eqref{1/5sigma_soln}. The seven-dimensional metric reads
\begin{equation}
ds^2_7=\frac{Br^{3/10}}{16h^2\sqrt{s(1-m^2r)}}\left[ds^2_{AdS_5}+\frac{r^{-3/2}}{16W\left[s(1-m^2r)\right]^{3/2}}dr^2+\frac{16\mathcal{C}^2W}{B}dz^2\right]\label{SO(2)diag_SO(4)_Met}
\end{equation}
with
\begin{equation}\label{SO(2)diag_SO(4)_Def_W}
W=B-\sqrt{sr(1-m^2r)}\, .
\end{equation}
The scalar field $\phi$ and the $SO(2)_{\text{diag}}$ gauge field are respectively given by
\begin{equation}\label{SO(2)diag_phi2_vector_Soln1}
\phi=-\frac{1}{2}\ln r-\ln m\qquad\text{ and }\qquad A_1=-\frac{1}{8mh}\left[2|\mathcal{C}|(1-2m^2r)+q\right].
\end{equation}
Since we have used $\sigma<0$, it follows that the constant $\mathcal{C}$ must be negative due to the condition \eqref{SO(2)xSO(2)_sign_Con}.
The Killing spinor takes the form of
\begin{equation}\label{SO(2)diag_eta_Soln}
\eta=Y_0e^{iqz}\frac{2^{1/8}r^{3/40}}{\left[s(1-m^2r)\right]^{\frac{1}{8}}}\begin{pmatrix} \sqrt{\sqrt{B}+sr^{1/4}\left[s(1-m^2r)\right]^{\frac{1}{4}}} \\ -\sqrt{\sqrt{B}-sr^{1/4}\left[s(1-m^2r)\right]^{\frac{1}{4}}} \end{pmatrix}.
\end{equation}
\indent The behaviours of the solution are very similar to those given in Section \ref{pSUGRA_Soln_Sec} and in \cite{Bah_M5}. However, the first equation in \eqref{SO(2)diag_phi2_vector_Soln1} implies that $r$ and $m$ must be positive in order to find a real solution for $\phi$. Therefore, there is no regular solution with $m<0$ as in Case II of Section \ref{pSUGRA_Soln_Sec}. For convenience, the remaining six possibilities with $m>0$ will be combined into four different cases as follows
\begin{eqnarray}
i)\qquad s&=&+1,\quad 0<B<\frac{1}{2m},\quad r\in(0,r_-)\cup (r_+,\frac{1}{m^2}),\quad\nonumber\\
ii)\qquad s&=&+1,\quad B=\frac{1}{2m},\quad r\in(0,r_\ast)\cup (r_\ast,\frac{1}{m^2}),\nonumber\\
iii)\qquad s&=&+1,\quad B>\frac{1}{2m},\quad r\in(0,\frac{1}{m^2}),\nonumber\\
iv)\qquad s&=&-1,\quad B>0,\quad r\in(\frac{1}{m^2},r_+)
\end{eqnarray}
in which 
\begin{equation}\label{SO(2)diag_r_roots}
r_{\pm}=\frac{1\pm\sqrt{1-4sB^2m^2}}{2m^2}
\end{equation}
are the two roots of $W=B-\sqrt{sr(1-m^2r)}=0$. For $B=\frac{1}{2m}$ with $s=+1$, the two roots are equal, $r_+=r_-=r_\ast=\frac{1}{2m^2}$. 

\begin{figure}[h!]
  \centering
  \begin{subfigure}[b]{0.326\linewidth}
    \includegraphics[width=\linewidth]{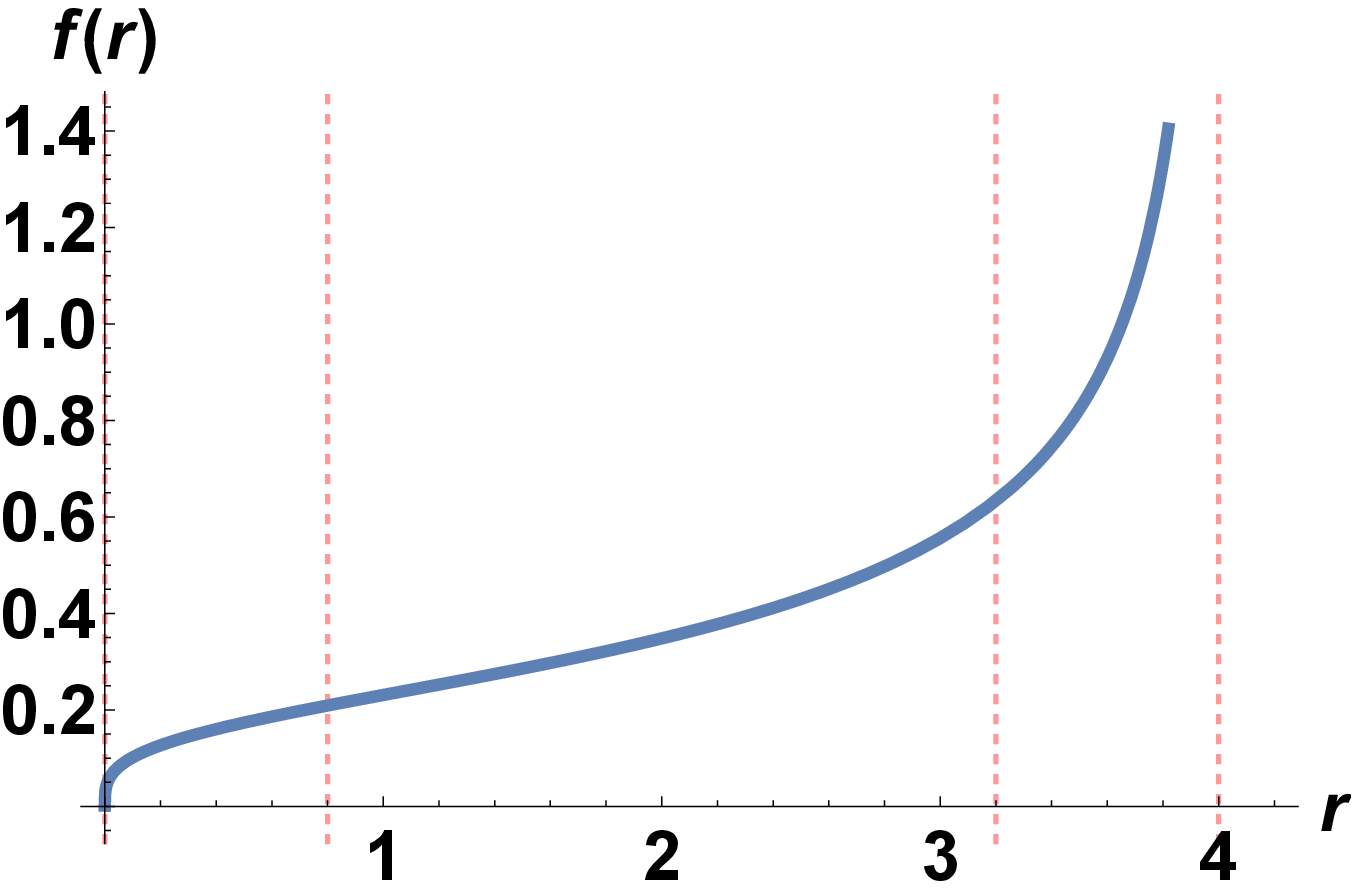}
\caption{$f$ solution}
  \end{subfigure}
  \begin{subfigure}[b]{0.326\linewidth}
    \includegraphics[width=\linewidth]{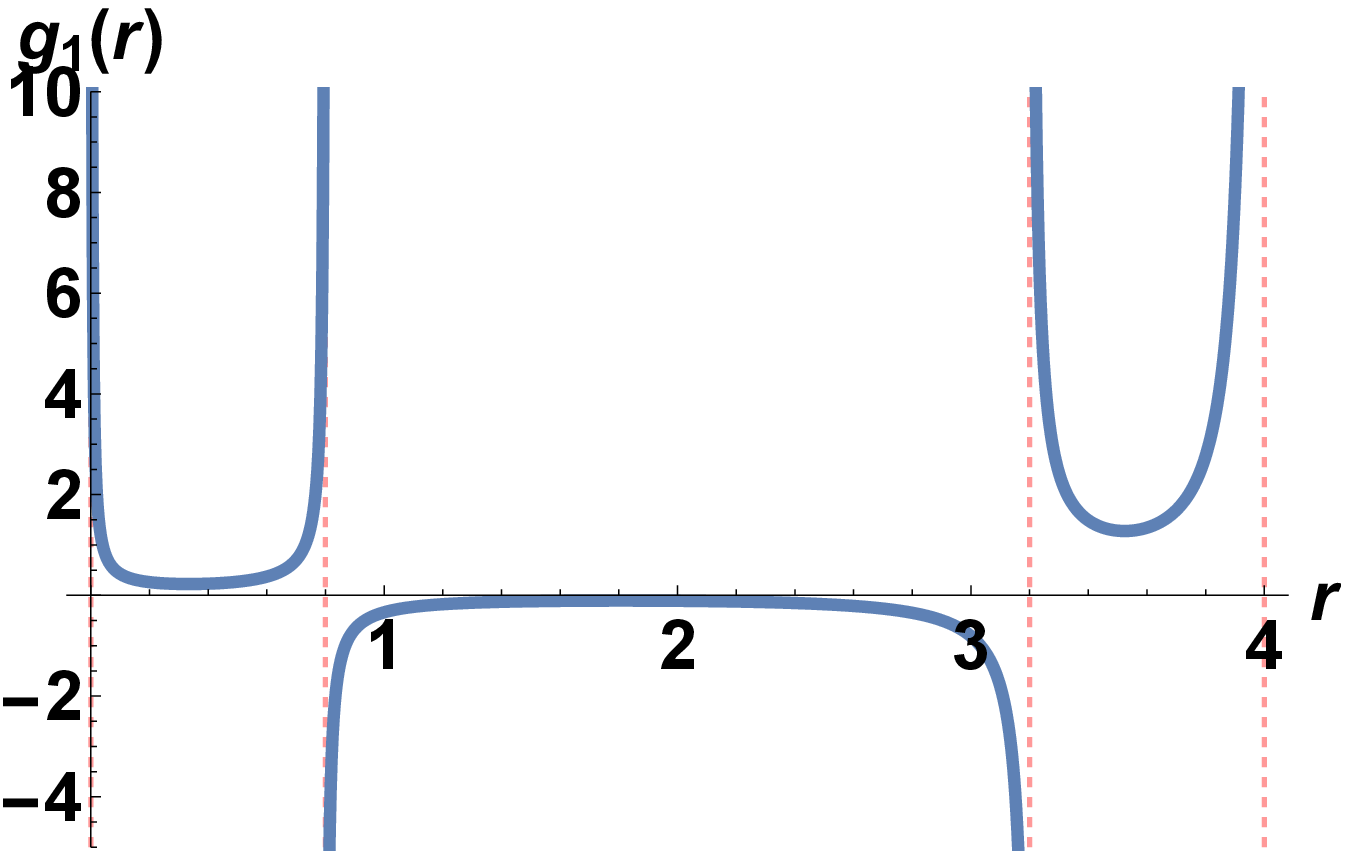}
\caption{$g_1$ solution}
  \end{subfigure}
  \begin{subfigure}[b]{0.326\linewidth}
    \includegraphics[width=\linewidth]{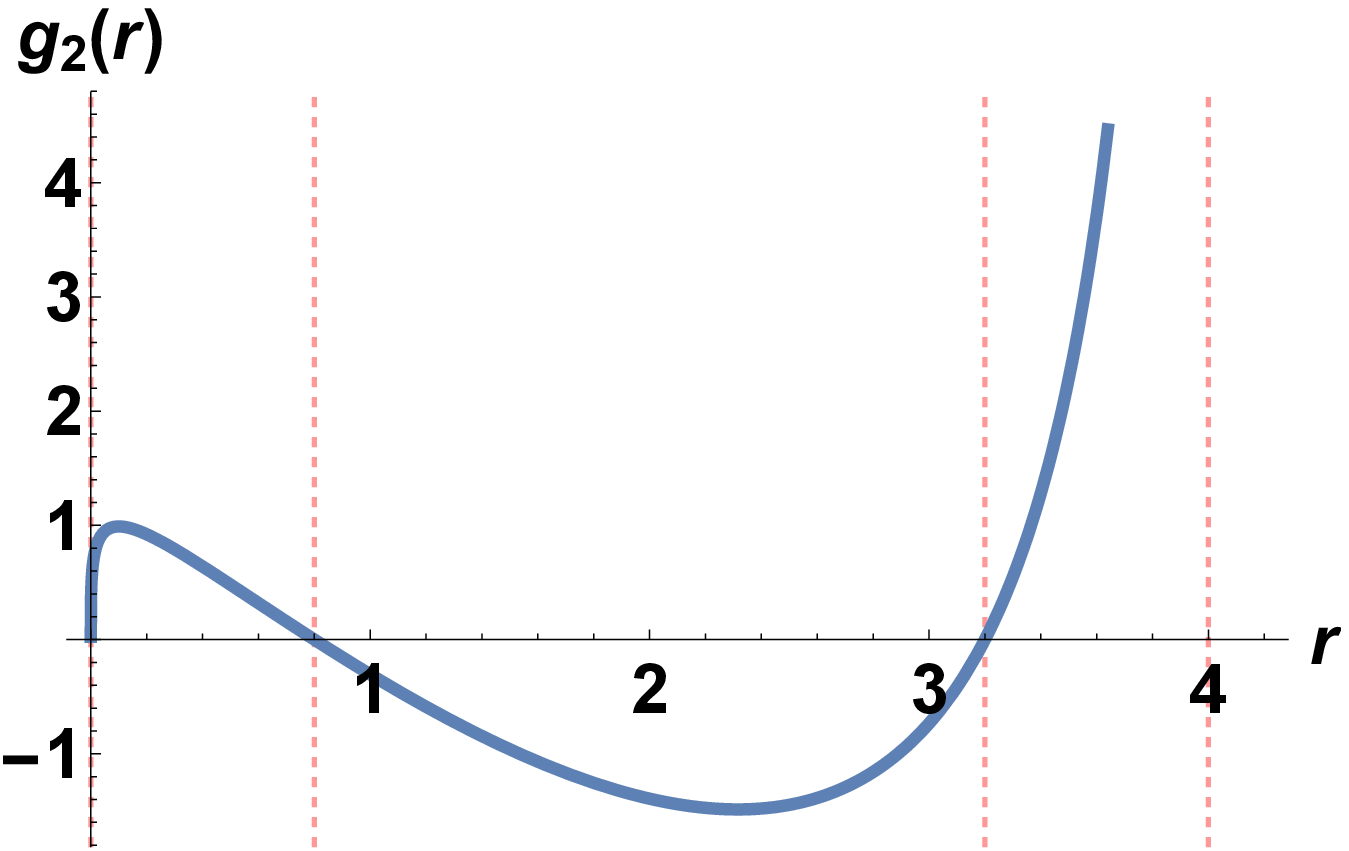}
\caption{$g_2$ solution}
  \end{subfigure}
  \caption{Numerical plots of the warp factors for $SO(2)_{\text{diag}}$ symmetric solution in $SO(4)$ gauge group in Case $i$ with $s=1$, $m=\frac{1}{2}$, $B=\frac{4}{5}$, $h=\frac{1}{2}$, and $\mathcal{C}=-1$. All the warp factors are positive in the ranges $0<r<r_{-}=0.8$ and $r_{+}=3.2<r<\frac{1}{m^2}=4$ with the four vertical red dashed lines representing the four boundaries.}
  \label{SO(2)diag_caseI+II_Soln}
\end{figure}

\begin{figure}[h!]
  \centering
  \begin{subfigure}[b]{0.326\linewidth}
    \includegraphics[width=\linewidth]{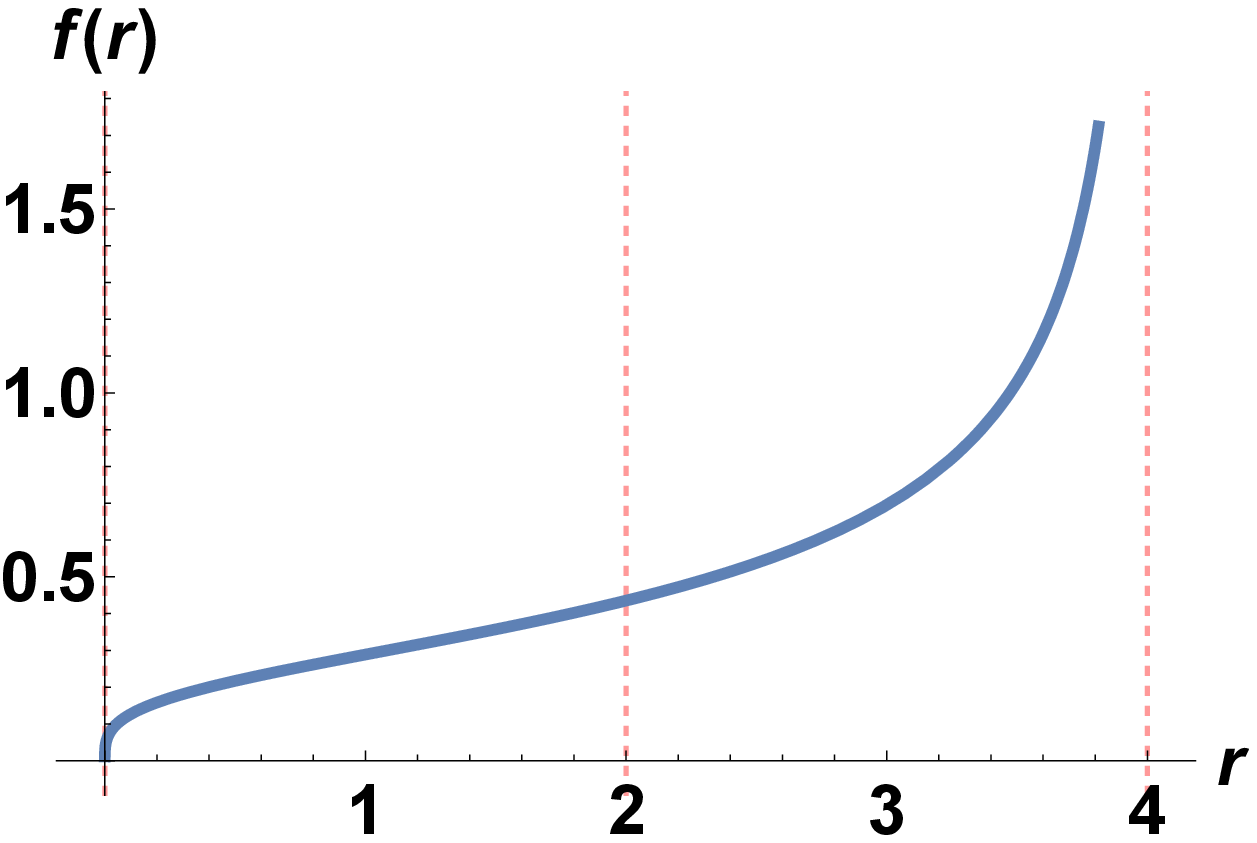}
\caption{$f$ solution}
  \end{subfigure}
  \begin{subfigure}[b]{0.326\linewidth}
    \includegraphics[width=\linewidth]{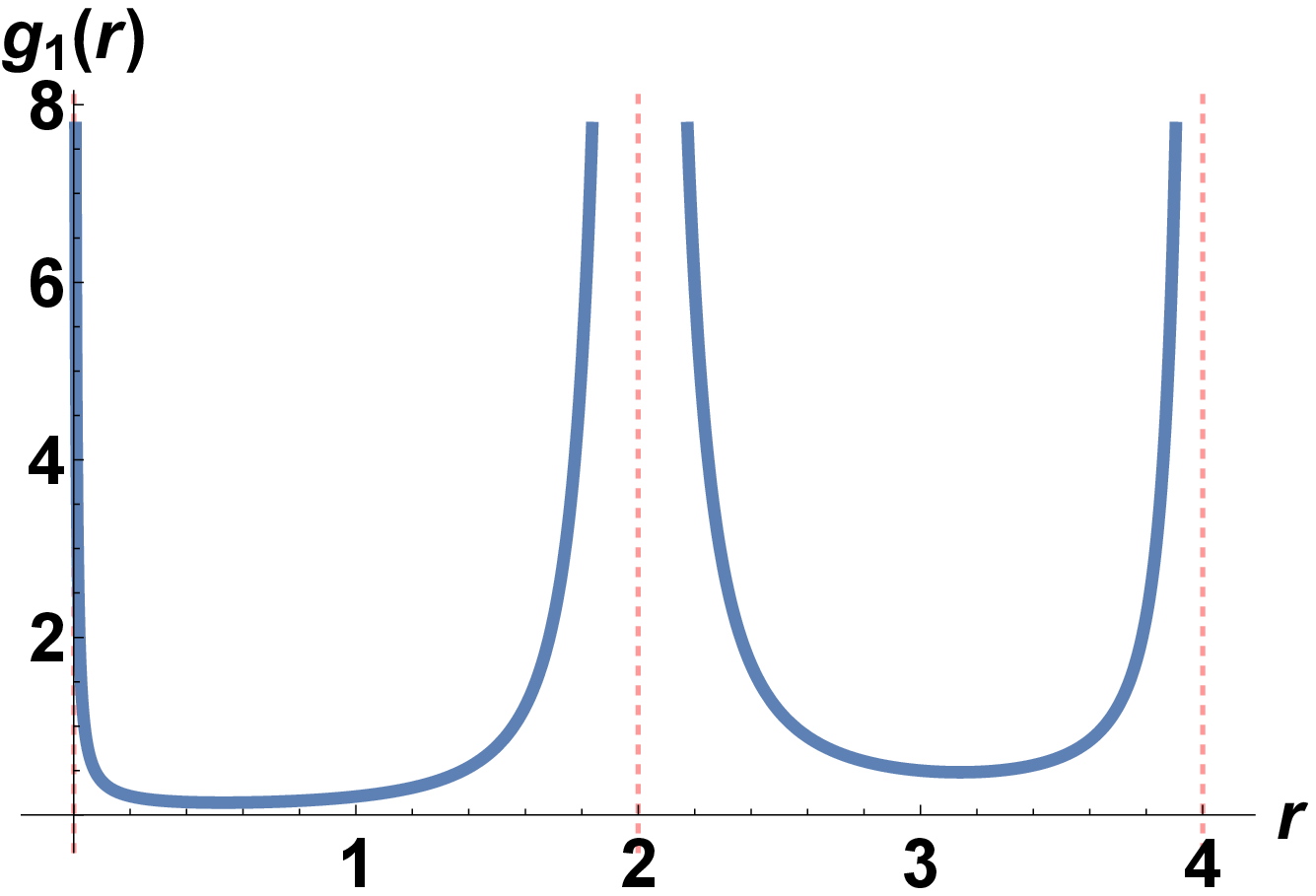}
\caption{$g_1$ solution}
  \end{subfigure}
  \begin{subfigure}[b]{0.326\linewidth}
    \includegraphics[width=\linewidth]{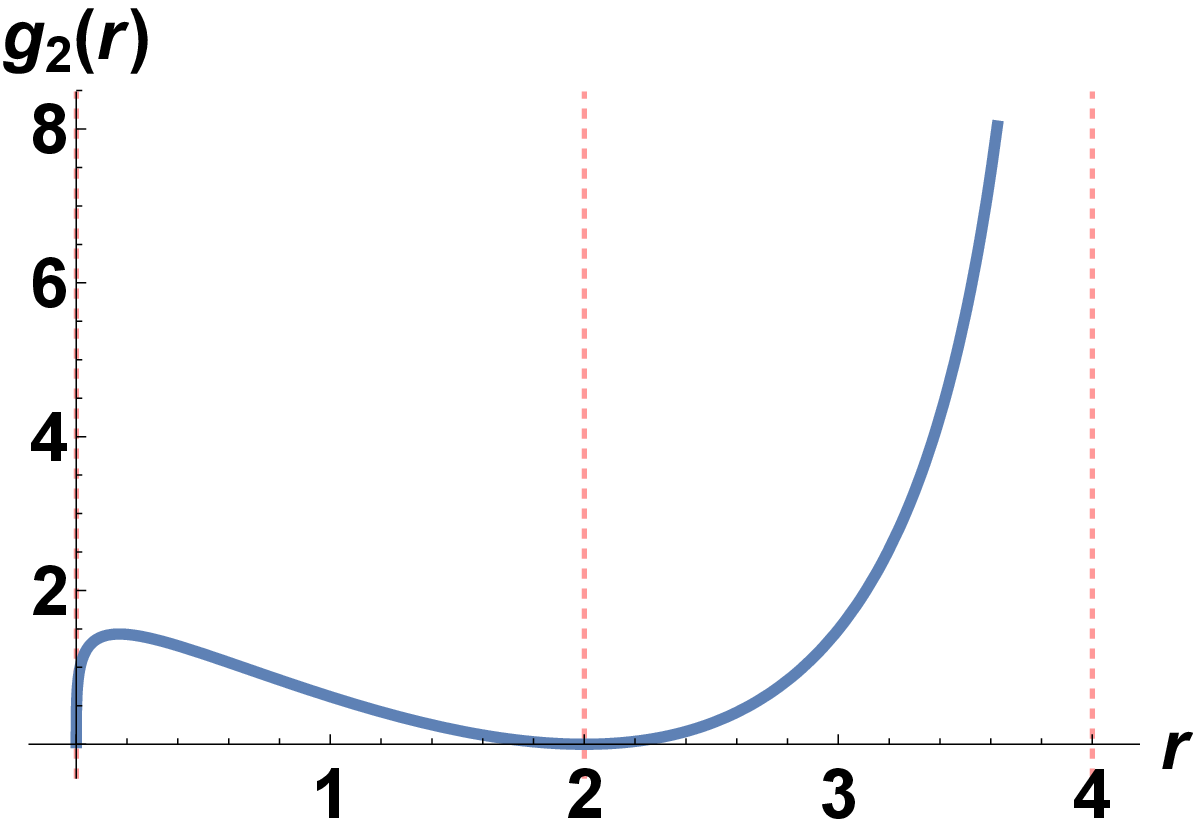}
\caption{$g_2$ solution}
  \end{subfigure}
  \caption{Numerical plots of the warp factors for $SO(2)_{\text{diag}}$ symmetric solution in $SO(4)$ gauge group in Case $ii$ with $s=1$, $m=\frac{1}{2}$, $B=1$, $h=\frac{1}{2}$, and $\mathcal{C}=-1$. The warp factors are positive in the ranges $0<r<r_\ast=2$ and $r_\ast=2<r<\frac{1}{m^2}=4$ with the three vertical red dashed lines representing the three boundaries.}
  \label{SO(2)diag_caseIII+IV_Soln}
\end{figure}

\begin{figure}[h!]
  \centering
  \begin{subfigure}[b]{0.326\linewidth}
    \includegraphics[width=\linewidth]{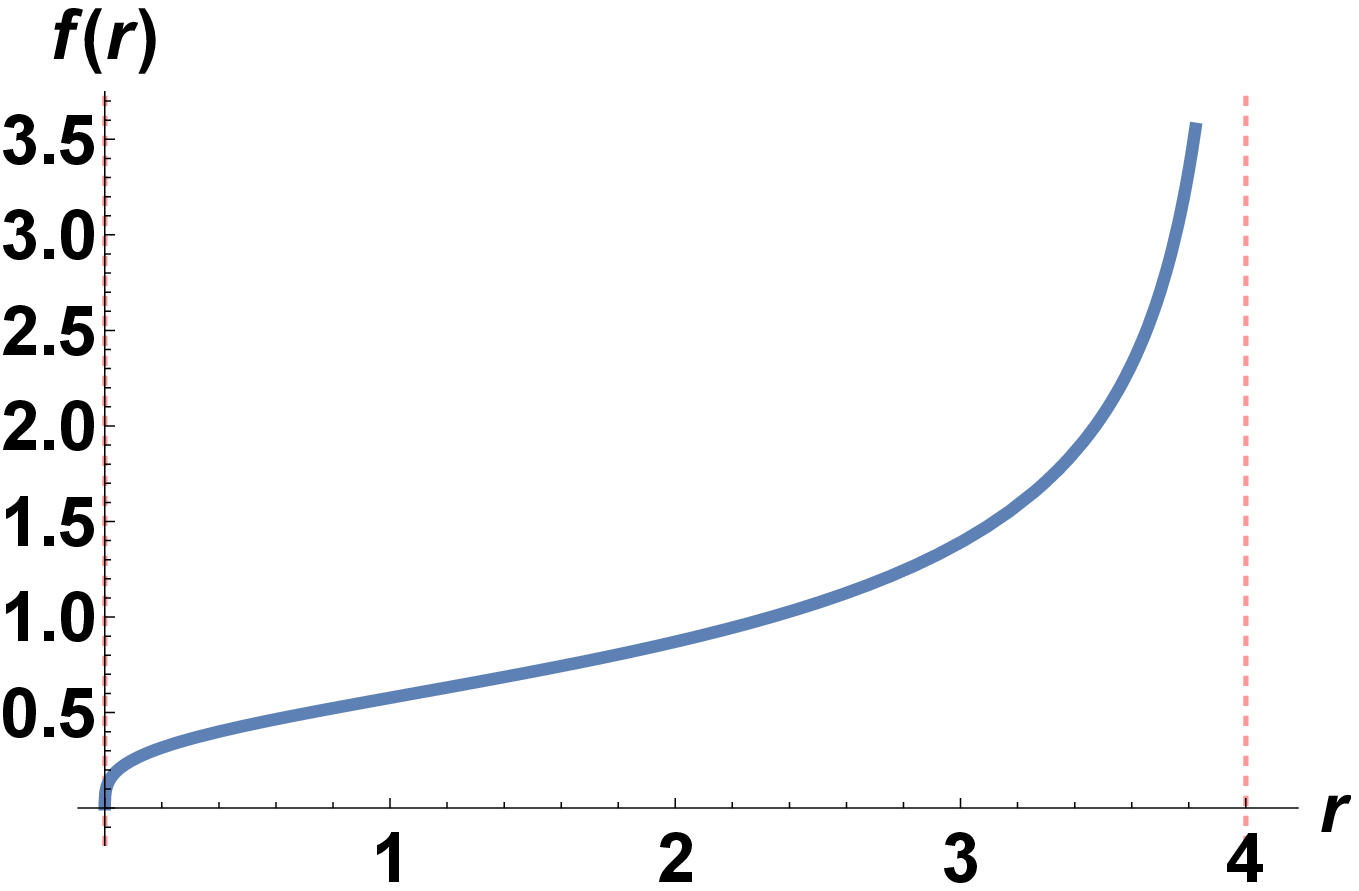}
\caption{$f$ solution}
  \end{subfigure}
  \begin{subfigure}[b]{0.326\linewidth}
    \includegraphics[width=\linewidth]{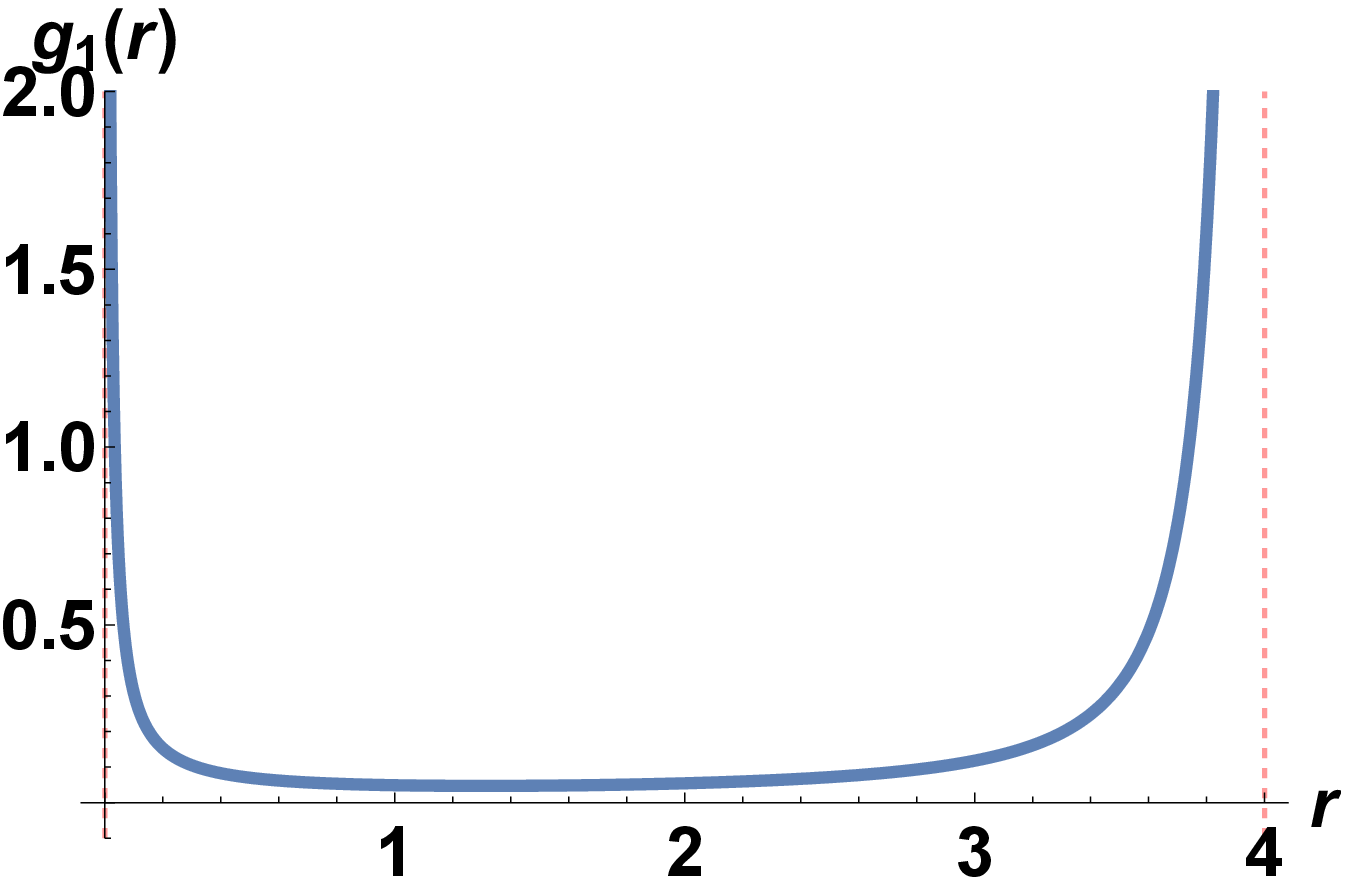}
\caption{$g_1$ solution}
  \end{subfigure}
  \begin{subfigure}[b]{0.326\linewidth}
    \includegraphics[width=\linewidth]{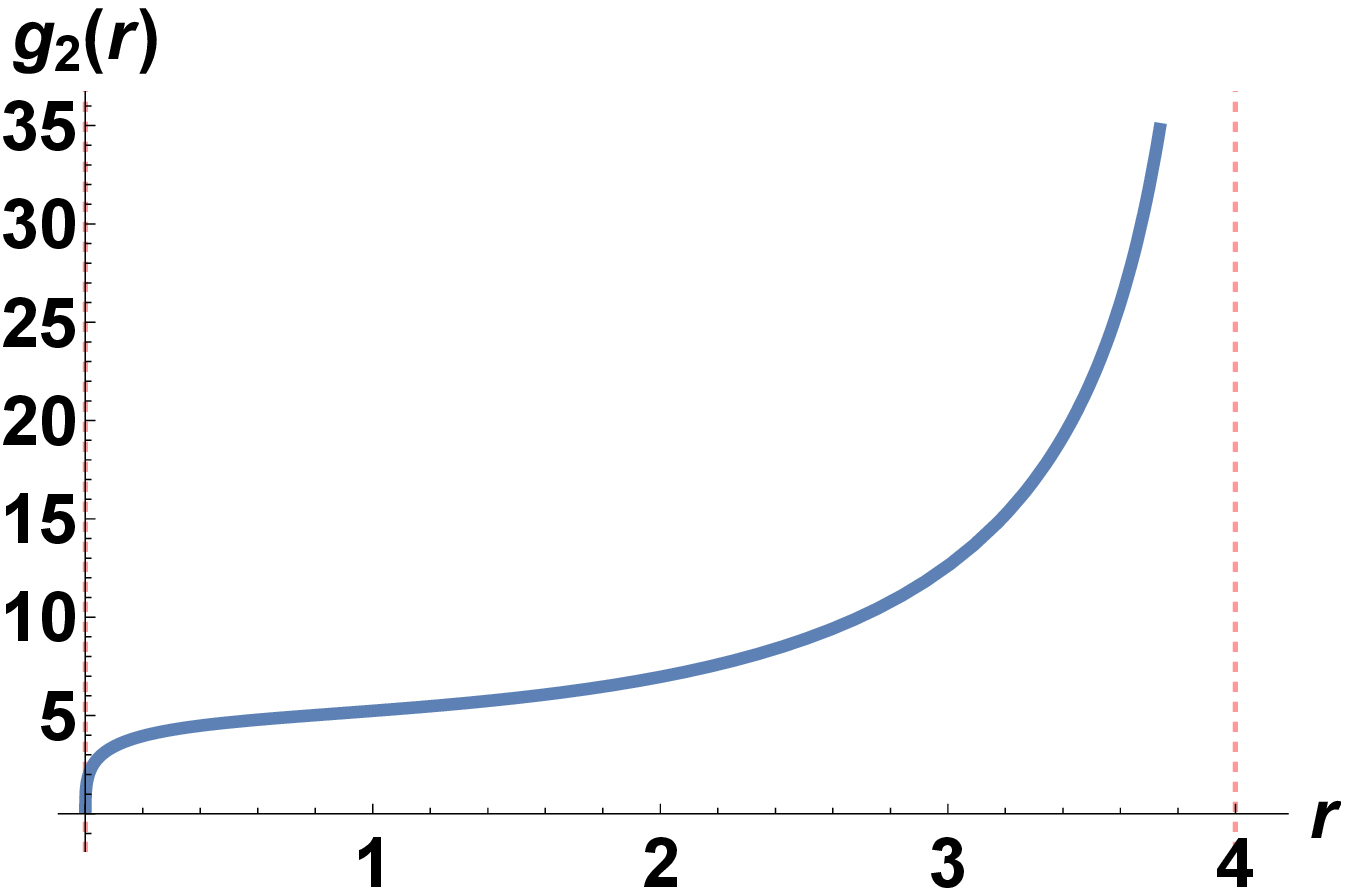}
\caption{$g_2$ solution}
  \end{subfigure}
  \caption{Numerical plots of the warp factors for $SO(2)_{\text{diag}}$ symmetric solution in $SO(4)$ gauge group in Case $iii$ with $s=1$, $m=\frac{1}{2}$, $B=2$, $h=\frac{1}{2}$, and $\mathcal{C}=-1$. The warp factors are positive in the range $0<r<\frac{1}{m^2}=4$ with the two vertical red dashed lines representing the two boundaries.}
  \label{SO(2)diag_caseV_Soln}
\end{figure}

\begin{figure}[h!]
  \centering
  \begin{subfigure}[b]{0.326\linewidth}
    \includegraphics[width=\linewidth]{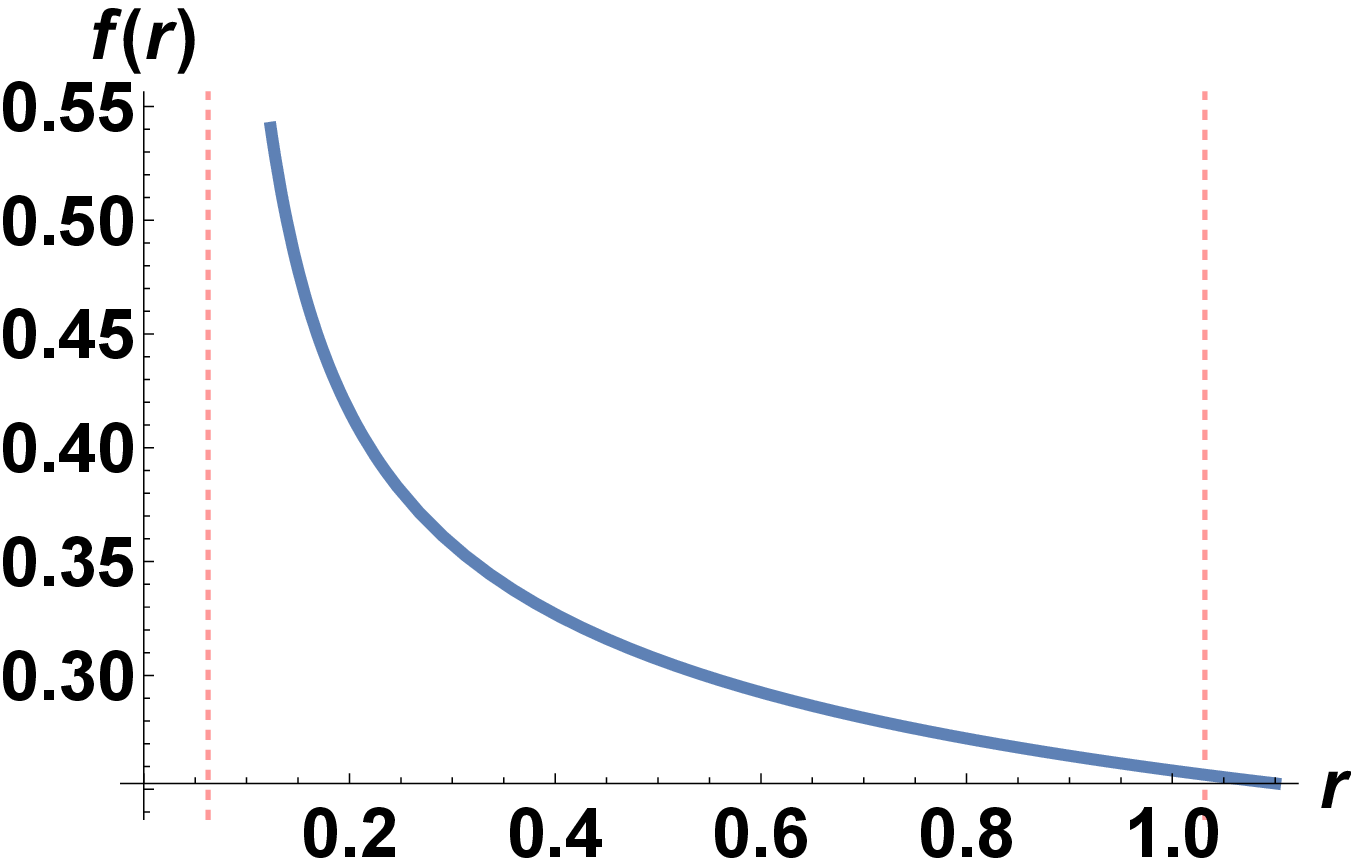}
\caption{$f$ solution}
  \end{subfigure}
  \begin{subfigure}[b]{0.326\linewidth}
    \includegraphics[width=\linewidth]{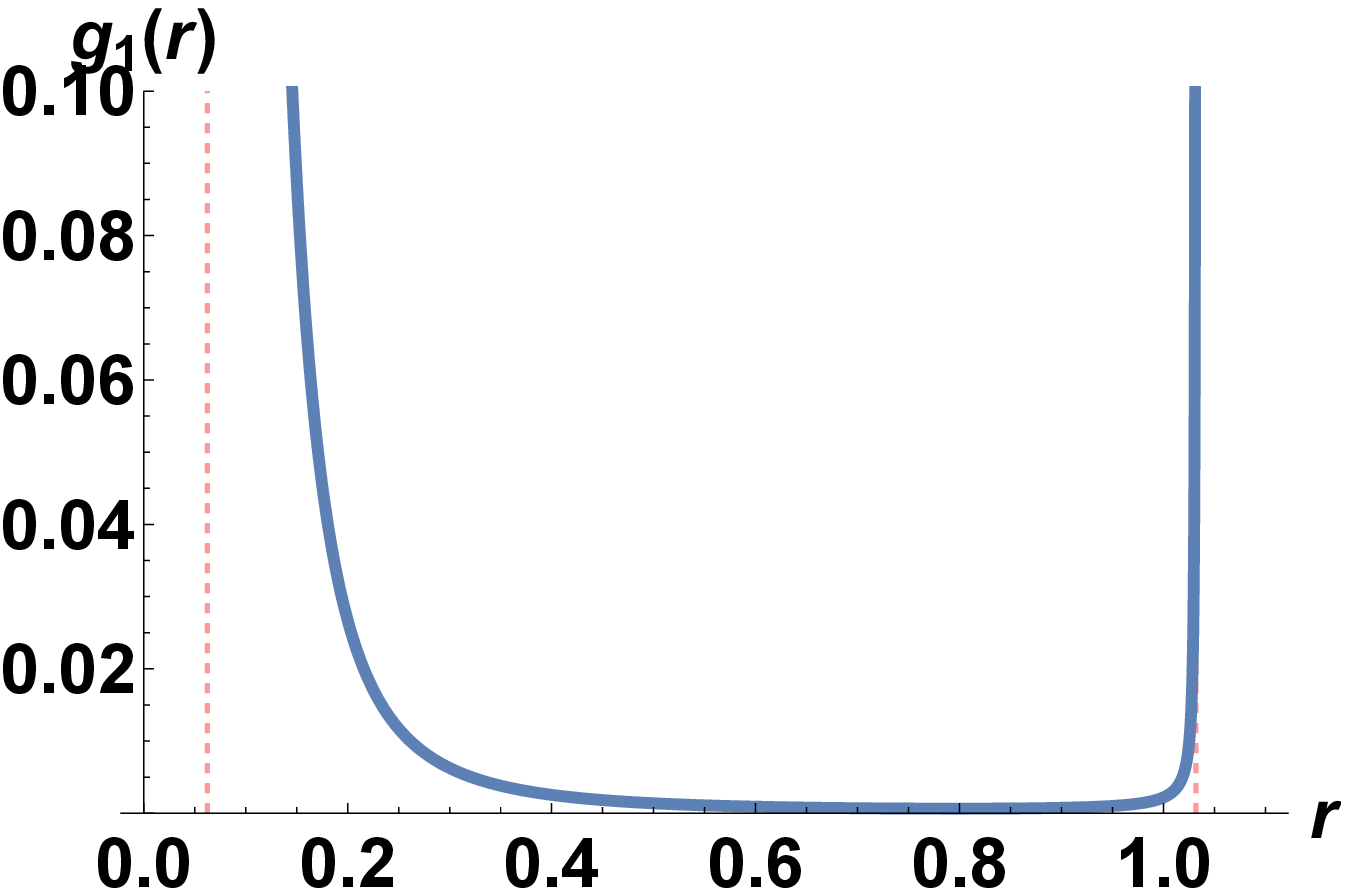}
\caption{$g_1$ solution}
  \end{subfigure}
  \begin{subfigure}[b]{0.326\linewidth}
    \includegraphics[width=\linewidth]{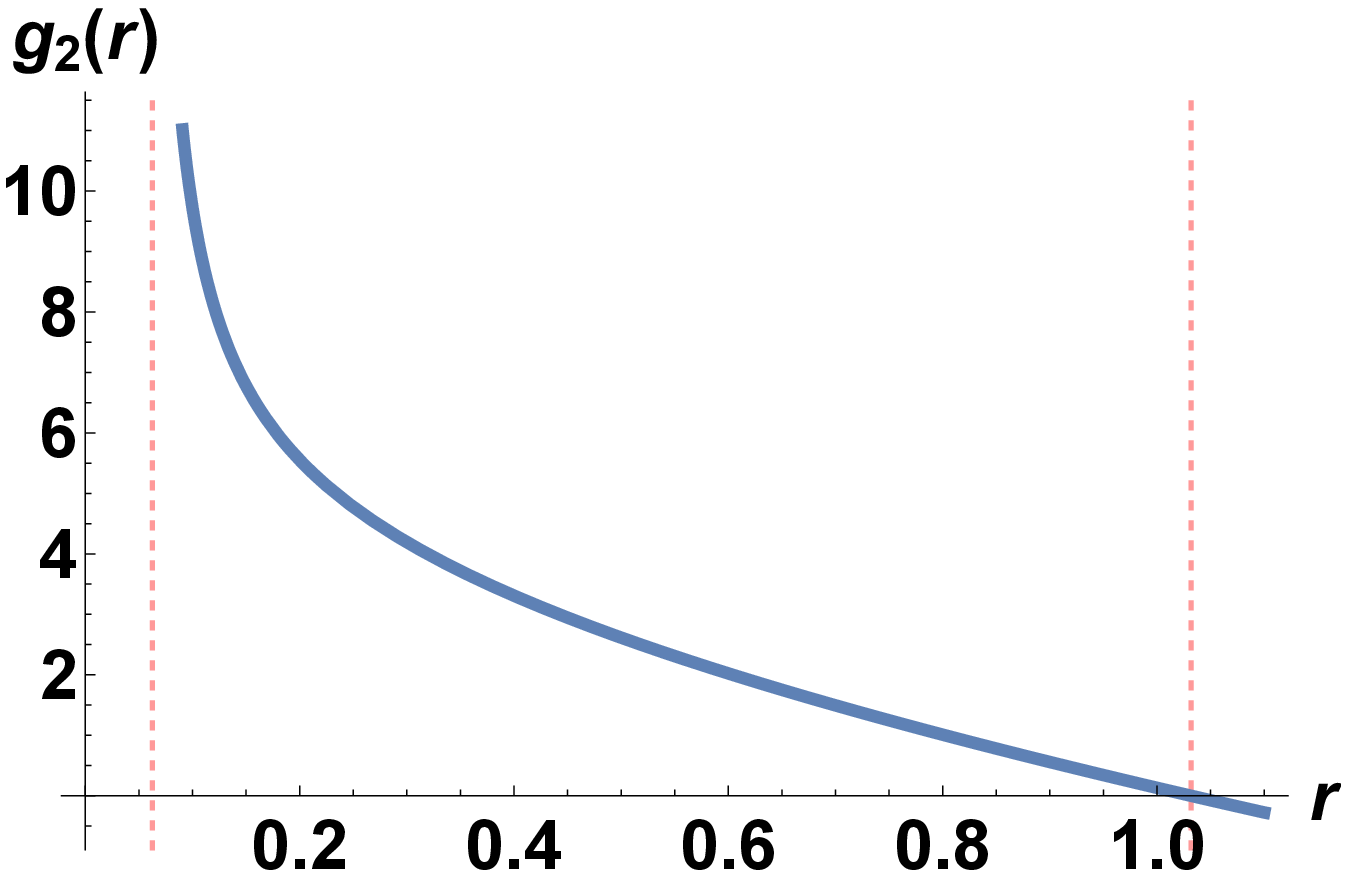}
\caption{$g_2$ solution}
  \end{subfigure}
  \caption{Numerical plots of the warp factors for $SO(2)_{\text{diag}}$ symmetric solution in $SO(4)$ gauge group in Case $iv$ with $s=-1$, $m=4$, $B=4$, $h=\frac{1}{2}$, and $\mathcal{C}=-1$. The warp factors are positive in the range $\frac{1}{m^2}=0.0625<r<r_+=1.03$ with the two vertical red dashed lines representing the two boundaries.}
  \label{SO(2)diag_caseVI_Soln}
\end{figure}

Examples of numerical solutions for the warp factors in these four cases are given in Figures \ref{SO(2)diag_caseI+II_Soln}, \ref{SO(2)diag_caseIII+IV_Soln}, \ref{SO(2)diag_caseV_Soln}, and \ref{SO(2)diag_caseVI_Soln}, respectively.

The endpoint behaviours are also similar to the $SO(2)_R$ symmetric solution. As $r\rightarrow0$ or $r\rightarrow\frac{1}{m^2}$, the seven-dimensional metric is conformal to a product of $AdS_5$ and a cylinder. The explicit form of the metric in these limits can be found in the same way as in Case I with $r\rightarrow 0$ and Case III with $r\rightarrow\frac{1}{m^2}$. 

The more interesting behaviours as $r\rightarrow r_{\pm}$ are in Case $i$ and Case $iv$. In terms of the new radial coordinate $R=\sqrt{r_\pm-r}$, the seven-dimensional metric, as $r\rightarrow r_{\pm}$, is approximately given by
\begin{equation}\label{SO(2)diag_spindle_metric}
ds_7^2\approx\frac{B\,r_{\pm}^{3/10}}{16h^2\sqrt{s(1-m^2r_{\pm})}}\left[ds^2_{AdS_5}+\frac{dR^2+16\mathcal{C}^2\left[1-2m^2r_{\pm}\right]^2R^2dz^2}{-4W'(r_{\pm})r_{\pm}^{3/2}(1-m^2r_{\pm})^{3/2}}\right].
\end{equation}
The $z$ circle shrinks smoothly near these endpoints $r= r_{\pm}$ if we impose
\begin{equation}\label{SO(2)diag_smooth_con}
|\mathcal{C}|=\frac{1}{4l\left|1-2m^2r_{\pm}\right|}=\frac{1}{4l\sqrt{1-4sB^2m^2}},\qquad l=1,2,3,\ldots \, .
\end{equation}
Unlike in the previous solutions, this condition can be written explicitly in terms of $B$ and $m$ since $r_\pm$ takes a much simpler form than $r_{(\pm_1,\pm_2)}$ and $r_1$. 

In order for the $SO(2)_{\text{diag}}$ gauge field to vanish at $r= r_{\pm}$, we fix the constant $q$ such that
\begin{equation}
q=\mp2|\mathcal{C}||1-2m^2r_\pm|=\mp\frac{1}{2l}\qquad \textrm{and}\qquad A_1=\frac{m|\mathcal{C}|}{2h}(r-r_\pm)
\end{equation}
which gives rise to the Killing spinor $\eta$ at $r= r_{\pm}$ of the form
\begin{equation}\label{SO(2)diag_spindle_eta}
\eta=2^{-3/8}Y_0e^{\mp\frac{iz}{2l}}r^{1/5}_{\pm}\begin{footnotesize}\begin{pmatrix} s+1 \\ s-1 \end{pmatrix}\end{footnotesize}.
\end{equation}
\indent Finally, it should be pointed out that if $B=\frac{1}{2m}$, we find that $|\mathcal{C}|$ diverges due to $\sqrt{1-4B^2m^2}=0$. This implies that the $z$ circle does not shrink smoothly at $r=r_\ast=\frac{1}{2m^2}$ in Case $ii$ as in Cases IV and V of the $SO(2)_R$ symmetric solution. As $r\rightarrow r_\ast$, the seven-dimensional metric is approximately given by
\begin{equation}\label{SO(2)diaf_rstar_metric}
ds^2_7\approx\frac{1}{32(2^{4/5})h^2m^{8/5}}\left[2ds^2_{AdS_5}+\frac{dR^2+64\mathcal{C}^2m^4dz^2}{R^2}\right]
\end{equation}
with the new radial coordinate $R$ given by $r=\frac{3}{4m}-\frac{1}{R}$ and $R\rightarrow+\infty$. In order for the $SO(2)_{\text{diag}}$ gauge field to vanish at $r=r_\ast$, we have to set $q=0$ resulting in the $z$-independent Killing spinor as in Case IV and Case V of the pure supergravity solution.

%%%%%%%%%%%%%%%%%%%%%%%%%%%%%%%%%%%%%%%%%%%%%%%%%%%%%%%%%%%%%%%%%%%%%%%%%%%%%%%%%%%%%%%%%%%%%%%%%%%%%%%%%%%%%%%%%%%%%%%%%%%%%%%%%%%%%%%%%
\section{$SO(2,2)$ gauge group}\label{SO(2,2)_Sec}
In this section, we consider $N=2$ gauged supergravity with non-compact $SO(2,2)\sim SO(2,1)\times SO(2,1)$ gauge group. The embedding of this gauge group in $SO(3,n)$ requires at least $n=3$ vector multiplets. We will work with the minimal number of $n=3$ and choose the following $SO(2,2)$ structure constants
\begin{equation}\label{SO(2,2)_fIJK}
f_{IJK}=(\widetilde{g}_1\varepsilon_{\bar{i}\bar{j}\bar{k}},-\widetilde{g}_2\varepsilon_{\bar{r}\bar{s}\bar{t}}).
\end{equation}
Indices $\bar{i},\bar{j},...=1,2,6$ and $\bar{r},\bar{s},...=3,4,5$ are two sets of $SO(2,1)$ indices raised and lowered by $\eta_{\bar{i}\bar{j}}=\text{diag}(-1,-1,1)$ and $\eta_{\bar{r}\bar{s}}=\text{diag}(-1,1,1)$, respectively. Moreover, $\widetilde{g}_1$ and $\widetilde{g}_2$ are coupling constants for the two $SO(2,1)$ subgroups respectively generated by the generators
\begin{eqnarray}
T_1&=&Y_{23},\qquad\ T_2\ =\ -Y_{13},\qquad T_6\ =\ J^{(1)}_{12},\\
T_3&=&J^{(2)}_{12},\qquad  T_4\ =\ Y_{32},\qquad\quad T_5\ =\ -Y_{31}\, .
\end{eqnarray}
These satisfy $SO(2,1)\times SO(2,1)$ algebra
\begin{equation}
[T_{\bar{i}},T_{\bar{j}}]=-{\varepsilon_{\bar{i}\bar{j}}}^{\bar{k}}T_{\bar{k}},\qquad [T_{\bar{r}},T_{\bar{s}}]={\varepsilon_{\bar{r}\bar{s}}}^{\bar{t}}T_{\bar{t}},\qquad [T_{\bar{i}},T_{\bar{r}}]=0\, .
\end{equation}
The compact $SO(2)\times SO(2)$ subgroup is generated by $T_3$ and $T_6$. In this case, $T_6$ generates $SO(2)_R$ subgroup of the $SO(3)_R$ R-symmetry while $T_3$ generates the other $SO(2)$ in the $SO(3)$ symmetry of the vector multiplets. We now repeat the same procedure of finding $AdS_5\times \Sigma$ solutions as in the previous section. Unlike the $SO(4)$ gauge group, this gauged supergravity does not admit any supersymmetric $AdS_7$ vacua. The maximally supersymmetric vacua are given by half-supersymmetric domain walls dual to non-conformal $N=(1,0)$ field theories in six dimensions \cite{7D_noncompact}. In this case, $AdS_5\times \Sigma$ solutions would describe conformal fixed points in four dimensions of six-dimensional $N=(1,0)$ field theories on a half-spindle. 

%%%%%%%%%%%%%%%%%%%%%%%%%%%%%%%%%%%
\subsection{$SO(2)\times SO(2)$ symmetric solution}\label{SO(2,2)_SO(2)diag_Sec}
As in Section \ref{SO(4)_Sec}, supersymmetric $AdS_5\times \Sigma$ solutions preserving $SO(2)\times SO(2)\subset SO(2,1)\times SO(2,1)\sim SO(2,2)$ can be found by using the same metric ansatz \eqref{7Dmetrix}, $SO(2)\times SO(2)$ gauge fields \eqref{SO(2)xSO(2)_Gaugefield}, and $SO(2)\times SO(2)$ singlet scalar $\phi$ corresponding to the non-compact generator $Y_{33}$. As in $SO(4)$ gauge group, we can consistently set $C_{(3)}=0$ due to $F^I_{(2)}\wedge F^I_{(2)}=0$.

With the coset representative \eqref{L_SO3d}, the scalar vielbein is still given by \eqref{SO(4)_SO(2)xSO(2)_P_Q}. However, the composite connection for $SO(2)_R$ is now given by
\begin{equation}
Q^{ij}_{(1)}=\widetilde{g}_1A_2\varepsilon^{ij3}dz\, .
\end{equation}
The $C$-functions take the form of
\begin{equation}\label{SO(2,2)_SO(2)xSO(2)_Cfns}
C=-3\sqrt{2}\,\widetilde{g}_1\sinh\phi,\qquad
C^{ir}=\sqrt{2}\,\widetilde{g}_1\cosh{\phi}\,\delta^i_3\delta^r_3.
\end{equation}
By the same analysis, we can derive a similar set of BPS equations with $A_1$ and $A_2$ interchanged. General solutions for $\phi$, $f$, $g_1$, $\widehat{A}_2$, and $g_2$ are given in \eqref{phi_gen_sol}, \eqref{f_gen_sol}, \eqref{g1_gen_sol}, \eqref{Gen_A1_soln}, and \eqref{g2_gen_sol}, respectively with $\lambda=-1$. As in the case of $SO(4)$ gauge group, we are not able to determine an analytic solution for the second $SO(2)$ gauge field in this case given by $A_1$. Therefore, to explicitly write down a complete solution and discuss some properties of the solution, we further simplify the BPS equations by choosing $a_1=-a_2=b$ as in the $SO(4)$ case. It should be noted here that setting $a_1=a_2=b$ is also possible in this case but does not lead to any new solutions. 

Repeating the same analysis as in the $SO(4)$ case, we can eventually solve all the BPS conditions and find the $SO(2)\times SO(2)$ symmetric solution
\begin{eqnarray}
\phi&=&\sinh^{-1}\left[\frac{\widetilde{g}_1e^{-\frac{5\sigma}{2}}}{16h}\right],\label{SO(2,2)_SO(2)xSO(2)_phi_Soln}\\
f&=&\frac{2\sqrt{bh}e^\sigma}{\sqrt{s}(256h^2e^{5\sigma}+\widetilde{g}_1^2)^{1/4}},\\
g_1&=&\frac{51,200\sqrt{b\,h^9}\,e^{6\sigma}\,(\sigma')^2}{(256h^2e^{5\sigma}+\widetilde{g}_1^2)^2\left[32\sqrt{b\,h^5}-\sqrt{s}e^{-5\sigma}(256h^2e^{5\sigma}+\widetilde{g}_1^2)^{1/4}\right]},\\
g_2&=&\frac{256c_1^2h^2\widetilde{g}_1^2e^\sigma\left[32\sqrt{b\,h^5}-\sqrt{s}e^{-5\sigma}(256h^2e^{5\sigma}+\widetilde{g}_1^2)^{1/4}\right]}{\sqrt{s}(256h^2e^{5\sigma}+\widetilde{g}_1^2)^{1/4}},\\
A_1&=&-c_1\widetilde{g}_1e^{-5\sigma}\sqrt{256h^2e^{5\sigma}+\widetilde{g}_1^2}+c_2,\label{SO(2,2)_SO(2)xSO(2)_A1_soln}\\
A_2&=&c_1\left(192h^2+\widetilde{g}_1^2e^{-5\sigma}\right)-\frac{2q}{\widetilde{g}_1}\label{SO(2,2)_SO(2)xSO(2)_A2_soln}
\end{eqnarray}
together with the sign condition $\text{sign}(c\widetilde{g}_1\sigma')=+1$. The explicit form of the Killing spinor is given by
\begin{equation}
\eta=\frac{Y_0\,e^{iqz-\frac{\sigma}{4}}}{s^{1/8}(256h^2e^{5\sigma}+\widetilde{g}_1^2)^{\frac{1}{16}}}\begin{pmatrix} \sqrt{4\sqrt{2}(h^5b)^{\frac{1}{4}}+s^{1/4}e^{-\frac{5\sigma}{2}}(256h^2e^{5\sigma}+\widetilde{g}_1^2)^{\frac{1}{8}}} \\ -\sqrt{4\sqrt{2}(h^5b)^{\frac{1}{4}}-s^{1/4}e^{-\frac{5\sigma}{2}}(256h^2e^{5\sigma}+\widetilde{g}_1^2)^{\frac{1}{8}}}\end{pmatrix}
\end{equation}
with $Y_0$ being a constant.

The analysis of the regularity of the solution proceeds as in the previous cases. We first take the solution for the dilaton as in \eqref{1/5sigma_soln} and introduce the following parameters
\begin{equation}
B=\frac{8h^{2}\sqrt{b}}{\sqrt{s}},\qquad m=\frac{\widetilde{g}_1}{16h},\qquad \mathcal{C}=32\,\widetilde{g}_1h^2c_1\, .
\end{equation}
The seven-dimensional metric reads
\begin{equation}
ds^2_7=\frac{Br^{1/20}}{16h^2(1+m^2r)^{1/4}}\left[ds^2_{AdS_5}+\frac{r^{-5/4}}{64W(1+m^2r)^{7/4}}dr^2+\frac{4\mathcal{C}^2W}{B}dz^2\right]\label{SO(2,2)_SO(2)xSO(2)_Met}
\end{equation}
with
\begin{equation}\label{SO(2,2)_SO(2)xSO(2)_Def_W}
W=B-r^{3/4}(1+m^2r)^{1/3}
\end{equation}
while the solutions for $\phi$, $A_1$, and $A_2$ are given by
\begin{eqnarray}
\phi&=&\sinh^{-1}\left[m\sqrt{r}\right],\\
 A_1&=&\frac{|\mathcal{C}|}{2h}\sqrt{r(1+m^2r)}+c_2,\\ 
A_2&=&-\frac{1}{8mh}\left[|\mathcal{C}|(4m^2r+3)-q\right].
\end{eqnarray}
The Killing spinor $\eta$ takes the form
\begin{equation}\label{SO(2)diag_eta_Soln}
\eta=Y_0e^{iqz}\frac{r^{1/80}}{(1+m^2r)^{\frac{1}{16}}}\begin{pmatrix} \sqrt{\sqrt{B}+r^{3/8}(1+m^2r)^{\frac{1}{8}}} \\ -\sqrt{\sqrt{B}-r^{3/8}(1+m^2r)^{\frac{1}{8}}} \end{pmatrix}.
\end{equation}
It should be noted that since we have chosen sign$(\sigma)=-1$, we need to impose the condition sign$(c_1\widetilde{g}_1)=-1$ resulting in sign$(\mathcal{C})=-1$. 

There is only one possible range for the radial coordinate $r$ in order to obtain a regular solution 
\begin{equation}\label{SO(2,2)_SO(2)xSO(2)_Pos_Range}
m\neq0,\qquad B>0,\qquad 0<r<r_2
\end{equation}
where $r_2$ is determined from $W(r_2)=0$ as
\begin{equation}\label{SO(2,2)_SO(2)xSO(2)_r1}
r_2=-\frac{1}{4m^2}-\frac{1}{2}\sqrt{X_2}+\frac{1}{2}\sqrt{\frac{3}{4m^4}-X_2+\frac{1}{4m^6\sqrt{X_2}}}
\end{equation}
with
\begin{equation}
X_2=\frac{1}{4m^4}-\frac{4(2/3)^{1/3}B^{8/3}}{(\sqrt{81+768B^4m^6}-9)^{1/3}}+\frac{B^{4/3}(\sqrt{81+768B^4m^6}-9)^{1/3}}{18^{1/3}m^2}\, .
\end{equation}
There are two possibilities with $m>0$ and $m<0$. However, these are related to each other by a sign change in the $\phi$ solution. In the following, we will choose $m>0$ for definiteness. An example of numerical plots for the three warp factors is given in Figure \ref{SO(2,2)_SO(2)xSO(2)_plots}.

\begin{figure}[h!]
  \centering
  \begin{subfigure}[b]{0.326\linewidth}
    \includegraphics[width=\linewidth]{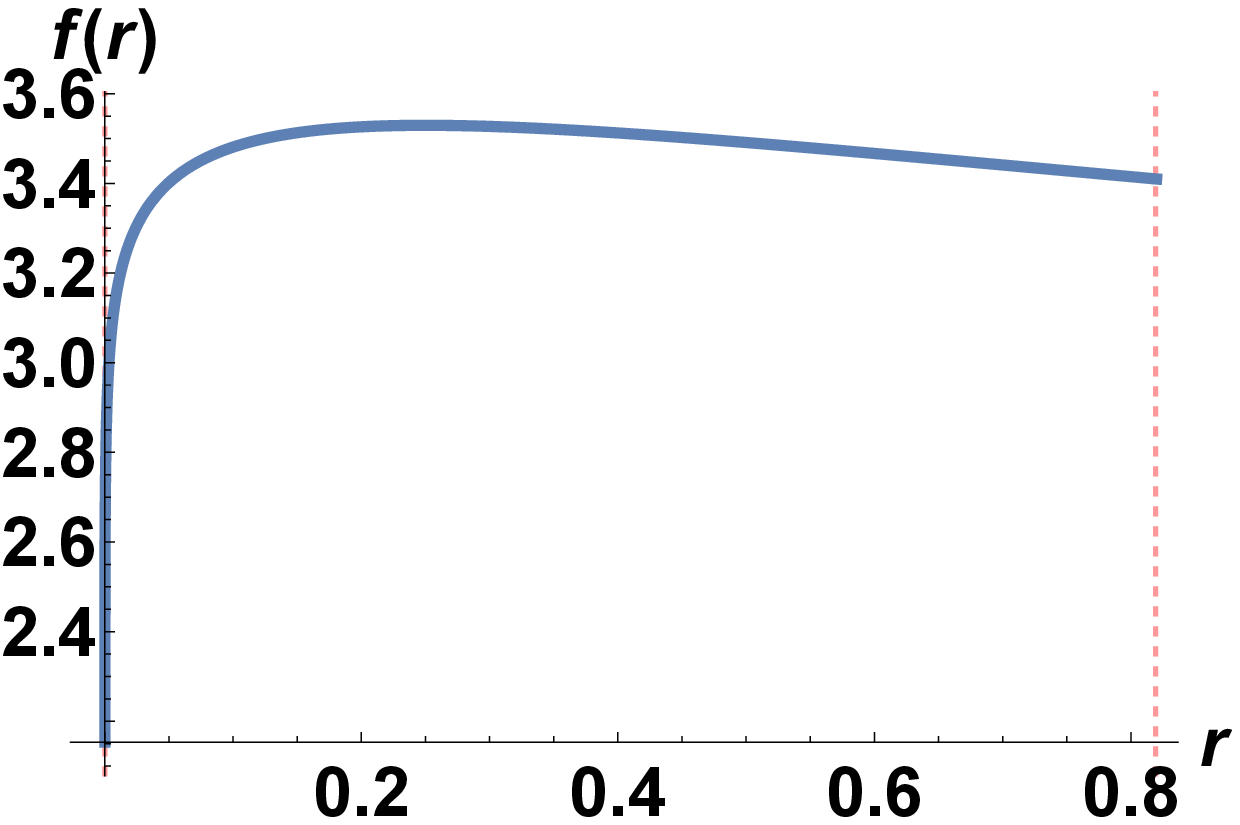}
\caption{$f$ solution}
  \end{subfigure}
  \begin{subfigure}[b]{0.326\linewidth}
    \includegraphics[width=\linewidth]{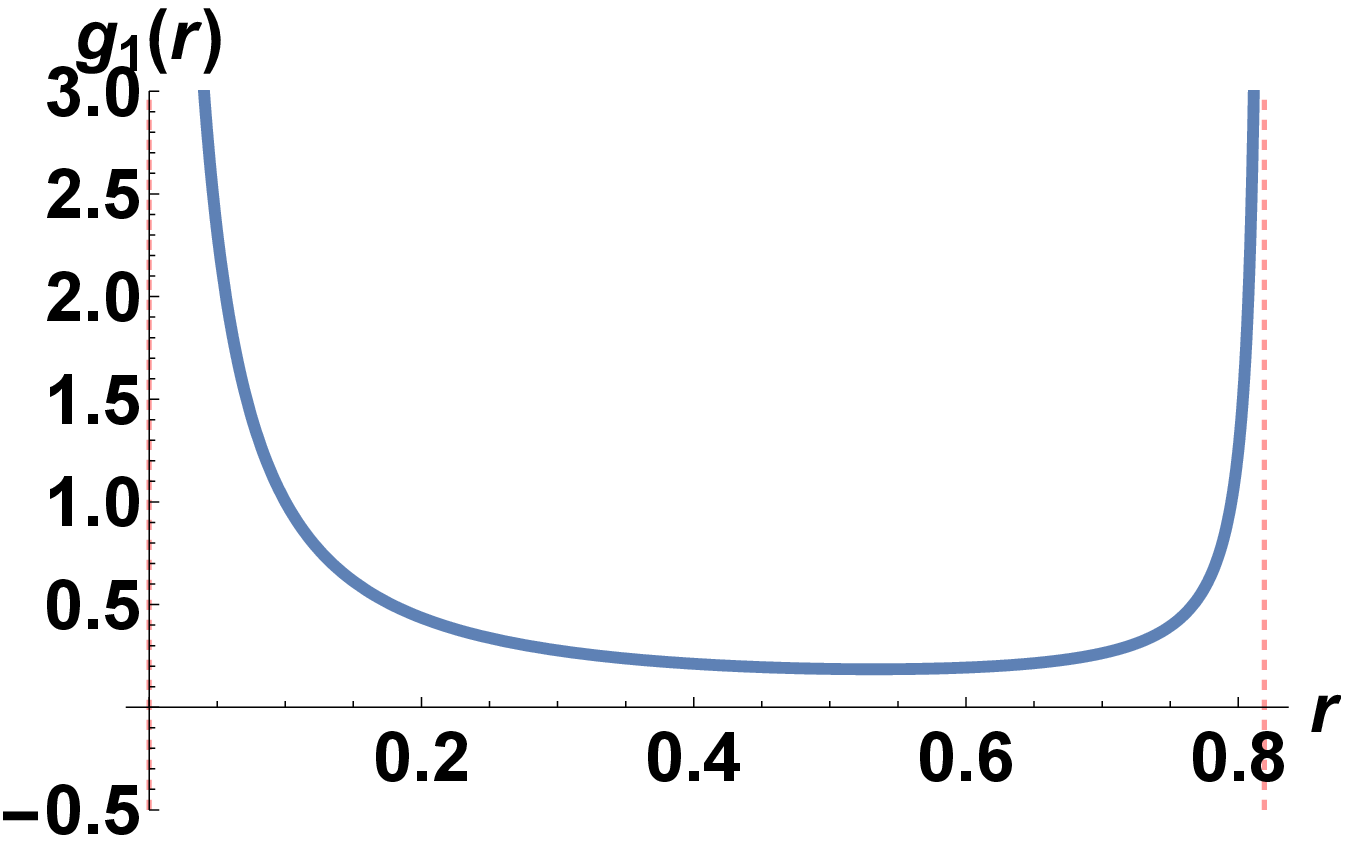}
\caption{$g_1$ solution}
  \end{subfigure}
  \begin{subfigure}[b]{0.326\linewidth}
    \includegraphics[width=\linewidth]{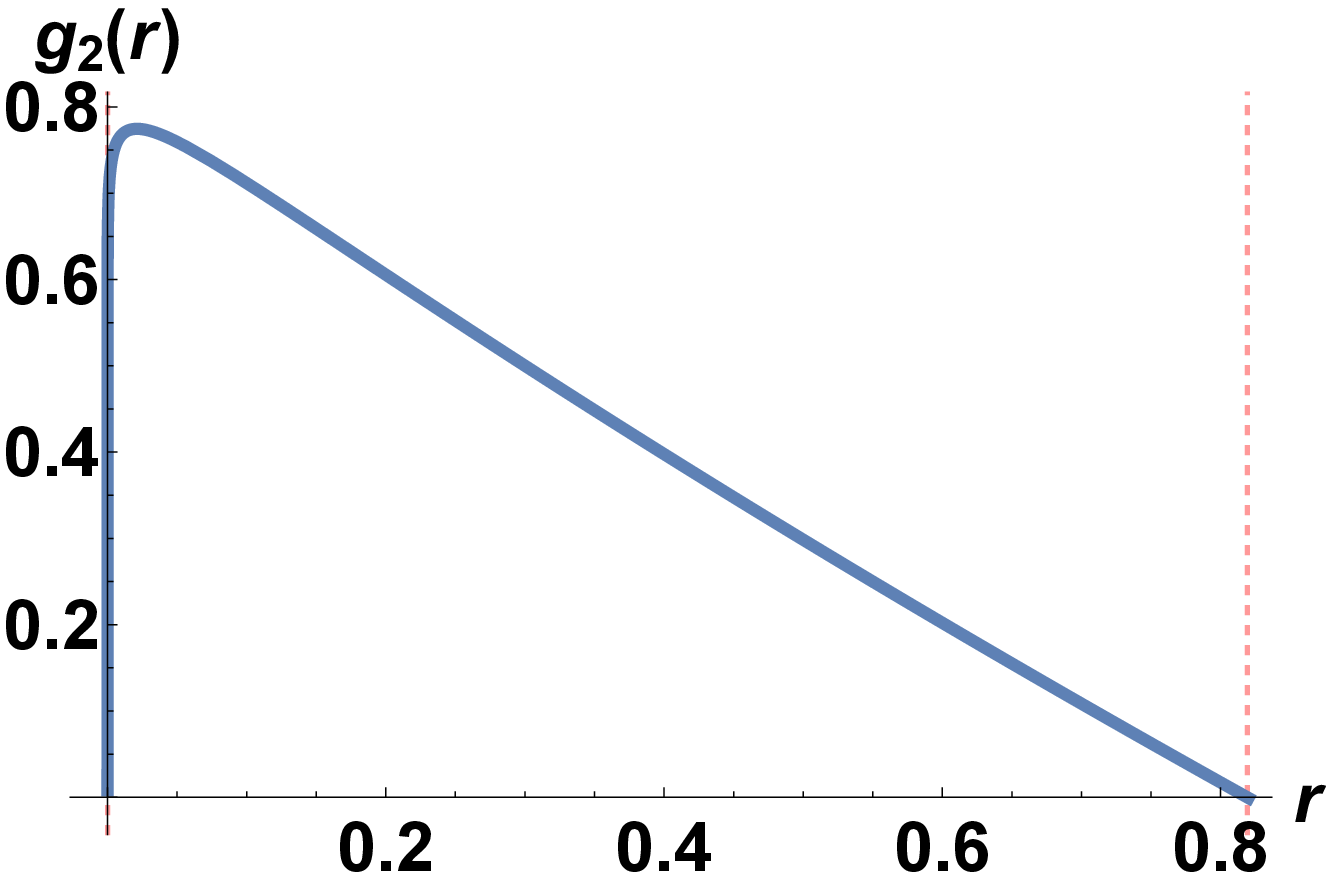}
\caption{$g_2$ solution}
  \end{subfigure}
  \caption{Numerical plots of the warp factors for $SO(2)\times SO(2)$ symmetric solution in $SO(2,2)$ gauge group with $m=1$, $B=1$, $h=\frac{1}{8}$, and $\mathcal{C}=-\frac{1}{4}$. The solution is regular in the range $0<r<r_2=0.819$ with the two vertical red dashed lines representing the two boundaries.}
  \label{SO(2,2)_SO(2)xSO(2)_plots}
\end{figure}

This is very similar to Case II in Section \ref{pSUGRA_Soln_Sec}. As $r\rightarrow 0$, the seven-dimensional metric is again conformal to a product of $AdS_5$ and a cylinder. With the new radial coordinate $R=r^{3/8}$, the metric near $R=0$ is given in \eqref{CaseIw0_metric}. As $r\rightarrow r_2$, the seven-dimensional metric is approximately given by
\begin{equation}\label{SO(2,2)_SO(2)xSO(2)_spindle_metric}
ds_7^2\approx\frac{B\,r_2^{1/20}}{16h^2(1+m^2r_2)^{1/4}}\left[ds^2_{AdS_5}+\frac{dR^2+4\mathcal{C}^2\left[3+4m^2r_2\right]^2R^2dz^2}{-16W'(r_2)r_2^{5/4}(1+m^2r_2)^{7/4}}\right]
\end{equation}
with the coordinate $R$ defined by $R=\sqrt{r_2-r}$. The $z$ circle shrinks smoothly giving rise to an $\mathbb{R}^2/\mathbb{Z}_l$ orbifold at $r=r_2$ after imposing the condition
\begin{equation}\label{SO(2,2)_SO(2)xSO(2)_smooth_con}
|\mathcal{C}|=\frac{1}{2l\left[3+4m^2r_2\right]},\qquad l=1,2,3,\ldots \,.
\end{equation}
Since the explicit form of $3+4m^2r_2$ derived from \eqref{SO(2,2)_SO(2)xSO(2)_r1} is rather complicated, we only numerically show that this function is always greater than $3$ in the regularity range in Figure \ref{SO(2,2)_SO(2)xSO(2)_sur}.

\begin{figure}[h!]
  \centering
    \includegraphics[width=0.56\linewidth]{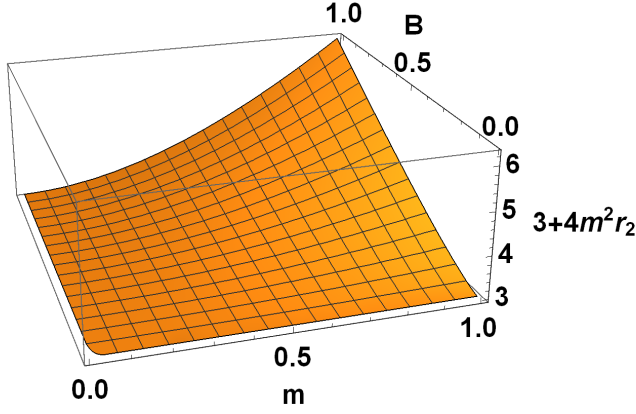}
  \caption{A numerical plot of the function $3+4m^2r_2$ appearing in the condition \eqref{SO(2,2)_SO(2)xSO(2)_smooth_con}.}
  \label{SO(2,2)_SO(2)xSO(2)_sur}
\end{figure}

Fixing the constants
\begin{equation}
q=|\mathcal{C}|\left[3+4m^2r_2\right]=\frac{1}{2l},\qquad c_2=-\frac{|\mathcal{C}|}{2h}\sqrt{r_2(1+m^2r_2)}=-\frac{|\mathcal{C}|B}{2h},
\end{equation}
we obtain the $SO(2)\times SO(2)$ gauge fields
\begin{equation}
A_1=\frac{|\mathcal{C}|}{2h}\left[\sqrt{r(1+m^2r)}-\sqrt{r_2(1+m^2r_2)}\right],\qquad A_2=\frac{m|\mathcal{C}|}{2h}(r_2-r)
\end{equation}
which vanish at $r=r_2$. The Killing spinor at the endpoint $r=r_2$ is explicitly given by
\begin{equation}\label{SO(2)diag_eta_Soln}
\eta=\sqrt{2}Y_0e^{\frac{iz}{2l}}r_2^{1/5}\begin{footnotesize}\begin{pmatrix} 1 \\ 0 \end{pmatrix}\end{footnotesize}.
\end{equation}

%%%%%%%%%%%%%%%%%%%%%%%%%%%%%%%%%%%%%%%
\subsection{$SO(2)_{\text{diag}}$ symmetric solution}\label{SO(2,2)_SO(2)diag_Sec}
To find solutions preserving $SO(2)_{\text{diag}}$ symmetry generated by $T_3+ T_6$ in $SO(2,2)$ gauge group, we use the following ansatz for vector fields
\begin{equation}\label{SO(2)diag_Gaugefield}
A^{I}_{(1)}=A_2\left(\frac{\widetilde{g}_1}{\widetilde{g}_2}\delta^I_3+\delta^I_6\right)dz\, .
\end{equation}
There are three $SO(2)_{\text{diag}}$ singlets corresponding to non-compact generators $Y_{33}$ together with $\hat{Y}_1$ and $\hat{Y}_2$ given in \eqref{hatY12}. As in $SO(4)$ gauge group, solving the vector field equations \eqref{Vec_eq} requires $\widetilde{g}_2=\pm\widetilde{g}_1$. We again choose $\widetilde{g}_2=\widetilde{g}_1$ giving rise to non-vanishing components of the dressed field strength tensors in \eqref{SO(4)_SO(2)diag_F1F2}.

Repeating the same procedure as in Section \ref{SO(4)_SO(2)diag_Sec}, we can solve all the BPS conditions and find the solution
\begin{eqnarray}
\phi&=&\frac{5\sigma}{2}+\ln\left[-\frac{16h}{\widetilde{g}_1}\right],\qquad  \varphi_1=\varphi_2=0,\label{SO(2,2)_SO(2)diag_scalar soln}\\
f&=&-\frac{2\sqrt{2}\sqrt{-\widetilde{g}_1b\,h}e^\sigma}{\sqrt{s(256h^2e^{5\sigma}+\widetilde{g}_1^2)}},\\
g_1&=&\frac{204,800\sqrt{2}\sqrt{-\widetilde{g}_1b\,h^9}\,e^{6\sigma}\,(\sigma')^2}{(256h^2e^{5\sigma}+\widetilde{g}_1^2)^2\left[32\sqrt{2}\sqrt{-\widetilde{g}_1b\,h^5}+e^{-5\sigma}\sqrt{s(256h^2e^{5\sigma}+\widetilde{g}_1^2)}\right]},\\
g_2&=&-\frac{1,024c^2h^2e^\sigma\left[32\sqrt{2}\sqrt{-\widetilde{g}_1b\,h^5}+e^{-5\sigma}\sqrt{s(256h^2e^{5\sigma}+\widetilde{g}_1^2)}\right]}{\sqrt{s(256h^2e^{5\sigma}+\widetilde{g}_1^2)}},\\
A_2&=&c\left(128h^2+\widetilde{g}_1^2e^{-5\sigma}\right)-\frac{2q}{\widetilde{g}_1}
\end{eqnarray}
together with the Killing spinor
\begin{equation}
\eta=\frac{Y_0\,e^{iqz+\frac{\sigma}{4}}}{\left[2s(256h^2e^{5\sigma}+\widetilde{g}_1^2)\right]^{\frac{1}{8}}}\begin{pmatrix} \sqrt{8(-\widetilde{g}_1b\,h^5)^{\frac{1}{4}}+e^{-\frac{5\sigma}{2}}\left[2s(256h^2e^{5\sigma}+\widetilde{g}_1^2)\right]^{\frac{1}{4}}} \\ -\sqrt{8(-\widetilde{g}_1b\,h^5)^{\frac{1}{4}}-e^{-\frac{5\sigma}{2}}\left[2s(256h^2e^{5\sigma}+\widetilde{g}_1^2)\right]^{\frac{1}{4}}}\end{pmatrix}
\end{equation}
and the sign condition $\text{sign}(c\widetilde{g}_1\sigma')=+1$. We also note that the $\phi$ solution in \eqref{SO(2,2)_SO(2)diag_scalar soln} requires $\frac{16h}{\widetilde{g}_1}<0$. 
Defining the following parameters
\begin{equation}\label{SO(2)diag_SO(4)_Cons}
B=-\frac{2\sqrt{2}h\sqrt{-\widetilde{g}_1b\,h}}{\sqrt{s}},\qquad m=\frac{\widetilde{g}_1}{16h},\qquad \mathcal{C}=32\widetilde{g}_1 h^2c
\end{equation}
and using the solution for $\sigma$ from \eqref{1/5sigma_soln}, we find the seven-dimensional metric given by
\begin{equation}
ds^2_7=\frac{Br^{3/10}}{16h^2\sqrt{1+m^2r}}\left[ds^2_{AdS_5}+\frac{r^{-3/2}}{16W(1+m^2r)^{3/2}}dr^2+\frac{16\mathcal{C}^2W}{B}dz^2\right]\label{SO(2)diag_SO(2,2)_Met}
\end{equation}
with
\begin{equation}\label{SO(2)diag_SO(2,2)_Def_W}
W=B-\sqrt{r(1+m^2r)}\, .
\end{equation}
The scalar $\phi_2$ and the gauge field can be written as
\begin{equation}\label{SO(2)diag_phi2_vector_Soln}
\phi_2=-\frac{1}{2}\ln r-\ln\left[-m\right]\qquad\text{ and }\qquad A_2=-\frac{1}{8mh}\left[2|\mathcal{C}|(1+2m^2r)+q\right].
\end{equation}
Furthermore, since we have chosen $\sigma<0$, the constant $\mathcal{C}$ must be negative. Finally, the Killing spinor reads
\begin{equation}\label{SO(2)diag_eta_Soln}
\eta=Y_0e^{iqz}\frac{2^{1/8}r^{3/40}}{(1+m^2r)^{\frac{1}{8}}}\begin{pmatrix} \sqrt{\sqrt{B}+\left[r(1+m^2r)\right]^{\frac{1}{4}}} \\ -\sqrt{\sqrt{B}-\left[r(1+m^2r)\right]^{\frac{1}{4}}} \end{pmatrix}.
\end{equation}

As in the $SO(2)\times SO(2)$ symmetric solution, there is only one possible range of the radial coordinate $r$ in order to obtain regular solutions. This is given by
\begin{equation}\label{SO(2,2)_SO(2)diag_Pos_Range}
m<0,\qquad B>0,\qquad 0<r<r_3,\qquad r_3=\frac{-1+\sqrt{1+4B^2m^2}}{2m^2}
\end{equation}
with $r_3$ determined from $W(r_3)=0$. An example of numerical plots for the three warp factors is given in Figure \ref{SO(2,2)_SO(2)diag_plots}.

\begin{figure}[h!]
  \centering
  \begin{subfigure}[b]{0.326\linewidth}
    \includegraphics[width=\linewidth]{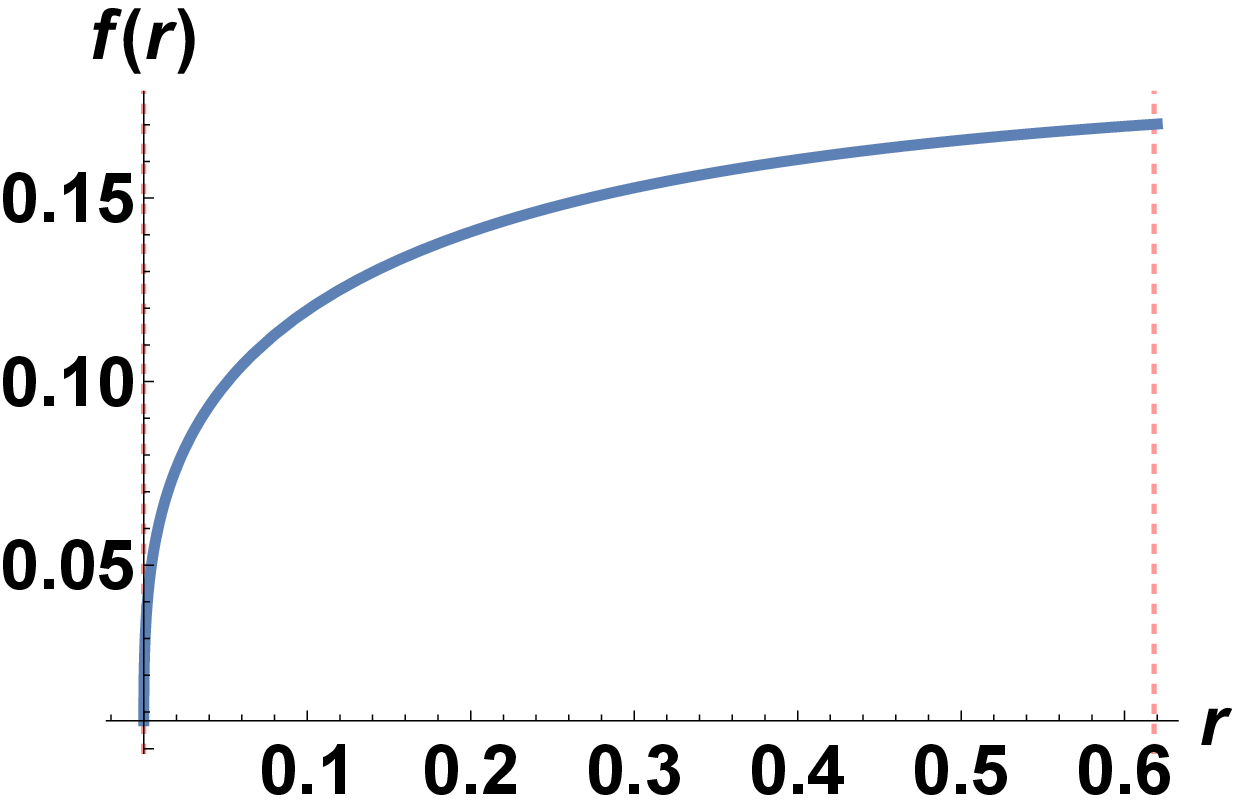}
\caption{$f$ solution}
  \end{subfigure}
  \begin{subfigure}[b]{0.326\linewidth}
    \includegraphics[width=\linewidth]{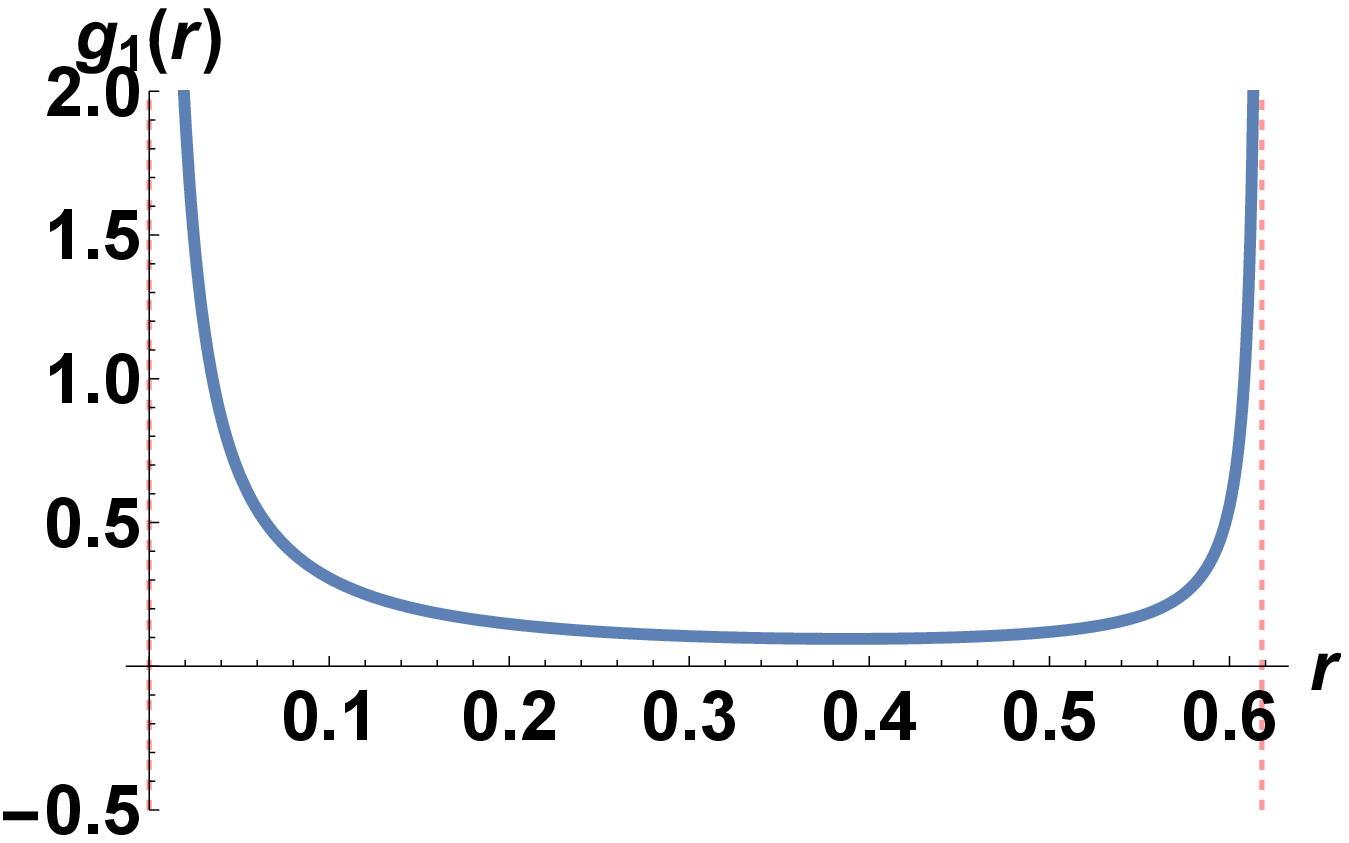}
\caption{$g_1$ solution}
  \end{subfigure}
  \begin{subfigure}[b]{0.326\linewidth}
    \includegraphics[width=\linewidth]{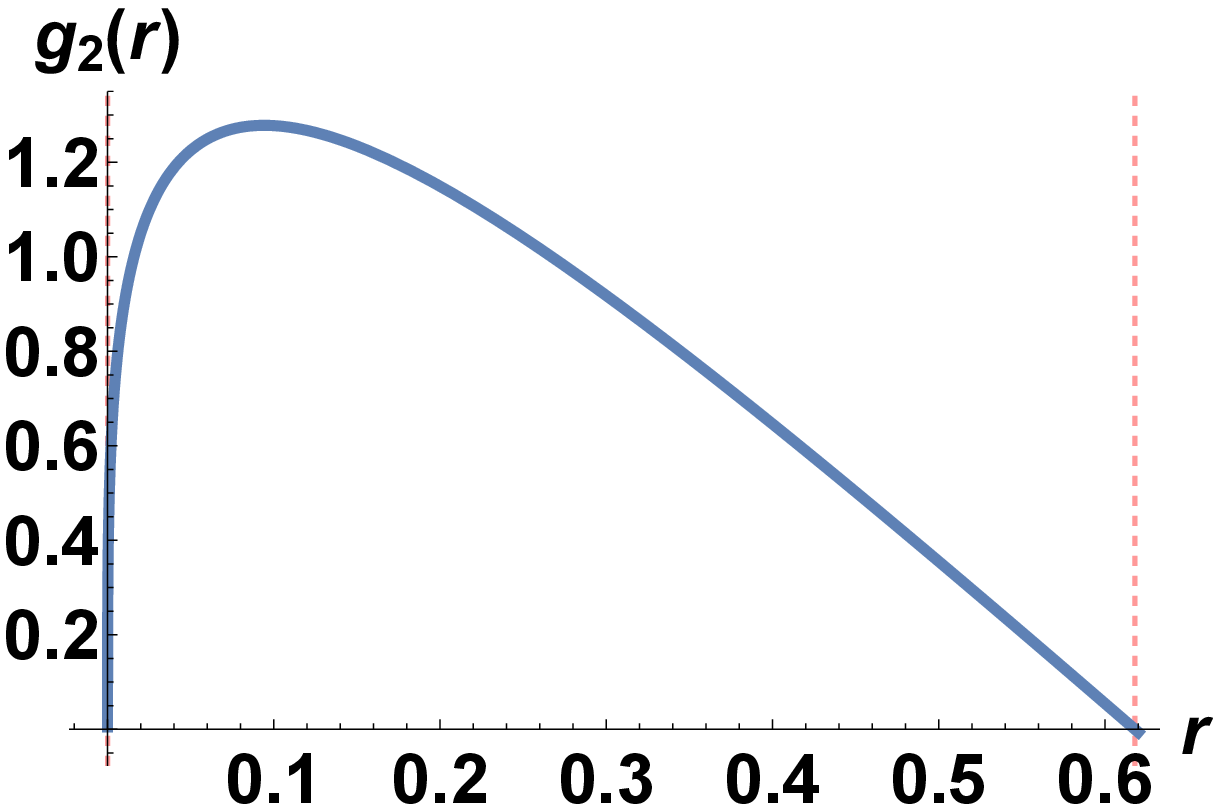}
\caption{$g_2$ solution}
  \end{subfigure}
  \caption{Numerical plots of the warp factors for $SO(2)_{\text{diag}}$ symmetric solution in $SO(2,2)$ gauge group with $m=-1$, $B=1$, $h=\frac{1}{2}$, and $\mathcal{C}=-1$. The solution is regular in the range $0<r<r_3=0.618$ with the two vertical red dashed lines representing the two boundaries.}
  \label{SO(2,2)_SO(2)diag_plots}
\end{figure}

As $r\rightarrow0$, the seven-dimensional metric becomes a conformal rescaling of a product between $AdS_5$ and a cylinder as in Case I of the $SO(4)$ gauge group. As $r\rightarrow r_3$, the metric is approximately given by
\begin{equation}
ds_7^2\approx\frac{B\,r_3^{3/10}}{16h^2\sqrt{1+m^2r_3}}\left[ds^2_{AdS_5}+\frac{dR^2+16\mathcal{C}^2\left[1+2m^2r_{3}\right]^2R^2dz^2}{-4W'(r_{3})r_{3}^{3/2}(1+m^2r_{3})^{3/2}}\right]
\end{equation}
with the new radial coordinate $R$ defined by $R=\sqrt{r_3-r}$. The $z$ circle shrinks smoothly giving rise to an $\mathbb{R}^2/\mathbb{Z}_l$ orbifold at $r=r_3$ for
\begin{equation}
|\mathcal{C}|=\frac{1}{4l\left[1+2m^2r_3\right]}=\frac{1}{4l\sqrt{1+4B^2m^2}},\qquad l=1,2,3,\ldots \, .
\end{equation}
With the constant $q$ chosen to be
\begin{equation}
q=-2|\mathcal{C}|\left[1+2m^2r_2\right]=-\frac{1}{2l},
\end{equation}
the $SO(2)_{\textrm{diag}}$ gauge field takes the form
\begin{equation}
A_2=\frac{m|\mathcal{C}|}{2h}(r_3-r).
\end{equation}
The Killing spinor at the endpoint $r=r_2$ is given by
\begin{equation}\label{SO(2)diag_eta_Soln}
\eta=2^{5/8}Y_0e^{\frac{-iz}{2l}}r_2^{1/5}\begin{footnotesize}\begin{pmatrix} 1 \\ 0 \end{pmatrix}\end{footnotesize}.
\end{equation}

%%%%%%%%%%%%%%%%%%%%%%%%%%%%%%%%%%%%%%%%%%%%%%%%%%%%%%%%%%%%%%%%%%%%%%%%%%%%%%%%%%%%%%%%%%%%%%%%%%%%%%%%%%%%%%%%%%%%%%%%%%%%%%%%%%%%%%%%%
\section{$SO(3,1)$ gauge group}\label{SO(3,1)_Sec}
In this section, we consider non-compact $SO(3,1)$ gauge group which can also be embedded in $SO(3,3)$. The gauge structure constants are chosen to be
\begin{equation}
f_{IJK}=-\widetilde{g}(\varepsilon_{ijk},\varepsilon_{rsi}),\qquad r,s,\ldots=1,2,3
\end{equation}
with a coupling constant $\widetilde{g}$. The $SO(3,1)$ generators can be written as $T_I=(T_i,\hat{T}_r)$ with the explicit form in terms of $SO(3,3)$ generators given by
\begin{eqnarray}
T_1&=&J^{(1)}_{23}-J^{(2)}_{23},\quad T_2\ =\ J^{(1)}_{31}+J^{(2)}_{31},\quad T_3\ =\ J^{(1)}_{12}-J^{(2)}_{12},\nonumber\\
\hat{T}_1&=&Y_{23}+Y_{32},\quad\ \, \hat{T}_2\ =\ Y_{13}-Y_{31},\quad\ \, \hat{T}_3\ =\ -Y_{12}-Y_{21}
\end{eqnarray}
satisfying $SO(3,1)$ algebra
\begin{equation}
[T_i,T_j]={\varepsilon_{ij}}^kT_k,\qquad [T_i,\hat{T}_r]={\varepsilon_{ir}}^s\hat{T}_s,\qquad [\hat{T}_r,\hat{T}_s]={\varepsilon_{rs}}^iT_i\, .
\end{equation}
$T_{i}$ are compact generators of the maximal compact subgroup $SO(3)\subset SO(3,1)$ while $\hat{T}_r$ are non-compact generators. 

We now look for supersymmetric $AdS_5\times \Sigma$ solutions preserving $SO(2)\subset SO(3)$ symmetry by using the same analysis as in the previous two gauge groups. The metric is still given by \eqref{7Dmetrix}, and the non-vanishing $SO(2)$ gauge field takes the form of
\begin{equation}\label{SO(3,1)_Gaugefield}
A^{3}_{(1)}=A(r)dz\, .
\end{equation}
There are three $SO(2)$ singlet scalars corresponding to the non-compact generators
\begin{equation}\label{SO(3,1)_singlet}
\bar{Y}_1=Y_{11}-Y_{22},\qquad \bar{Y}_2=Y_{33}, \qquad \bar{Y}_3=Y_{12}+Y_{21},
\end{equation}
and the coset representative is given by
\begin{equation}\label{SO(3,1)_coset_rep}
L=e^{\phi_1\bar{Y}_1}e^{\phi_2\bar{Y}_2}e^{\phi_3\bar{Y}_3}\, .
\end{equation}

With all these, the vector field equation \eqref{Vec_eq} gives rise to
\begin{eqnarray}
\phi'_3&=&\frac{\phi'_1}{2}\tanh{2\phi_1}\sinh{4\phi_3},\\
A''&\hspace{-0.2cm}=&\hspace{-0.2cm}-\left[\frac{5f'}{2f}-\frac{g'_1}{2g_1}-\frac{g'_2}{2g_2}+\sigma'+2\phi'_2\tanh{2\phi_2}\right]A',\label{SO(3,1)_G_Eq2}\\
A''&\hspace{-0.2cm}=&\hspace{-0.2cm}-\left[\frac{5f'}{2f}-\frac{g'_1}{2g_1}-\frac{g'_2}{2g_2}+\sigma'+2\phi'_2\,\text{coth}\,{2\phi_2}\right]A'.\label{SO(3,1)_G_Eq3}
\end{eqnarray}
Consistency between equations \eqref{SO(3,1)_G_Eq2} and \eqref{SO(3,1)_G_Eq3} requires $\phi'_2=0$ for $A'\neq0$. We will write the constant $\phi_2$ as $c$. This leads to a single differential equation for $A$
\begin{equation}
A''=-\left[\frac{5f'}{2f}-\frac{g'_1}{2g_1}-\frac{g'_2}{2g_2}+\sigma'\right]A
\end{equation}
with the solution 
\begin{equation}
A'=be^{-\sigma}\sqrt{g_1g_2}f^{-\frac{5}{2}}
\end{equation}
for an integration constant $b$. The dressed field strength tensors are now given by
\begin{equation}
\mathbf{F}_1=b\cosh{c}e^{-\sigma}f^{-\frac{5}{2}}\qquad\text{ and }\qquad\mathbf{F}_2=b\sinh{c}e^{-\sigma}f^{-\frac{5}{2}}.
\end{equation}
\indent In this case, non-vanishing conponenets of the $C$-functions read 
\begin{eqnarray}
C&\hspace{-0.2cm}=&\hspace{-0.2cm}-3\sqrt{2}\,\widetilde{g}(\cosh{c}-\sinh{c}\sinh{2\phi_1}\cosh{2\phi_3}),\nonumber\\
C^{11}&\hspace{-0.2cm}=&\hspace{-0.2cm}-C^{22}=-\sqrt{2}\,\widetilde{g}\sinh{c}\cosh{2\phi_1},\nonumber\\
C^{12}&\hspace{-0.2cm}=&\hspace{-0.2cm}C^{21}=-\sqrt{2}\,\widetilde{g}\sinh{c}\sinh{2\phi_1}\sinh{2\phi_3},\nonumber\\
C^{33}&\hspace{-0.2cm}=&\hspace{-0.2cm}\sqrt{2}\,\widetilde{g}(\sinh{c}-\cosh{c}\sinh{2\phi_1}\cosh{2\phi_3}).
\end{eqnarray}
The scalar vielbein and the $SO(2)$ composite connection are given by
\begin{equation}
P^{ir}_{(1)}=\begin{pmatrix}	\phi'_1\cosh{2\phi_3} & \phi'_3 & 0 \\
						\phi'_3 & -\phi'_1\cosh{2\phi_3} & 0 \\
						0 & 0 & 0
			\end{pmatrix} dr
\end{equation}
and
\begin{equation}
Q^{ij}_{(1)}=\varepsilon^{ij3}(\phi'_1\sinh{2\phi_3}dr-\widetilde{g}Adz).
\end{equation}

With the supersymmetry parameter \eqref{7DKilling} subject to the projector \eqref{ProjCon}, the supersymmetry transformations $\delta\lambda^{a1}=0$ and $\delta\lambda^{a2}=0$ give the same BPS condition while $\delta\lambda^{a3}=0$ gives another equation. These equations take the form
\begin{eqnarray}
0&=&\widetilde{g}\sinh{c}e^{-\frac{\sigma}{2}}\left[\cosh{2\phi_1}+i \sinh{2\phi_1}\sinh{2\phi_3}\right](i\sigma^2\eta)\nonumber\\&& +\frac{1}{\sqrt{g_1}}\left[\phi'_1\cosh{2\phi_3}-i \phi'_3\right]\sigma^3\eta,\label{SO(3,1)_vBPS1}\\
0&=&\widetilde{g}e^{-\frac{\sigma}{2}}[\sinh{c}-\cosh{c}\sinh{2\phi_1}\cosh{2\phi_3}]\eta-e^{\frac{\sigma}{2}}be^{-\sigma}\sinh{c}f^{-\frac{5}{2}}\sigma^3\eta\, .\label{SO(3,1)_vBPS2}\qquad
\end{eqnarray}
Solving these conditions gives $\phi_1=0$, $c=0$ and $\phi'_3=0$. Since the constant $\phi_3$ does not appear in other equations, we can simply set $\phi_3=0$ without losing any generality. However, with $\phi_1=\phi_2=\phi_3=0$, all matter fields from the vector multiplets vanish. The resulting solution turns out to be the same as the $SO(2)_R$ symmetric solution given in pure $N=2$ gauged supergravity discussed in Section \ref{pSUGRA_Soln_Sec}. 

%%%%%%%%%%%%%%%%%%%%%%%%%%%%%%%%%%%%%%%%%%%%%%%%%%%%%%%%%%%%%%%%%%%%%%%%%%%%%%%%%%%%%%%%%%%%%%%%%%%%%%%%%%%%%%%%%%%%%%%%%%%%%%%%%%%%%%%%%
\section{Conclusions}\label{Conclus}
We have found supersymmetric $AdS_5\times \Sigma$ solutions in which $\Sigma$ is a topological disk with a non-trivial $U(1)$ holonomy at the boundary or a half-spindle from matter-coupled $N=2$ gauged supergravity in seven dimensions with $SO(4)$ and $SO(2,2)$ gauge groups. These solutions preserve eight supercharges and $SO(2)_R$, $SO(2)\times SO(2)$ and $SO(2)_{\text{diag}}$ symmetries. These solutions represent a new class of supersymmetric solutions of $N=2$ gauged supergravity in seven dimensions and might be useful in holographic studies. We have also extensively discussed various possible ranges of the radial coordinate in which the resulting solutions are regular. The $SO(2)_R$ symmetric solution requires vanishing of all fields from the vector multiplets and can be considered as a solution of pure $N=2$ gauged supergravity with $SO(3)$ gauge group.  

For $SO(3,1)$ gauge group, we have found only solutions with $SO(2)_R\subset SO(3)_R$ symmetry. Furthermore, we have also considered other possible gauge groups namely $SL(3,\mathbb{R})$, $SO(2,1)$, and $SO(2, 2)\times SO(2, 1)\sim SO(2, 1)\times SO(2, 1)\times SO(2, 1)$ gauge groups. However, all of these gauge groups do not lead to $AdS_5\times \Sigma$ solutions with non-vanishing fields from vector multiplets. On the other hand, these gauge groups do admit a solution with $SO(2)_R$ symmetry which appears to be a universal solution to all gauge groups.

Similar to the result of \cite{Ferrero_M5}, the solutions should be dual to $N=1$ SCFTs in four dimensions obtained from compactifications on a half-spindle of six-dimensional $N=(1,0)$ SCFTs in the case of $SO(4)$ gauge group or $N=(1,0)$ non-conformal field theory in the case of $SO(2,2)$ gauge group. Solutions of pure $N=2$ and $SO(4)$ gauged supergravity with equal $SO(3)$ coupling constants, $\widetilde{g}_1=\widetilde{g}_2$, can be embedded in eleven-dimensional supergravity via consistent truncations constructed in \cite{Pure_Red_Ans} and \cite{7D_from_11D}. In this case, the uplifted solutions would describe M5-branes wrapped on a half-spindle as in \cite{Ferrero_M5}, but the brane configurations preserve only $\frac{1}{4}$ of the maximal supersymmetry rather than $\frac{1}{2}$. For $SO(2,2)$ gauge group, an embedding in ten-dimensional type I or heterotic theories via a truncation on a hyperbolic space $H^{2,2}$ might be obtained by extending the result of \cite{Pope_SO2_2_7D} to $SO(2,2)$ gauged supergravity with a non-vanishing topological mass for the three-form field. In this case, the uplifted solutions would describe D5-branes or NS5-branes wrapped on half-spindles.  

It would be interesting to explicitly identify the four-dimensional $N=1$ SCFTs that are dual to the supergravity solutions found in this paper. Uplifting the solutions in $SO(2,2)$ gauge group and in $SO(4)$ gauge group with different $SO(3)$ coupling constants to ten or eleven dimensions is of particular interest in the holographic context and could lead to new configurations of D5/NS5-branes or M5-branes wrapped on half-spindles. This could be done along the line of \cite{Henning_dualize_IIA_IIB} in which the embedding of maximal $N=4$ gauged supergravity with various gauge groups has been given by using $SL(5)$ exceptional field theory. In the present case of half-maximal gauged supergravity, the results on $SO(3,n)$ double field theory and non-geometric fluxes in \cite{Dibitetto_non_Geometric,7D_N2_Malek,EFT_Heterotic_Malek,Henning_Malek_AdS7_6} would be very useful. We hope to come back to these issues in future works.  

\begin{acknowledgments}
This work is supported by the Second Century Fund (C2F), Chulalongkorn University. P. K. is supported by The Thailand Research Fund (TRF) under grant RSA6280022.
\end{acknowledgments}

%%%%%%%%%%%%%%%%%%%%%%%%%%%%%%%%%%%%%%%%%%%%%%%%%%%%%%%%%%%%%%%%
\appendix
\section{Derivation of BPS equations in $SO(4)$ gauge group}
In this appendix, we analyze first-order BPS conditions derived from fermionic supersymmetry transformations. We begin with the first condition $\delta\psi_\mu^a=0$ along $m$, $\rho$, $r$, and $z$ directions. These are respectively given by
\begin{eqnarray}
0&=&2\partial_m\epsilon^a-\frac{1}{\sqrt{f}}\Gamma_m\Gamma^{\hat{\rho}}\epsilon^a+\frac{f'}{2f\sqrt{g_1}}\Gamma_m\Gamma^{\hat{r}}\epsilon^a-\left[\frac{\sqrt{2}}{30}e^{-\frac{\sigma}{2}}C+\frac{4}{5}he^{2\sigma}\right]\Gamma_m\epsilon^a\nonumber\\&&+\frac{i}{5}e^{\frac{\sigma}{2}}\mathbf{F}_1{(\sigma^3)^a}_b\Gamma_m\Gamma^{\hat{r}\hat{z}}\epsilon^b,\label{BPS1}\\
0&=&2\partial_\rho\epsilon^a+\frac{f'}{2f\sqrt{g_1}}\Gamma_\rho\Gamma^{\hat{r}}\epsilon^a-\left[\frac{\sqrt{2}}{30}e^{-\frac{\sigma}{2}}C+\frac{4}{5}he^{2\sigma}\right]\Gamma_\rho\epsilon^a\nonumber\\&&+\frac{i}{5}e^{\frac{\sigma}{2}}\mathbf{F}_1{(\sigma^3)^a}_b\Gamma_\rho\Gamma^{\hat{r}\hat{z}}\epsilon^b,\label{BPS2}\\
0&=&2\partial_r\epsilon^a-\sqrt{g_1}\left[\frac{\sqrt{2}}{30}e^{-\frac{\sigma}{2}}C+\frac{4}{5}he^{2\sigma}\right]\Gamma^{\hat{r}}\epsilon^a-\frac{4i}{5}\sqrt{g_1}e^{\frac{\sigma}{2}}\mathbf{F}_1{(\sigma^3)^a}_b\Gamma^{\hat{z}}\epsilon^b,\label{BPS3}\\
0&=&2\partial_z\epsilon^a-\frac{g'_2}{2\sqrt{g_1g_2}}\Gamma^{\hat{r}\hat{z}}\epsilon^a+i\widetilde{g}_1A_1{(\sigma^3)^a}_b\epsilon^b-\sqrt{g_2}\left[\frac{\sqrt{2}}{30}e^{-\frac{\sigma}{2}}C+\frac{4}{5}he^{2\sigma}\right]\Gamma^{\hat{z}}\epsilon^a\nonumber\\&&+\frac{4i}{5}\sqrt{g_2}e^{\frac{\sigma}{2}}\mathbf{F}_1{(\sigma^3)^a}_b\Gamma^{\hat{r}}\epsilon^b\, .\label{BPS4}
\end{eqnarray}
The next condition $\delta\chi^a=0$ yields
\begin{equation}
0\ =\ -\frac{\sigma'}{2\sqrt{g_1}}\Gamma^{\hat{r}}\epsilon^a+\left[\frac{\sqrt{2}}{30}e^{-\frac{\sigma}{2}}C-\frac{16}{5}he^{2\sigma}\right]\epsilon^a-\frac{i}{5}e^{\frac{\sigma}{2}}\mathbf{F}_1{(\sigma^3)^a}_b\Gamma^{\hat{r}\hat{z}}\epsilon^b\, .\label{BPS5}
\end{equation}
From the supersymmetry transformation of the gaugini, only $\delta\lambda^{a3}=0$ gives rise to non-trivial conditions given by
\begin{equation}
0\ =\ \frac{\phi'}{\sqrt{g_1}}{(\sigma^3)^a}_b\Gamma^{\hat{r}}\epsilon^b-\frac{1}{\sqrt{2}}e^{-\frac{\sigma}{2}}C^{33}{(\sigma^3)^a}_b\epsilon^b+ie^{\frac{\sigma}{2}}\mathbf{F}_2\Gamma^{\hat{r}\hat{z}}\epsilon^a\, .\label{BPS6}
\end{equation}
In these equations, for convenience, we have introduced the notations $\mathbf{F}_1$ and $\mathbf{F}_2$ for non-vanishing components of the dressed field strength tensors via 
\begin{equation}\label{Gen_form_DF}
F^i_{(2)}={L_I}^iF^I_{(2)}=\mathbf{F}_1\delta^i_3\,e^{\hat{r}}\wedge e^{\hat{z}}\qquad\text{and}\qquad 
F^r_{(2)}={L_I}^rF^I_{(2)}=\mathbf{F}_2\delta^r_3\,e^{\hat{r}}\wedge e^{\hat{z}}\, .
\end{equation}
Explicitly, in the case of $SO(2)\times SO(2)$ symmetric solutions, these are given by
\begin{eqnarray}
& &\mathbf{F}_1=\frac{1}{\sqrt{g_1g_2}}\left(A'_1\cosh{\phi}+A'_2\sinh{\phi}\right)\nonumber \\
\textrm{and}\qquad & &\mathbf{F}_2=\frac{1}{\sqrt{g_1g_2}}\left(A'_1\sinh{\phi}+A'_2\cosh{\phi}\right).
\end{eqnarray}
\indent To solve the resulting BPS equations, we follow \cite{Bah_M5} and use the supersymmetry parameters of the form
\begin{equation}\label{7DKilling}
\epsilon^a=n^a\,\vartheta\otimes\eta\, .
\end{equation}
$n^a$ are two components of a constant object in the doublet representation of $SO(3)_R$ while $\eta$ is a two-component spinor depending on $r$ and $z$ coordinates. On the other hand, $\vartheta$ is a four-component Killing spinor on $AdS_5$ satisfying
\begin{equation}\label{5DKilling}
\nabla_{\hat{\alpha}}^{AdS_5}\vartheta=\frac{s}{2}\gamma_{\hat{\alpha}}\vartheta
\end{equation}
where $\hat{\alpha}=(\hat{m},\hat{\rho})=0,1,...,4$ is a flat space-time index on $AdS_5$, $s=\pm1$ is an arbitrary sign, and $\gamma_{\hat{\alpha}}$ are five-dimensional $4\times4$ gamma matrices satisfying the Clifford algebra 
\begin{equation}
\{\gamma_{\hat{\alpha}},\gamma_{\hat{\beta}}\}=2\eta_{\hat{\alpha}\hat{\beta}}, \qquad\eta_{\hat{\alpha}\hat{\beta}}=\text{diag}(-++++).
\end{equation}
In terms of $\gamma_{\hat{\alpha}}$, we further decompose the seven-dimensional gamma matrices as
\begin{equation}\label{7D_gamma_Dec}
\Gamma_{\hat{\alpha}}=\gamma_{\hat{\alpha}}\otimes\sigma^3,\qquad\quad \Gamma_{\hat{r}}=\mathds{1}_4\otimes\sigma^1, \qquad\quad \Gamma_{\hat{z}}=\mathds{1}_4\otimes\sigma^2
\end{equation}
in which $\mathds{1}_n$ is an $n\times n$ identity matrix.

With the supersymmetry parameter \eqref{7DKilling}, the first two BPS equations given in \eqref{BPS1} and \eqref{BPS2} reduce to a single equation 
\begin{eqnarray}
0&\hspace{-0.2cm}=&\hspace{-0.2cm}\frac{s}{\sqrt{f}}(\gamma_{\hat{\alpha}}\otimes\mathds{1}_2)\epsilon^a+\frac{f'\Gamma_{\hat{\alpha}}\Gamma^{\hat{r}}\epsilon^a}{2f\sqrt{g_1}}-\left[\frac{\sqrt{2}}{30}e^{-\frac{\sigma}{2}}C+\frac{4}{5}he^{2\sigma}\right]\Gamma_{\hat{\alpha}}\epsilon^a+\frac{i}{5}e^{\frac{\sigma}{2}}\mathbf{F}_1{(\sigma^3)^a}_b\Gamma_{\hat{\alpha}}\Gamma^{\hat{r}\hat{z}}\epsilon^b\nonumber\\
&\hspace{-0.2cm}=&\hspace{-0.2cm}n^a(\gamma_{\hat{\alpha}}\vartheta)\otimes\left\{\frac{s}{\sqrt{f}}\eta+\frac{f'}{2f\sqrt{g_1}}(i\sigma^2\eta)-\left[\frac{\sqrt{2}}{30}e^{-\frac{\sigma}{2}}C+\frac{4}{5}he^{2\sigma}\right](\sigma^3\eta)-\frac{1}{5}e^{\frac{\sigma}{2}}\mathbf{F}_1\eta\right\}\nonumber\\&&
\end{eqnarray}
in which we have expressed all the supersymmetry parameters in terms of two- and four-component spinors and imposed the projector
\begin{equation}\label{ProjCon}
{(\sigma^3)^a}_bn^b=n^a\, .
\end{equation}
\indent By using the same procedure in the remaining conditions, we find the following set of BPS equations on the two-component spinor $\eta$
\begin{eqnarray}
0&=&\frac{s}{\sqrt{f}}\eta+\frac{f'}{2f\sqrt{g_1}}(i\sigma^2\eta)-\left[\frac{\sqrt{2}}{30}e^{-\frac{\sigma}{2}}C+\frac{4}{5}he^{2\sigma}\right](\sigma^3\eta)-\frac{1}{5}e^{\frac{\sigma}{2}}\mathbf{F}_1\eta,\quad\label{nBPSAdS5}\\
0&=&2\partial_r\eta-\sqrt{g_1}\left[\frac{\sqrt{2}}{30}e^{-\frac{\sigma}{2}}C+\frac{4}{5}he^{2\sigma}\right]\sigma^1\eta-\frac{4}{5}\sqrt{g_1}e^{\frac{\sigma}{2}}\mathbf{F}_1(i\sigma^2\eta),\quad\label{nBPS3}\\
0&=&-2i\partial_z\eta+\widetilde{g}_1A_1\eta-\frac{g'_2}{2\sqrt{g_1g_2}}\sigma^3\eta+\sqrt{g_2}\left[\frac{\sqrt{2}}{30}e^{-\frac{\sigma}{2}}C+\frac{4}{5}he^{2\sigma}\right](i\sigma^2\eta)\nonumber\\&&+\frac{4}{5}\sqrt{g_2}e^{\frac{\sigma}{2}}\mathbf{F}_1\sigma^1\eta,\label{nBPS4}\\
0&=&-\frac{\sigma'}{2\sqrt{g_1}}\sigma^1\eta+\left[\frac{\sqrt{2}}{30}e^{-\frac{\sigma}{2}}C-\frac{16}{5}he^{2\sigma}\right]\eta+\frac{1}{5}e^{\frac{\sigma}{2}}\mathbf{F}_1\sigma^3\eta,\label{nBPS5}\\
0&=&\frac{\phi'}{\sqrt{g_1}}\sigma^1\eta-\frac{1}{\sqrt{2}}e^{-\frac{\sigma}{2}}C^{33}\eta-e^{\frac{\sigma}{2}}\mathbf{F}_2\sigma^3\eta\, .\label{nBPS6}
\end{eqnarray}
\indent As in \cite{Bah_M5}, we assume a definite charge under the $U(1)_z$ isometry for the two-component spinor leading to the ansatz for $\eta$ of the form
\begin{equation}\label{eta_ansatz}
\eta(r,z)=e^{iqz}\widehat{\eta}(r)
\end{equation}
with a constant $q$. With this explicit form of $\eta$, it follows that the combination $(-2i\partial_z+\widetilde{g}_1A_1)\eta=(2q+\widetilde{g}_1A_1)\eta$ in \eqref{nBPS4} is invariant under the transformations
\begin{equation}
A_1\rightarrow A_1-\frac{2\alpha_0}{\widetilde{g}_1},\qquad \eta\rightarrow e^{i\alpha_0z}\eta
\end{equation}
where $\alpha_0$ is an arbitrary constant. It is then convenient to define
\begin{equation}\label{AandA}
\widetilde{g}_1\widehat{A}_1=2q+\widetilde{g}_1A_1
\end{equation}
with $\widehat{A}'_1=A'_1$.
\\
\indent With suitable left-multiplications by Pauli matrices and additions of \eqref{nBPS5} to \eqref{nBPSAdS5} and \eqref{nBPS4}, we find the following set of algebraic equations
\begin{eqnarray}
0&\hspace{-0.1cm}=&\hspace{-0.1cm}\frac{s}{\sqrt{f}}\eta+\frac{1}{2\sqrt{g_1}}\left[\frac{f'}{f}-\sigma'\right](i\sigma^2\eta)-4he^{2\sigma}\sigma^3\eta,\label{keyBPS1}\\
0&\hspace{-0.1cm}=&\hspace{-0.1cm}\frac{\widetilde{g}_1\widehat{A}_1}{\sqrt{g_2}}\sigma^1\eta+\frac{1}{2\sqrt{g_1}}\left[\frac{g'_2}{g_2}-\sigma'\right](i\sigma^2\eta)-4he^{2\sigma}\sigma^3\eta+e^{\frac{\sigma}{2}}\mathbf{F}_1\eta,\qquad\label{keyBPS2}\\
0&\hspace{-0.1cm}=&\hspace{-0.1cm}\frac{5\sigma'}{2\sqrt{g_1}}\sigma^3\eta-e^{-\frac{\sigma}{2}}\left[\frac{C}{3\sqrt{2}}-16he^{\frac{5\sigma}{2}}\right](i\sigma^2\eta)+e^{\frac{\sigma}{2}}\mathbf{F}_1\sigma^1\eta,\label{keyBPS3}\\
0&\hspace{-0.1cm}=&\hspace{-0.1cm}\frac{\phi'}{\sqrt{g_1}}\sigma^3\eta-\frac{1}{\sqrt{2}}e^{-\frac{\sigma}{2}}C^{33}(i\sigma^2\eta)+e^{\frac{\sigma}{2}}\mathbf{F}_2\sigma^1\eta\, .\label{keyBPS4}
\end{eqnarray}
These equations are of the form $M^{(x)}\eta=0$ where $x=1,2,3,4$ respectively labels the BPS equations in \eqref{keyBPS1} to \eqref{keyBPS4}. The four $2\times2$ matrices $M^{(x)}$ can be parametrized by
\begin{equation}
M^{(x)}=X^{(x)}_0\mathds{1}_2+X^{(x)}_1\sigma^1+X^{(x)}_2(i\sigma^2)+X^{(x)}_3\sigma^3\, .
\end{equation}
Following \cite{Bah_M5}, for each matrix $M^{(x)}$, we define the two-component vectors
\begin{equation}
v^{(x)}=\begin{pmatrix} X^{(x)}_1+X^{(x)}_2 \\ -X^{(x)}_0-X^{(x)}_3 \end{pmatrix},\qquad w^{(x)}=\begin{pmatrix} X^{(x)}_0-X^{(x)}_3 \\ -X^{(x)}_1+X^{(x)}_2 \end{pmatrix}
\end{equation}
together with
\begin{equation}
\mathcal{A}^{xy}=\text{det}(v^{(x)}|w^{(y)}),\qquad\mathcal{B}^{xy}=\text{det}(v^{(x)}|v^{(y)}),\qquad\mathcal{C}^{xy}=\text{det}(w^{(x)}|w^{(y)}).
\end{equation}
The notation $(a|b)$ denotes a $2\times2$ matrix obtained from a juxtaposition of the two-column vectors $a$ and $b$. As pointed out in \cite{Bah_M5}, the vanishing of $\mathcal{A}^{xy}$, $\mathcal{B}^{xy}$, and $\mathcal{C}^{xy}$ gives a number of necessary conditions for the existence of a non-trivial solution for $\eta$. We will separately determine the conditions from the supergravity and vector multiplets by splitting the index $x=(\bar{x},4)$ with $\bar{x}=1,2,3$. 

Starting from the first matrix $\mathcal{A}^{xy}$, the vanishing of the diagonal components $\mathcal{A}^{\bar{x}\bar{x}}$ gives the following conditions
\begin{eqnarray}
0&=&\frac{1}{f}-16h^2e^{4\sigma}+\frac{1}{4g_1}\left[\frac{f'}{f}-\sigma'\right]^2,\label{DABPS1}\\
0&=&e^{\sigma}(\mathbf{F}_1)^2+\frac{1}{4g_1}\left[\frac{g'_2}{g_2}-\sigma'\right]^2-16h^2e^{4\sigma}-\frac{(\widetilde{g}_1\widehat{A}_1)^2}{g_2},\label{DABPS2}\\
0&=&\frac{25\sigma'^2}{4g_1}+e^{\sigma}(\mathbf{F}_1)^2-e^{-\sigma}\left[\frac{C}{3\sqrt{2}}-16he^{\frac{5\sigma}{2}}\right]^2.\label{DABPS3}
\end{eqnarray}
The vanishing of the off-diagonal symmetric components $\mathcal{A}^{\bar{x}\bar{y}}+\mathcal{A}^{\bar{y}\bar{x}}$ ($\bar{x}\neq \bar{y}$) yields 
\begin{eqnarray}
0&\hspace{-0.1cm}=&\hspace{-0.1cm}\frac{1}{4g_1}\left[\frac{f'}{f}-\sigma'\right]\left[\frac{g'_2}{g_2}-\sigma'\right]+\frac{s e^{\frac{\sigma}{2}}}{\sqrt{f}}\mathbf{F}_1-16h^2e^{4\sigma},\label{SABPS1}\\
0&\hspace{-0.1cm}=&\hspace{-0.1cm}20he^{2\sigma}\sigma'-e^{-\frac{\sigma}{2}}\left[\frac{C}{3\sqrt{2}}-16he^{\frac{5\sigma}{2}}\right]\left[\frac{f'}{f}-\sigma'\right],\label{SABPS2}\\
0&\hspace{-0.1cm}=&\hspace{-0.1cm}\frac{20he^{2\sigma}\sigma'}{\sqrt{g_1}}-\frac{e^{-\frac{\sigma}{2}}}{\sqrt{g_1}}\left[\frac{C}{3\sqrt{2}}-16he^{\frac{5\sigma}{2}}\right]\hspace{-0.1cm}\left[\frac{g'_2}{g_2}-\sigma'\right]-\frac{2\widetilde{g}_1e^{\frac{\sigma}{2}}}{\sqrt{g_2}}\widehat{A}_1\mathbf{F}_1\qquad\label{SABPS3}
\end{eqnarray}
while the off-diagonal antisymmetric components $\mathcal{A}^{\bar{x}\bar{y}}-\mathcal{A}^{\bar{y}\bar{x}}=0$ gives
\begin{eqnarray}
0&=&8he^{\frac{5\sigma}{2}}\mathbf{F}_1-\frac{8she^{2\sigma}}{\sqrt{f}}+\frac{\widetilde{g}_1}{\sqrt{g_1g_2}}\left[\frac{f'}{f}-\sigma'\right]\widehat{A}_1,\label{AABPS1}\\
0&=&e^{\frac{\sigma}{2}}\left[\frac{f'}{f}-\sigma'\right]\mathbf{F}_1+\frac{5s}{\sqrt{f}}\sigma',\label{AABPS2}\\
0&=&\frac{e^\sigma}{\sqrt{g_1}}\left[\frac{g'_2}{g_2}+4\sigma'\right]\mathbf{F}_1+\frac{2\widetilde{g}_1}{\sqrt{g_2}}\left[\frac{C}{3\sqrt{2}}-16he^{\frac{5\phi}{2}}\right]\widehat{A}_1\, .\label{AABPS3}
\end{eqnarray}
\indent Matrices $\mathcal{B}^{xy}$ and $\mathcal{C}^{xy}$ are both antisymmetric in $xy$ indices. It turns out that the resulting BPS conditions arising from the combinations $\mathcal{B}^{xy}\pm\mathcal{C}^{xy}=0$ take a simpler form. The combinations $\mathcal{B}^{\bar{x}\bar{y}}+\mathcal{C}^{\bar{x}\bar{y}}=0$ give the following conditions
\begin{eqnarray}
0&=&\frac{e^{\frac{\sigma}{2}}}{\sqrt{g_1}}\left[\frac{f'}{f}-\sigma'\right]\mathbf{F}_1-\frac{s}{\sqrt{f\,g_1}}\left[\frac{g'_2}{g_2}-\sigma'\right]+\frac{8\widetilde{g}_1he^{2\sigma}}{\sqrt{g_2}}\widehat{A}_1,\label{ABCBPS1}\\
0&=&4he^{3\sigma}\mathbf{F}_1+\frac{s}{\sqrt{f}}\left[\frac{C}{3\sqrt{2}}-16he^{\frac{5\phi}{2}}\right],\label{ABCBPS2}\\
0&=&\frac{5\widetilde{g}_1\sigma'}{\sqrt{g_1g_2}}\widehat{A}_1+2\left[\frac{C}{3\sqrt{2}}-12he^{\frac{5\phi}{2}}\right]\mathbf{F}_1\label{ABCBPS3}
\end{eqnarray}
while the combinations $\mathcal{B}^{\bar{x}\bar{y}}-\mathcal{C}^{\bar{x}\bar{y}}=0$ lead to
\begin{eqnarray}
0&=&\frac{2he^{2\sigma}}{\sqrt{g_1}}\left[\frac{f'}{f}-\frac{g'_2}{g_2}\right]+\frac{s\widetilde{g}_1}{\sqrt{f\,g_2}}\widehat{A}_1,\label{SBCBPS1}\\
0&=&\frac{2se^{\frac{\sigma}{2}}}{\sqrt{f}}\mathbf{F}_1-\frac{5\sigma'}{2g_1}\left[\frac{f'}{f}-\sigma'\right]+8he^{\frac{3\sigma}{2}}\left[\frac{C}{3\sqrt{2}}-16he^{\frac{5\phi}{2}}\right],\label{SBCBPS2}\\
0&=&2e^{\sigma}(\mathbf{F}_1)^2-\frac{5\sigma'}{2g_1}\left[\frac{g'_2}{g_2}-\sigma'\right]+8he^{\frac{3\sigma}{2}}\left[\frac{C}{3\sqrt{2}}-16he^{\frac{5\phi}{2}}\right].\label{SBCBPS3}
\end{eqnarray}
There are, in total, fifteen algebraic conditions obtained from the BPS equations of the supergravity multiplet in \eqref{keyBPS1} to \eqref{keyBPS3}.  

Extending this procedure to the BPS equation \eqref{keyBPS4} from vector multiplets gives additional BPS conditions derived from the vanishing of $\mathcal{A}^{x4}$, $\mathcal{B}^{\bar{x}4}$, and $\mathcal{C}^{\bar{x}4}$. The first condition obtained from $\mathcal{A}^{44}=0$ takes the form
\begin{equation}
0=\frac{(\phi')^2}{g_1}+e^\sigma(\mathbf{F}_2)^2-\frac{e^{-\sigma}}{2}(C^{33})^2.\label{A44_BPS_con}
\end{equation}
The symmetric part $\mathcal{A}^{\bar{x}4}+\mathcal{A}^{4\bar{x}}=0$ gives 
\begin{eqnarray}
0&\hspace{-0.1cm}=&\hspace{-0.1cm}16he^{\frac{5\sigma}{2}}\phi'-\sqrt{2}C^{33}\left[\frac{f'}{f}-\sigma'\right],\label{SABPS4}\\
0&\hspace{-0.1cm}=&\hspace{-0.1cm}\frac{16he^{\frac{5\sigma}{2}}\phi'}{\sqrt{g_1}}-\frac{\sqrt{2}C^{33}}{\sqrt{g_1}}\left[\frac{g'_2}{g_2}-\sigma'\right]-\frac{4\widetilde{g}_1e^{\sigma}}{\sqrt{g_2}}\mathbf{F}_2\widehat{A}_1,\label{SABPS5}\\
0&\hspace{-0.1cm}=&\hspace{-0.1cm}\frac{5\sigma'\phi'}{g_1}-\sqrt{2}e^{-\sigma}C^{33}\left[\frac{C}{3\sqrt{2}}-16he^{\frac{5\phi}{2}}\right]+2e^\sigma\mathbf{F}_1\mathbf{F}_2\label{SABPS6}
\end{eqnarray}
while the antisymmetric part $\mathcal{A}^{\bar{x}4}-\mathcal{A}^{4\bar{x}}=0$ results in
\begin{eqnarray}
0&=&\frac{2s\phi'}{\sqrt{f}}+e^{\frac{\sigma}{2}}\mathbf{F}_2\left[\frac{f'}{f}-\sigma'\right],\label{AABPS4}\\
0&=&\frac{2e^{\sigma}\phi'}{\sqrt{g_1}}\mathbf{F}_1+\frac{e^{\sigma}}{\sqrt{g_1}}\left[\frac{g'_2}{g_2}-\sigma'\right]\mathbf{F}_2+\frac{\sqrt{2}\widetilde{g}_1C^{33}}{\sqrt{g_2}}\widehat{A}_1,\label{AABPS5}\\
0&=&C^{33}\mathbf{F}_1-\sqrt{2}\left[\frac{C}{3\sqrt{2}}-16he^{\frac{5\phi}{2}}\right]\mathbf{F}_2\, .\label{AABPS6}
\end{eqnarray}
Moreover, the combinations $\mathcal{B}^{\bar{x}4}+\mathcal{C}^{\bar{x}4}=0$ and $\mathcal{B}^{\bar{x}4}-\mathcal{C}^{\bar{x}4}=0$ respectively lead to the following sets of extra conditions 
\begin{eqnarray}
0&=&\frac{2\widetilde{g}_1\phi'}{\sqrt{g_1g_2}}\widehat{A}_1+\sqrt{2}C^{33}\mathbf{F}_1+8he^{\frac{5\sigma}{2}}\mathbf{F}_2,\label{ABCBPS4}\\
0&=&2\phi'\mathbf{F}_1-5\sigma'\mathbf{F}_2,\label{ABCBPS5}\\
0&=&\frac{\sqrt{2}sC^{33}}{\sqrt{f}}+8he^{3\sigma}\mathbf{F}_2\label{ABCBPS6}
\end{eqnarray}
and
\begin{eqnarray}
0&=&\frac{\phi'}{g_1}\left[\frac{f'}{f}-\sigma'\right]-\frac{2se^{\frac{\sigma}{2}}}{\sqrt{f}}\mathbf{F}_2-4\sqrt{2}he^{\frac{3\sigma}{2}}C^{33},\label{SBCBPS4}\\
0&=&\frac{\phi'}{g_1}\left[\frac{g'_2}{g_2}-\sigma'\right]-2e^{\sigma}\mathbf{F}_1\mathbf{F}_2-4\sqrt{2}he^{\frac{3\sigma}{2}}C^{33},\label{SBCBPS5}\\
0&=&5\sqrt{2}\sigma'C^{33}-4\phi'\left[\frac{C}{3\sqrt{2}}-16he^{\frac{5\phi}{2}}\right].\label{SBCBPS6}
\end{eqnarray}
Finally, the BPS equation \eqref{nBPS3} will be used to determine the explicit form of $\eta(r,z)$. 

We now turn to the vector fields and determine the explicit forms of $A'_1$ and $A'_2$ which are relevant for solving the BPS conditions obtained previously. With only the $SO(2)\times SO(2)$ singlet scalar non-vanishing, we find that $\ast P^{ir}_{(1)}{f_{IJ}}^K{L_r}^IL_{Ki}=0$. Together with $C_{(3)}=0$, the field equation \eqref{Vec_eq} reduces to
\begin{eqnarray}
A''_1&=&-2\phi'A'_2-\left(\frac{5f'}{2f}-\frac{g'_1}{2g_1}-\frac{g'_2}{2g_2}+\sigma'\right)A'_1,\label{SO(2)xSO(2)_Diff_eq1}\\
A''_2&=&-2\phi'A'_1-\left(\frac{5f'}{2f}-\frac{g'_1}{2g_1}-\frac{g'_2}{2g_2}+\sigma'\right)A'_2\, .\label{SO(2)xSO(2)_Diff_eq2}
\end{eqnarray}
The most general solution to these equations is given by 
\begin{eqnarray}
& &A'_1=\frac{e^{-\sigma-2\phi}}{2}(a_1+a_2e^{4\phi})\sqrt{g_1g_2}f^{-\frac{5}{2}}\nonumber \\
\textrm{and} \qquad & & A'_2=\frac{e^{-\sigma-2\phi}}{2}(a_1-a_2e^{4\phi})\sqrt{g_1g_2}f^{-\frac{5}{2}}\label{Gen_SO(2)xSO(2)_Ansatz}
\end{eqnarray}
where $a_1$ and $a_2$ are constants.

%%%%%%%%%%%%%%%%%%%%%%%%%%%%%%%%%%%%%%%%%%%%%%%%%%%%%%%%%%%%%%%%%%%%%%%%%%%%%%%%%%%%%%%%%%%%%%%%%%%%%%%%%%%%%%%%%%%%%%%%%%%%%%%%%%%%%%%%%

\end{document}